\documentclass[fleqn,usenatbib]{mnras}

\usepackage{newtxtext,newtxmath}

\usepackage[T1]{fontenc}
\usepackage{ae,aecompl}

\usepackage{graphicx}	
\usepackage{amsmath}	
\usepackage{amssymb}	

\usepackage{bm, xcolor}
\usepackage[all]{hypcap}
\usepackage{epsfig}
\usepackage{times}

\usepackage{etoolbox}
\makeatletter
\patchcmd\@combinedblfloats{\box\@outputbox}{\unvbox\@outputbox}{}{%
    \errmessage{\noexpand\@combinedblfloats could not be patched}%
}%
\makeatother

\newcommand\Msun{\,\rmn{M}_{\sun}}
\newcommand\Zsun{\,\rmn{Z}_{\sun}}
\newcommand\Gyr{\,\rmn{Gyr}}
\newcommand\Myr{\,\rmn{Myr}}
\newcommand{\yr}{\,\rmn{yr}}
\newcommand{\pyr}{\,\rmn{yr}^{-1}}

\newcommand\kpc{\,\rmn{kpc}}

\newcommand\Mag{\,\rmn{mag}}
\newcommand\Mvir{M_\rmn{vir}}

\newcommand{\K}        {\,{\rm K}}
\newcommand{\cmcubed}  {\,{\rm cm}^{-3}}

\newcommand{\cMpc}     {\,{\rm cMpc}}

\newcommand{\Mcstar}{M_\rmn{c,\ast}}
\newcommand{\Mtoomre}{M_\rmn{T}}

\newcommand{\Lstar}{L^\ast}
\newcommand{\SFR}{\rmn{SFR}}
\newcommand{\SigmaG}{\Sigma_\rmn{gas}}
\newcommand{\SigmaSFR}{\Sigma_\rmn{SFR}}
\newcommand{\MVbright}{M_\rmn{V}^\rmn{brightest}}

\title[YSC populations in E-MOSAICS]{Young star cluster populations in the E-MOSAICS simulations}

\author[J. Pfeffer et al.]{Joel Pfeffer,$^{1}$\thanks{E-mail: \href{j.l.pfeffer@ljmu.ac.uk}{j.l.pfeffer@ljmu.ac.uk}}
Nate Bastian,$^{1}$
J. M. Diederik Kruijssen,$^{2}$
Marta Reina-Campos,$^{2}$ \newauthor
Robert A. Crain$^{1}$
and Christopher Usher$^{1}$
\\
$^{1}$Astrophysics Research Institute, Liverpool John Moores University, 146 Brownlow Hill, Liverpool L3 5RF, UK\\
$^{2}$Astronomisches Rechen-Institut, Zentrum f\"{u}r Astronomie der Universit\"{a}t Heidelberg, M\"{o}nchhofstra\ss e 12-14, 69120 Heidelberg, Germany\\
}

\date{Accepted 2019 September 25. Received 2019 September 24; in original form 2019 May 9}

\pubyear{2019}

\begin{document}
\label{firstpage}
\pagerange{\pageref{firstpage}--\pageref{lastpage}}
\maketitle

\begin{abstract}
We present an analysis of young star clusters (YSCs) that form in the E-MOSAICS cosmological, hydrodynamical simulations of galaxies and their star cluster populations. 
Through comparisons with observed YSC populations, this work aims to test models for YSC formation and obtain an insight into the formation processes at work in part of the local galaxy population.
We find that the models used in E-MOSAICS for the cluster formation efficiency and high-mass truncation of the initial cluster mass function ($\Mcstar$) both quantitatively reproduce the observed values of cluster populations in nearby galaxies.
At higher redshifts ($z \geq 2$, near the peak of globular cluster formation) we find that, at a constant star formation rate (SFR) surface density, $\Mcstar$ is larger than at $z=0$ by a factor of four due to the higher gas fractions in the simulated high-redshift galaxies.
Similar processes should be at work in local galaxies, offering a new way to test the models.
We find that cluster age distributions may be sensitive to variations in the cluster formation rate (but not SFR) with time, which may significantly affect their use in tests of cluster mass loss.
By comparing simulations with different implementations of cluster formation physics, we find that (even partially) environmentally-independent cluster formation is inconsistent with the brightest cluster-SFR and specific luminosity-$\SigmaSFR$ relations, whereas these observables are reproduced by the fiducial, environmentally-varying model. 
This shows that models in which a constant fraction of stars form in clusters are inconsistent with observations.
\end{abstract}




\section{Introduction}

Star clusters are a natural by-product of the star formation process \citep[for recent reviews, see][]{Longmore_et_al_14, Kruijssen_14, Adamo_and_Bastian_18, Krumholz_McKee_and_Bland-Hawthorn_18}. 
Young star clusters (YSCs) are observed in all star-forming galaxies for which they can be resolved \citep[e.g.][]{Larsen_and_Richtler_99}; with the resolving power of the Hubble Space Telescope they can be detected out to distances of $\sim$100 Mpc \citep[e.g.][]{Adamo_et_al_10, Fedotov_et_al_11}.
This observability makes YSCs important tracers of the star formation process in galaxies.
The most massive YSCs are also thought to be young analogues of globular clusters (GCs) \citep*{Portegies-Zwart_McMillan_and_Gieles_10, Kruijssen_15, Forbes_et_al_18_review}, therefore understanding the formation of YSCs may reveal important clues about the formation of GCs.

In recent years, observational studies have established the close connection between the properties of YSC populations and the intensity of star formation in their host galaxies \citep[for a recent review, see][]{Adamo_and_Bastian_18}.
The fraction of stars formed in bound clusters, i.e. the cluster formation efficiency \citep[CFE or $\Gamma$,][]{Bastian_08}, correlates with the star formation rate (SFR) surface density $\Sigma_\rmn{SFR}$ \citep{Goddard_Bastian_and_Kennicutt_10, Adamo_Ostlin_and_Zackrisson_11, Silva-Villa_et_al_13, Adamo_et_al_15, Johnson_et_al_16}.
At the low-mass end of their mass range, YSCs are observed to have a power-law mass function with an exponent $\beta \approx -2$ \citep{Zhang_and_Fall_99, Bik_et_al_03, Gieles_et_al_06b, McCrady_and_Graham_07, Dowell_Buckalew_and_Tan_08, Fall_and_Chandar_12, Baumgardt_et_al_13}. Both of these observations are consistent with being a natural outcome of star formation in a hierarchical gas distribution, with clusters forming in the densest regions of the gas \citep{Elmegreen_and_Efremov_97, Efremov_and_Elmegreen_98, Elmegreen_and_Elmegreen_01, Kruijssen_12}. 
However, at the high-mass end, evidence suggests that clusters form with a high-mass exponential truncation to the power-law mass function ($\Mcstar$) that scales with $\Sigma_\rmn{SFR}$ \citep{Gieles_et_al_06a, Gieles_09, Larsen_09, Portegies-Zwart_McMillan_and_Gieles_10,Johnson_et_al_17}.
The observed relation between the magnitude of the brightest cluster in a population ($\MVbright$) and the SFR of the galaxy \citep*{Billett_Hunter_and_Elmegreen_02, Larsen_02} also implies an upper truncation mass, as it cannot be simply explained by a statistical (size-of-sample) effect with a power-law mass function \citep{Bastian_08}.
Instead, the high-mass end of the initial cluster mass function is likely set by a combination of galactic dynamics and stellar feedback \citep{Reina-Campos_and_Kruijssen_17}.

The dependence of star cluster formation on galactic scale properties means that modelling the formation of realistic star cluster populations also requires modelling the formation and evolution of galaxies and their environment.
In part for this reason, simulations of YSC populations have lagged behind the progress of observations.
For computational reasons, most works modelling YSC populations focus on isolated or merging galaxies in idealised, non-cosmological simulations \citep[e.g.][]{Li_MacLow_Klessen_04, Bournaud_Duc_and_Emsellem_08, Kruijssen_et_al_11, Kruijssen_et_al_12, Renaud_Bournaud_and_Duc_15, Maji_et_al_17}. 
For the same reason, simulations in a cosmological context also largely focus on high-redshift conditions \citep[e.g.][]{Li_et_al_17, Kim_et_al_18}, which does not allow for direct comparisons to present-day galaxies.
Moreover, most studies do not investigate \textit{populations} of galaxies, meaning scaling relations between YSC and galaxy properties generally cannot be compared comprehensively to observations.

In this work we investigate the YSC populations in simulations from the MOdelling Star cluster population Assembly In Cosmological Simulations within EAGLE (E-MOSAICS) project \citep{P18,K19}.
E-MOSAICS couples the MOSAICS model for star cluster formation and evolution \citep{Kruijssen_et_al_11, P18} to the Evolution and Assembly of GaLaxies and their Environments (EAGLE) galaxy formation model \citep{C15,S15}, therefore capturing both the evolution of the galaxies and their environment, as well as the formation and evolution of their star cluster populations.
The E-MOSAICS project aims to test the origin and evolution of GC populations within a YSC-based cluster formation scenario \citep{P18, Pfeffer_et_al_19, Reina-Campos_et_al_18, Reina-Campos_et_al_19, Usher_et_al_18} and the use of star clusters as tracers of galaxy formation and assembly \citep{K19, K19b, Hughes_et_al_19}.
In the fiducial cluster formation model, star cluster populations are fully described by the local, environmentally-varying CFE \citep{Kruijssen_12} and cluster truncation mass \citep{Reina-Campos_and_Kruijssen_17}.
Though the analytical formulations of both models have previously been tested against observations \citep{Kruijssen_12, Silva-Villa_et_al_13, Adamo_et_al_15, Johnson_et_al_16, Reina-Campos_and_Kruijssen_17, Messa_et_al_18_II}, the local formulation of the models and their coupling to hydrodynamical simulations through the MOSAICS model has not been systematically compared with observed YSC populations.
The simulations allow for each component of the model to be switched off, such that their role in the formation of YSC populations, and the variance with galaxy properties, can be assessed.
This work also serves as a means to validate the E-MOSAICS cluster formation model and thereby motivate its application to GC populations.

This paper is structured as follows. 
In Section \ref{sec:methods}, we briefly summarize the E-MOSAICS simulations and introduce a new set of $12.5$ comoving Mpc (cMpc) periodic volumes, for which this paper presents the first results.
In Section \ref{sec:results}, we present the results from the simulations and compare them to observations, for the CFE-$\SigmaSFR$ relation (Section~\ref{sec:CFE}), $\Mcstar$-$\SigmaSFR$ relation (Section~\ref{sec:Mcstar}), power-law indices of the mass functions (Section~\ref{sec:MFslope}), specific luminosities (Section~\ref{sec:TLU}), $\MVbright$-SFR relations (Section~\ref{sec:MVbright}) and cluster age distributions (Section~\ref{sec:age_dist}).
Finally, we summarize and discuss our conclusions in Section \ref{sec:conclusions}.

\section{Methods} \label{sec:methods}

In this section we briefly describe the E-MOSAICS model and simulation suite, the selection of simulated galaxies and their star clusters and the methods for analysing the simulations. A full description of the MOSAICS model, the coupling of MOSAICS to the EAGLE model, along with extensive testing of the subgrid models, is given by \citet{P18}, and the extension to the full suite of 25 zoom-in simulations is presented in \citet{K19}.


\subsection{The E-MOSAICS simulations} \label{sec:EMOSAICS}

The E-MOSAICS project \citep{P18, K19} is a suite of cosmological, hydrodynamical simulations of galaxy formation in the $\Lambda$ cold dark matter cosmogony that couples the MOSAICS model for star cluster formation and evolution \citep{Kruijssen_et_al_11, P18} to the EAGLE model of galaxy formation and evolution \citep{S15, C15}. 
The simulations are run with a highly modified version of the $N$-body, smoothed particle hydrodynamics code \textsc{gadget3} \citep[last described by][]{Springel_05}.
Bound galaxies (subhaloes) were identified within the simulations using the \textsc{subfind} algorithm \citep{Springel_et_al_01, Dolag_et_al_09}, in the same manner as in the EAGLE simulations \citep[for details see][]{S15}.
EAGLE includes subgrid routines describing radiative cooling \citep{Wiersma_Schaye_and_Smith_09}, star formation \citep{Schaye_and_Dalla_Vecchia_08}, stellar evolution and mass loss \citep{Wiersma_et_al_09}, the seeding and growth of black holes (BHs) via gas accretion and BH-BH mergers \citep{Rosas_Guevara_et_al_15}, and feedback associated with star formation and BH growth \citep{Booth_and_Schaye_09}.  
As current cosmological simulations lack the resolution and physics necessary to compute the feedback efficiencies from first principles, the stellar and active galactic nuclei feedback parameters are calibrated such that the simulations of cosmologically representative volumes reproduce the galaxy stellar mass function, galaxy sizes and BH masses at $z \approx 0$.
The EAGLE simulations successfully reproduce a range of galaxy properties, including the stellar masses \citep{Furlong_et_al_15} and sizes \citep{Furlong_et_al_17} of galaxies, their luminosities and colours \citep{Trayford_et_al_15_short}, their cold gas properties \citep{Lagos_et_al_15_short, Lagos_et_al_16,Bahe_et_al_16,Marasco_et_al_16,Crain_et_al_17}, and the properties of circumgalactic and intergalactic absorption systems \citep{Rahmati_et_al_15,Rahmati_et_al_16,Oppenheimer_et_al_16,Turner_et_al_16,Turner_et_al_17}.
The simulations also largely reproduce the cosmic star formation rate density and relation between specific star formation rate and galaxy mass \citep{Furlong_et_al_15}.
The simulations are therefore ideal for comparisons with observed galaxy populations.

In the MOSAICS model, star clusters are treated as a subgrid component of star particles in the simulation \citep{Kruijssen_et_al_11}.
Star clusters are therefore `attached' to star particles, such that they adopt the properties of the host particle (i.e. positions, velocities, ages, abundances).
In a newly formed star particle, a population of star clusters may be formed with properties that depend on the cluster formation model.
The model describes cluster formation in terms of two parameters: the CFE ($\Gamma$) and the high-mass exponential truncation of the \citet{Schechter_76} cluster mass function $\Mcstar$ (with a power-law index of $-2$ at lower masses).
Clusters are drawn from the mass function between masses of $10^2$ and $10^8 \Msun$, while only clusters with masses $>5\times10^3 \Msun$ are evolved to reduce the memory requirements of the simulations.
Each stellar particle forms (statistically) a fraction of its mass in bound clusters ($\Gamma$ times the particle mass).
Thus particles may host clusters more massive than the stellar mass of the particle, and the CFE and cluster mass function are only well sampled for an ensemble of star particles. However, the total cluster and field star mass is conserved on galactic scales (see \citealt{P18} for details).

In the E-MOSAICS suite, we consider four variations of the cluster formation model to assess the importance of each component.
The \textit{fiducial} cluster formation model is fully environmentally dependent. 
The CFE is determined by the local formulation of the \citet{Kruijssen_12} model, which varies as a function of the local natal gas pressure. 
The mass function truncation mass is determined by the local formulation of the \citet{Reina-Campos_and_Kruijssen_17} model, where $\Mcstar$ is related to the local \citet{Toomre_64} mass. 
The model assumes that $\Mcstar$ is proportional to the mass of the molecular cloud from which clusters form \citep{Kruijssen_14}.
As the simulations do not have the necessary physics and resolution to model molecular clouds, their (sub-grid) masses are calculated by assuming the local Toomre mass sets the maximum mass of molecular clouds, which may further decrease due to the effects of stellar feedback.
In the model, the truncation mass generally increases with the natal gas pressure, but decreases in regions with high Coriolis or centrifugal forces (i.e. near the centres of galaxies).

The three other cluster formation model variations then consider environmentally independent versions of the CFE and $\Mcstar$ \citep[see also][]{Reina-Campos_et_al_19}: (i) a constant CFE of $\Gamma = 0.1$ with a pure power-law mass function (i.e. $\Mcstar = \infty$; \textit{no formation physics} model); (ii) an environmentally varying CFE with $\Mcstar = \infty$ (\textit{CFE only} model); (iii) an environmentally dependent $\Mcstar$ with $\Gamma = 0.1$ (\textit{$\Mcstar$ only} model).

The simulations model several channels of mass loss for star clusters, namely stellar evolution, two-body relaxation, tidal shock driven mass loss and complete disruption by dynamical friction \citep[for details, see][]{Kruijssen_et_al_11, P18}.
Stellar evolutionary mass loss for clusters is proportional to that of the host stellar particle, calculated in the EAGLE model \citep{Wiersma_et_al_09}.
The mass loss rate from two-body relaxation is determined by the strength of the local tidal field, which is calculated via the eigenvalues of the tidal tensor at the location of the star particle.
The tidal shock mass loss is also calculated via the tidal tensors, based on the derivations of \citet*{Gnedin_Hernquist_and_Ostriker_99} and \citet{Prieto_and_Gnedin_08}. 
Star clusters that reach a mass below $100 \Msun$ are assumed to be fully disrupted.
Additionally, the removal of star clusters due to dynamical friction is treated in post-processing and applied at every snapshot (though this mechanism is mainly important for massive, old clusters and has little effect on young cluster populations). 

\begin{figure*}
  \includegraphics[width=\textwidth]{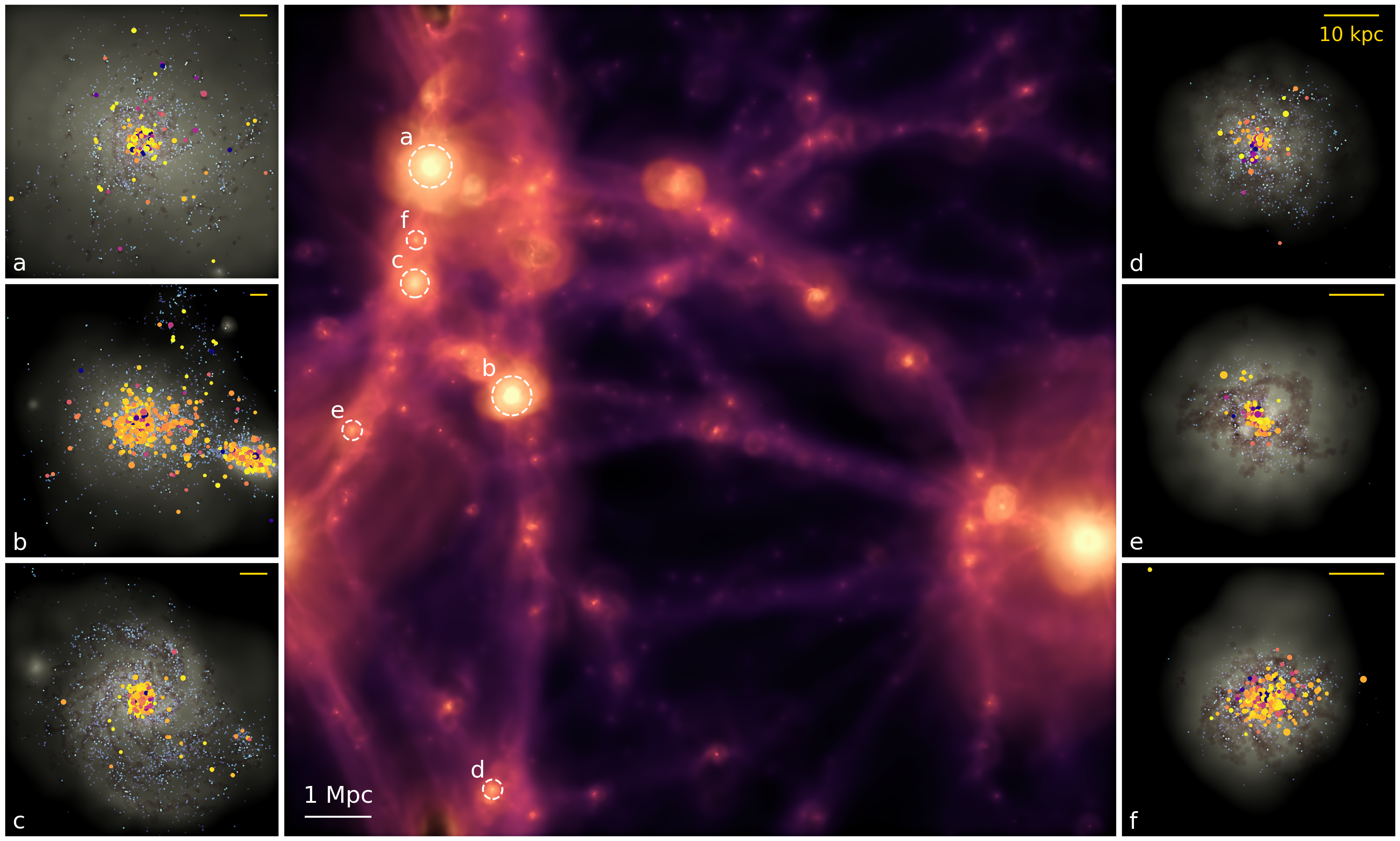}
  \caption{Visualization of the E-MOSAICS L012N0376 simulation at $z=0$. The main panel shows the gas surface density, coloured by temperature, for the entire volume. The side panels show three approximately Milky Way-mass ($M_\ast \approx 1.5$-$6\times 10^{10} \Msun$; left-hand side) and Large Magellanic Cloud-mass galaxies ($M_\ast \approx 5\times 10^{9} \Msun$; right-hand side). The side panels show mock optical images of the galaxies (grey-scale shows stellar density, small light blue points show young star particles, brown shows dense star-forming gas; rotated such that the discs are face on) and the location of young clusters (age $<300\Myr$, initial masses $>5\times 10^3 \Msun$), where symbol colours show cluster age (with dark blue to yellow colours spanning the age range $10^7$-$10^{8.5}$~yr) and symbol areas are proportional to cluster mass. The locations of the galaxies shown in the side panels are indicated in the main panel with dashed circles, where the radii of the circles show the virial radii ($r_{200}$) of the galaxies. The side panels show regions with side lengths of $L=100\kpc$ (left panels) and $50\kpc$ (right panels), with the exception of panel (b), which shows $L=160\kpc$, as the galaxy is undergoing a major merger (stellar masses of $2.8 \times 10^{10}$ and $1.3 \times 10^{10} \Msun$). Scale bars in the upper right corner of the side panels indicate a length of $10 \kpc$.}
  \label{fig:volume}
\end{figure*}

In this work, we use the 25 zoom-in simulations focussed on Milky Way-mass haloes \citep{P18,K19} and a new set of E-MOSAICS simulations of a $12.5 \cMpc$ periodic volume (L012N0376).
All simulations were performed using a \citet{Planck_2014_paperI} cosmology, the `recalibrated' EAGLE model \citep[see][]{S15} and initial baryonic particle masses of $\approx 2.25 \times 10^5 \Msun$.
The 25 zoom-in simulations of Milky Way-mass haloes ($\Mvir \approx 10^{12} \Msun$) were drawn from the high resolution EAGLE simulation of a $25 \cMpc$ volume (Recal-L025N0752) and resimulated in a zoom-in fashion with the E-MOSAICS model \citep[see][]{P18,K19}. 
Ten of these zoom-in simulations were run with all four MOSAICS model variations, while the other fifteen were run only with the fiducial model.
To increase the range of galaxy types and environments (mainly for galaxies with stellar masses $<10^{10} \Msun$), we also performed (for all four model variations) new simulations of a periodic volume with side length $L = 12.5 \cMpc$, using $2 \times 376^3$ particles (i.e. the EAGLE Recal-L012N0376 volume).
The E-MOSAICS L012N0376 volume, and six example galaxies from the fiducial cluster formation model, are visualized in Fig. \ref{fig:volume}.
In general, cluster formation is biased towards the centres of the galaxies ($\lesssim 10\kpc$ for Milky Way-mass galaxies), in regions where the natal gas densities of star formation are highest.


\subsection{Galaxy and star cluster selection} \label{sec:selection}

We select galaxies from bound subhaloes, including both central and satellite galaxies in a halo, from both the periodic volume and zoom-in simulations at redshifts $z=0$, $0.5$, $1$ and $2$.
For the zoom-in simulations, we only consider galaxies that fall within the high-resolution region of the simulations (those that reside in haloes with $<0.1$ per cent contamination by low-resolution particles at any snapshot).

Galaxies (and their bound particles) are then selected from the simulations for analysis in the following way.
\begin{itemize}
\item First, we select star particles within a $30 \kpc$ three-dimensional radius \citep[as for][]{S15} from the centre of potential of the galaxy (i.e. the position of the most bound particle in the subhalo). This focuses the particle selection on the main galaxy and helps to exclude particles being stripped from merging satellites.
Galaxies must have at least 20 star particles within this region, giving a minimum resolved stellar mass of $\approx 4 \times 10^6 \Msun$.
\item Next, galaxies are limited to having at least 10 young ($<300 \Myr$) star particles within a projected radius $R_\rmn{lim}$, where the galaxy is projected such that the disc is face on (using the angular momentum of the star particles).
This selection imposes a minimum total stellar mass of $>10^7 \Msun$ at $z=0$ and $>4\times 10^6 \Msun$ at $z=2$.
We calculate $R_\rmn{lim}$ as the minimum of $1.5 R_{1/2}$ (the projected half-mass radius) and the radius containing 68 per cent of the recent ($<300\Myr$) star formation in the galaxy.
These selections are made in order to approximate the typical footprints for observations of nearby star-forming galaxies \citep[e.g.][]{Adamo_et_al_15, Messa_et_al_18_I} and to limit the projected region such that area-averaged quantities (e.g. $\SigmaSFR$) are focussed on the main star-forming component of each galaxy, respectively.
The latter selection is important in galaxies with very centrally-concentrated star formation.
Note that, because of the scale-free nature of the interstellar medium (ISM) and star formation \citep[e.g.][]{Elmegreen_and_Falgarone_96, Elmegreen_02a}, there is no standard definition for the star-forming area of galaxies \citep[see also][for a discussion on appropriate areas]{Kruijssen_and_Bastian_16}.
Due to the heterogeneous coverage of observed galaxies, it is not possible to match the observational footprints directly \citep[e.g. see fig. 1 of][]{Larsen_02}. 
\item Finally, we select star-forming galaxies based on their specific star formation rate (sSFR) measured within $1.5 R_{1/2}$. 
Following \citet{Bourne_et_al_17}, we use their equation 6 to define the star-forming galaxy sequence as a function of redshift, but set a constant sSFR for galaxies with $M_\ast \leq 10^{10} \Msun$ and apply a vertical shift to lower sSFRs (by setting $b_0 = -10.2$ and $b_1 = 2.3$ in their equation 6) to match the EAGLE main sequence \citep[which predicts slightly lower sSFRs than observed, see][]{Furlong_et_al_15}.
We then select galaxies with sSFRs that do not fall more than $0.5$~dex below the sequence.
At $z=0$ and stellar masses of $M_\ast \leq 10^{10} \Msun$, this selects galaxies with $\mathrm{SFR}/M_\ast > 4 \times 10^{-10} \pyr$.
\end{itemize}

Star clusters are selected in galaxies following the same criteria as for the star particles to which they are attached.
With the exception of the CFE (which is calculated from the total initial mass in clusters) and when fitting initial cluster mass functions, we apply a mass limit for evolved clusters of $M>5\times10^3 \Msun$. Though this limit is necessary due to instantaneous disruption of low-mass clusters in the simulations, it is consistent with those imposed in observations of YSCs in nearby galaxies \citep[depending on distance to the galaxy and the upper age limit for clusters; e.g.][]{Annibali_et_al_11, Adamo_et_al_15, Johnson_et_al_17, Messa_et_al_18_I, Cook_et_al_19}.

For the $z=0$ snapshot, these selection criteria give us a sample of 153 galaxies with stellar masses between $2\times10^7$ and $4\times10^{10} \Msun$ (median $3.7\times10^8 \Msun$), SFRs between $8\times10^{-3}$ and $3 \Msun \pyr$ (median $0.04 \Msun \pyr$) and $\SigmaSFR$ between $10^{-4}$ and $0.3 \Msun \pyr \kpc^{-2}$ (median $2\times10^{-3} \Msun \pyr \kpc^{-2}$).
Of these, 39 galaxies have $>50$ YSCs (ages $<300\Myr$ and initial masses $>5\times10^3 \Msun$) within $R_\rmn{lim}$.


\subsection{Analysis} \label{sec:analysis}

All cluster and SFR-related quantities (SFR, $\SigmaSFR$) are calculated for clusters and star particles with ages $<300 \Myr$ at the time of the relevant snapshot, with the exception of Section \ref{sec:MVbright} (which investigates the $\MVbright$-SFR relation) and Section \ref{sec:age_dist} (which investigates cluster age distributions).
For clusters, this age limit is similar to observational studies for which YSC populations are typically only complete (in mass) below ages of a few hundred megayears (depending on the mass limit).
Star formation rates for the simulated galaxies are calculated directly from the mass in star particles formed over this time period.
Observational studies often use SFR tracers (e.g. H$\alpha$ or UV flux, stellar counts) which are sensitive to timescales $\lesssim 100 \Myr$ \citep{Kennicutt_and_Evans_12, Haydon_et_al_18}.

Projected galaxy quantities ($\SigmaSFR$, $\SigmaG$, $\Sigma_\ast$) are calculated within the surface area given by the projected radius $R_\rmn{lim}$ (i.e. area-weighted surface densities). 
This method follows most observational studies which use the same procedure, though it remains sensitive to the region over which the properties are measured (e.g. particularly if star formation is highly concentrated or substructured; see also \citealt{Johnson_et_al_16}, who apply a mass-weighted method, and Appendix~\ref{app:CFE-P-SigmaSFR}).

In Section \ref{sec:MVbright} we compare YSC properties against the SFR and sSFR of the galaxy. For this comparison, we calculate all properties within $1.5 R_{1/2}$ so as not to bias the sSFR measurement for cases with a very small $R_\rmn{lim}$.

In Sections \ref{sec:Mcstar} and \ref{sec:MFslope} we fit \citet{Schechter_76} mass functions to the simulated YSC populations.
We follow a similar analysis to that used in observational studies \citep[e.g.][]{Johnson_et_al_17, Messa_et_al_18_II}.
For each population, we use the Markov chain Monte Carlo (MCMC) code \textsc{pymc} \citep{PYMC} to sample the posterior probability distribution of the Schechter truncation mass.
For each population we sample the truncation mass in $\log$-space with a uniform prior between $5 \times 10^{3} \Msun$ and $10^{8} \Msun$ (the lowest mass YSC we consider and $\sim$30 times the mass of the most massive YSC at $z=0$ in our study, respectively).
When fitting for the initial cluster masses, we fix the power-law index of the mass function to $\beta = -2$, the input index in the cluster models. 
When fitting for the final (evolved) masses of clusters, we allow the power-law index to vary, and sample the index with a uniform prior between $-3$ and $-0.5$.
For each population, we perform 10,000 iterations with 1,000 burn-in steps.

For Sections \ref{sec:TLU} and \ref{sec:MVbright}, Johnson $U$ and $V$-band luminosities were calculated for clusters assuming simple stellar populations using the clusters' age, metallicity and mass in combination with the Flexible Stellar Population Synthesis (FSPS) model \citep*{Conroy_Gunn_and_White_09, Conroy_and_Gunn_10}, using the MILES spectral library \citep{Sanchez-Blazquez_et_al_06}, Padova isochrones \citep{Girardi_et_al_00,Marigo_and_Girardi_07,Marigo_et_al_08}, a \citet{Chabrier_03} initial stellar mass function and assuming the default FSPS parameters.
Mass-to-light ratios for the clusters were calculated by linearly interpolating the luminosities and relative stellar masses from the grid in ages and total metallicities $\log(Z/\Zsun)$.
Note that we do not include extinction in these estimates, as most observational studies correct for this effect.

\section{Results} \label{sec:results}


\subsection{Cluster formation efficiency} \label{sec:CFE}

\begin{figure}
  \includegraphics[width=84mm]{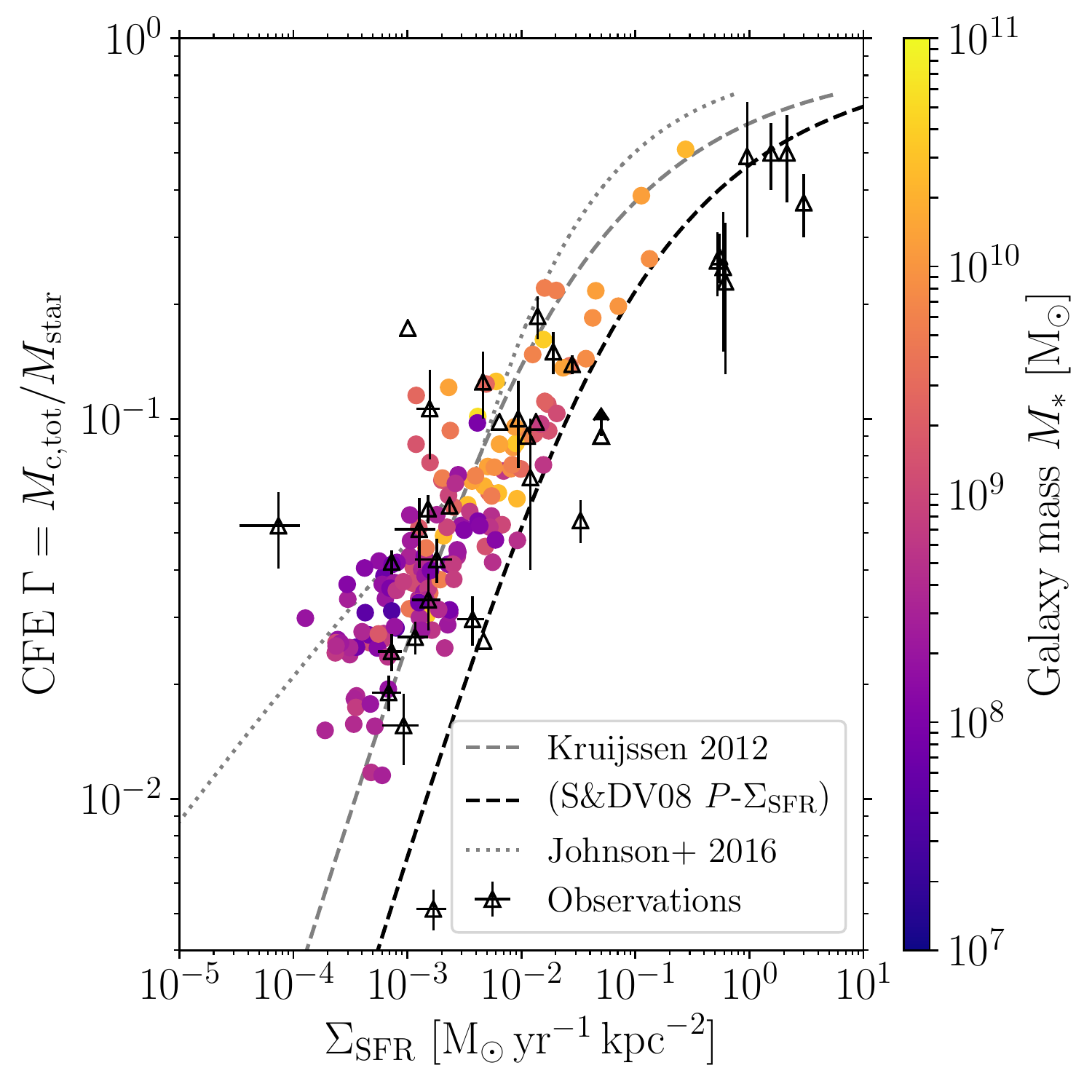}
  \caption{CFE as a function of the star formation rate surface density ($\SigmaSFR$) of the galaxy. For the simulations, each point shows the result for stars and star clusters younger than 300 Myr at $z=0$ for the fiducial cluster formation model. Points are coloured by the stellar mass of the galaxy. Open triangles show the CFEs for observed galaxies (see text). The grey dashed line shows the fiducial prediction of the \citet{Kruijssen_12} model (where $\SigmaSFR$ has been decreased by a factor of $1.65$ to convert from a \citealt{Salpeter_55} to \citealt{Chabrier_03} initial stellar mass function for the Kennicutt--Schmidt relation), while the black dashed line shows the same relation shifted to match the pressure-$\SigmaSFR$ relation adopted in the EAGLE simulations \citep{Schaye_and_Dalla_Vecchia_08, S15}.
The grey dotted line shows the same model but assuming a $\SigmaG$-$\SigmaSFR$ relation based on the \citet{Bigiel_et_al_08} observations \citep{Johnson_et_al_16}.}
  \label{fig:CFE}
\end{figure}

In Fig. \ref{fig:CFE}, we first test the CFE-$\SigmaSFR$ relation in the fiducial cluster formation model for stars and clusters younger than $300 \Myr$ at $z=0$.
Note that this is not strictly a prediction of the simulations, since the \citet{Kruijssen_12} model, for which the galaxy-scale version has previously been tested against observations, was adopted for the CFE model in the simulations. 
However, the section serves as a validation of the implementation in terms of local variables within the E-MOSAICS model and its extension to galaxy-wide scales within the simulations, and enables the testing of effects that may induce scatter in the relation. 
To calculate the CFE in the simulations, we sum the total mass of clusters formed\footnote{For a small number of particles where $\Mcstar < 100 \Msun$ (less than the lower mass for the cluster mass function) we assume that no clusters were formed.} in the star particles in the region of interest (i.e. the CFE at formation which includes any stochasticity in sampling initial cluster masses from the mass function, not the values calculated at the particle level from the natal gas pressure).
This value does not include the effect of cluster mass loss (which will lower the measured value of the CFE) or any observational uncertainties associated with cluster detection and measurement of the SFR. 
For the simulated galaxies, $\SigmaSFR$ is also measured over the same $300 \Myr$ timescale as the CFE.
Our results are not sensitive to the exact age limit used, as we find consistent results\footnote{Though with significantly fewer galaxies in the $1$-$10 \Myr$ age range, since this range is generally poorly resolved due to the resolution limits of the simulations.} when using ranges of $1$-$10 \Myr$, $1$-$100\Myr$ and $10$-$100 \Myr$.

Fig. \ref{fig:CFE} shows the galaxy-averaged CFE for all galaxies in the simulations with the fiducial cluster formation model.
For comparison, in the figure we also include observed CFEs from \citet{Goddard_Bastian_and_Kennicutt_10}, \citet{Adamo_Ostlin_and_Zackrisson_11}, \citet{Annibali_et_al_11}, \citet{Silva-Villa_and_Larsen_11}, \citet[using the results for ages $<100 \Myr$ and excluding galaxies with only upper limits for the CFE]{Cook_et_al_12}, \citet{Ryon_et_al_14}, \citet{Adamo_et_al_15}, \citet{Hollyhead_et_al_16}, \citet{Johnson_et_al_16}, \citet{Ginsburg_and_Kruijssen_18} and \citet{Messa_et_al_18_II}.
Note that the CFE for NGC 4449 is a lower limit \citep{Annibali_et_al_11}.
The simulations show a similar level of scatter in the CFE at a given $\SigmaSFR$ ($\sim0.25$ dex) to measurements of observed galaxies.
In the figure there is a galaxy mass gradient along the CFE-$\SigmaSFR$ relation, such that the more massive galaxies generally have a higher $\SigmaSFR$ and CFE.
This result is expected as, assuming pressure equilibrium in the galaxies, larger galaxy masses (therefore deeper potentials) result in higher characteristic ISM pressures, and thus higher $\SigmaSFR$ and CFE.
However, it is important to note that the volumes of the simulations are not large enough to capture rare objects, such as rapidly star-forming dwarf galaxies with high star-formation densities and CFEs \citep[e.g. blue compact dwarfs,][]{Adamo_Ostlin_and_Zackrisson_11}.
Additionally, the selection in sSFR for star-forming galaxies (Section \ref{sec:selection}) selects against high-mass galaxies with low star-formation densities and CFEs.

For the projected version of the \citet{Kruijssen_12} model (dashed lines in the figure), the gas surface density ($\SigmaG$, which is the fundamental quantity in the model setting the CFE, see e.g.\ \citealt{Kruijssen_and_Bastian_16} and \citealt{Ginsburg_and_Kruijssen_18}) is converted to a SFR surface density assuming the Kennicutt--Schmidt star formation relation \citep{Kennicutt_98}\footnote{Note that for consistency with the simulations, we adopt $\SigmaSFR = 1.515 \times 10^{-4} \Msun \pyr \kpc^{-2} \, (\SigmaG/1 \Msun \kpc^{-2})^{1.4}$, consistent with a \citet{Chabrier_03} initial stellar mass function \citep[see][]{S15}.}.
Additionally, star formation in EAGLE is implemented following the Kennicutt--Schmidt relation, rewritten as a pressure law \citep{S15}. 
Therefore, the naive expectation is that the simulations should broadly reproduce the (shifted) \citet{Kruijssen_12} relation in Fig. \ref{fig:CFE} (black dashed line), where the relation has been shifted to higher $\SigmaSFR$ by $\approx 0.6$~dex to account for the change from the \citet{Krumholz_and_McKee_05} $P$-$\SigmaSFR$ relation to the \citep{Schaye_and_Dalla_Vecchia_08} relation used in EAGLE.
At $\SigmaSFR \gtrsim 5\times 10^{-3} \Msun \pyr \kpc^{-2}$, the simulated galaxies broadly follow the expected relation (black dashed line), but are generally shifted to slightly lower $\SigmaSFR$, with most galaxies falling between the fiducial (grey dashed line) and shifted (black dashed line) \citet{Kruijssen_12} relations.
For the modelling in this work, the fundamental relation is between the CFE and natal gas pressure, and therefore some amount of uncertainty in the CFE-$\SigmaSFR$ relation arises simply due to the uncertainty in the $P$-$\SigmaSFR$ relation.
We discuss this point in further detail in Appendix \ref{app:CFE-P-SigmaSFR}, where we show that the offset from the expected CFE-$\SigmaSFR$ relation is due to galaxies being offset from the expected $P$-$\SigmaSFR$ (see also the discussion below).

At lower surface densities ($\SigmaSFR \lesssim 5 \times 10^{-3} \Msun \pyr \kpc^{-2}$), the simulations show a higher CFE at a given $\SigmaSFR$ than the \citet{Kruijssen_12} relation, for which the cause may be multifold.
Firstly, this can be caused by highly concentrated or substructured star formation, such that star and cluster formation largely occurs in a much smaller area compared to the area for which $\SigmaSFR$ is calculated, which will lower the measurement for $\SigmaSFR$ at a given CFE.
Secondly, at low $\SigmaSFR$ there is a physical effect that increases the CFE at lower metallicities (i.e. in lower mass galaxies) due to the metallicity-dependent density threshold for star formation implemented in EAGLE\footnote{Note that this effect of increasing CFE with metallicity is not expected to occur at high $\SigmaSFR$, since the densities of star-forming gas in this regime are well above the metallicity-dependent threshold.}.
This threshold is included to model the effect of the thermogravitational collapse of warm, photoionized interstellar gas into a cold, dense phase, which is expected to occur at lower densities and pressures in metal-rich gas \citep{Schaye_04}.
This higher density threshold at lower metallicities results (through the lower density limit for star formation imposed by the polytropic equation-of-state implemented at high gas densities) in higher pressures of star formation at a given $\SigmaSFR$, and therefore in higher CFEs \citep[see fig. 3 in][]{P18}.
Finally, variations in the natal pressure-$\SigmaSFR$ relation in the galaxies will, in turn, lead to variations in the CFE-$\SigmaSFR$ relation, through the dependence of the CFE on natal gas pressure. 
Such variations can be driven by random fluctuations within the galaxies (which may be most important at low $\SigmaSFR$), or physical variations due to differing contributions of the gravity of stars to the mid-plane gas pressure in galaxies (i.e. variations in $\phi_P$, see also Appendix~\ref{app:phiP}).
We test the importance of these effects in Appendix \ref{app:CFE-P-SigmaSFR}, finding the dominant effect to be the use of too large an area in the calculation for $\SigmaSFR$ (i.e. $\SigmaSFR$ is systematically underestimated).
This effect may be mitigated by calculating a mass-weighted surface density \citep[see][and Appendix~\ref{app:CFE-P-SigmaSFR}]{Johnson_et_al_16}, or by judicious aperture choice, focussing on the main region of star formation.
Since most studies apply the standard area-weighted calculations, we focus on that method in this paper.

\begin{figure}
  \includegraphics[width=84mm]{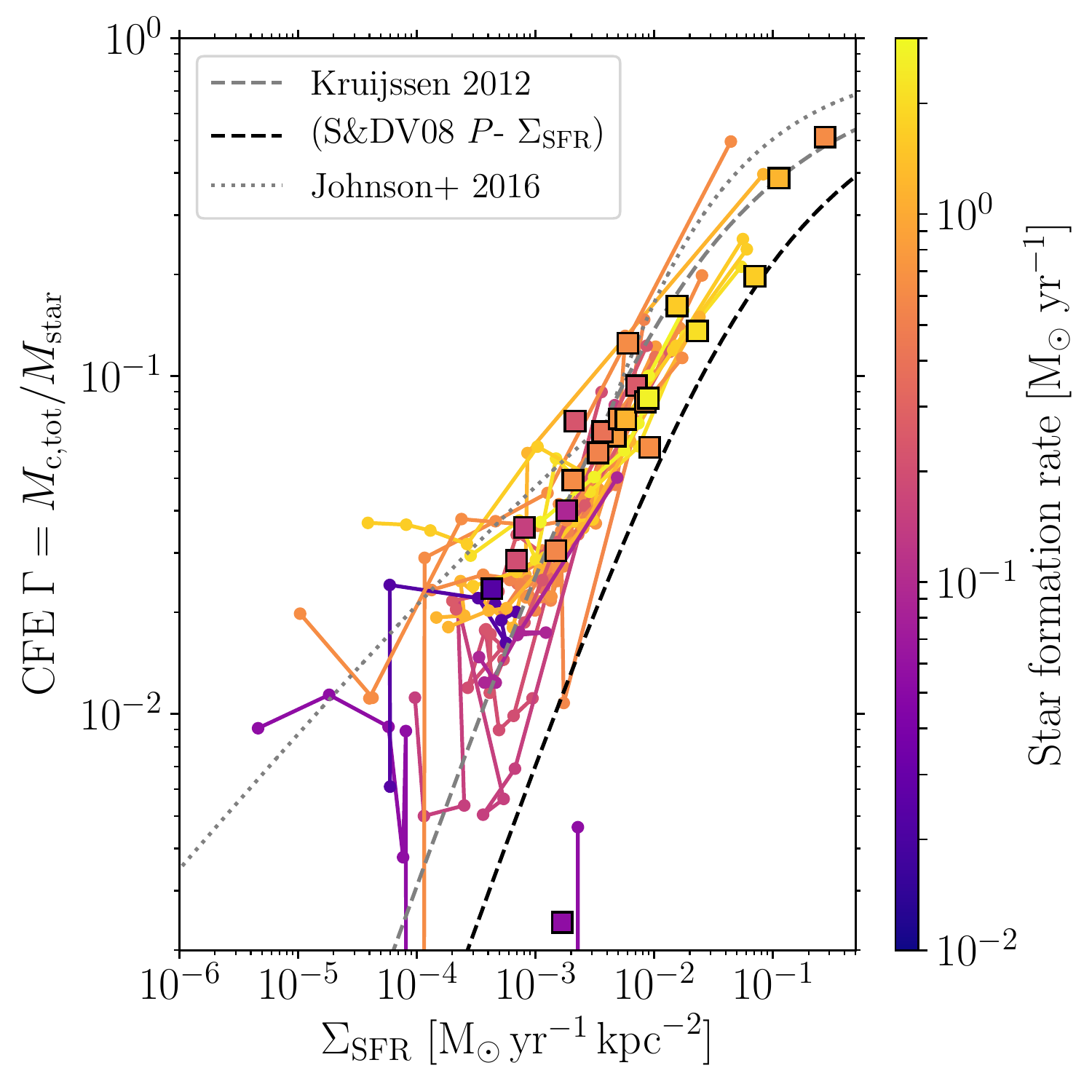}
  \caption{Radial CFE distributions (solid lines) in 2 kpc annuli, to a maximum of 16 kpc, for the target $\Lstar$ galaxies in 23 of the 25 zoom-in simulations. Each galaxy in the figure is coloured by its star formation rate. The squares show the values of the same galaxies measured within $R_\rmn{lim}$ (as in Fig. \ref{fig:CFE}). Dashed and dotted lines show the same relations as in Fig. \ref{fig:CFE}.}
  \label{fig:CFE_radius}
\end{figure}

The variation of the CFE at a given $\SigmaSFR$ can be further investigated by comparing the radial CFE distributions within the galaxies. 
In Fig. \ref{fig:CFE_radius} we show the radial CFE distributions in $2 \kpc$ annuli in 23 of the 25 $\Lstar$ galaxies (Milky Way-mass haloes; MW16 and MW22 are quenched and do not have young clusters, thus are excluded from the figure) from the zoom-in simulations \citep{K19}.
For this figure, we have not applied the limit on sSFR for the galaxies (Section \ref{sec:selection}) in order to sample a wide range of galactic environments.
The majority of measurements in the radial distributions fall along the \citet{Kruijssen_12} relation (as expected), and galaxies with higher SFRs generally show higher CFE and $\SigmaSFR$. 
However at low $\SigmaSFR$ ($\lesssim 10^{-3} \Msun \yr^{-1} \kpc^{-2}$), the simulations again show significant scatter from the \citet{Kruijssen_12} relation, with most points falling between the fiducial relation and the reinterpreted relation from \citet[which uses the Kruijssen model, but assumes a $\SigmaG$-$\SigmaSFR$ relation based on the observations of \citealt{Bigiel_et_al_08}, rather than the Kennicutt-Schmidt relation]{Johnson_et_al_16}. 
This deviation can be attributed to $\SigmaSFR$ being averaged over a larger area than for which star and cluster formation is occurring and variations in the natal pressure-$\SigmaSFR$ relation, since the natal pressure is approximately constant at a given CFE (see also Appendix \ref{app:CFE-P-SigmaSFR}).
Similarly, $\SigmaSFR$ for the innermost radial bin in MW13 (at $\Gamma \approx 0.5$) deviates significantly from both the `global' value (square symbol) due to very central star formation that dominates the cluster formation in the galaxy (for this galaxy $R_\rmn{lim} = 0.75 \kpc$).
One galaxy, MW05, has a CFE that is significantly below other galaxies at $\SigmaSFR \approx 2 \times 10^{-3} \Msun \pyr \kpc^{-2}$. 
The galaxy has a very low median cluster truncation mass at $z=0$ of $\Mcstar \approx 100 \Msun$, meaning many star particles with $\Mcstar < 100 \Msun$ form no clusters.
This is caused by the low natal gas pressure for star formation ($P/k < 10^{4} \K \cmcubed$) and a high stellar density (high $\phi_P$, see the discussion in Section \ref{sec:Mcstar}) in the galaxy at that epoch.


\subsection{Mass function truncation} \label{sec:Mcstar}

\begin{figure}
  \includegraphics[width=84mm]{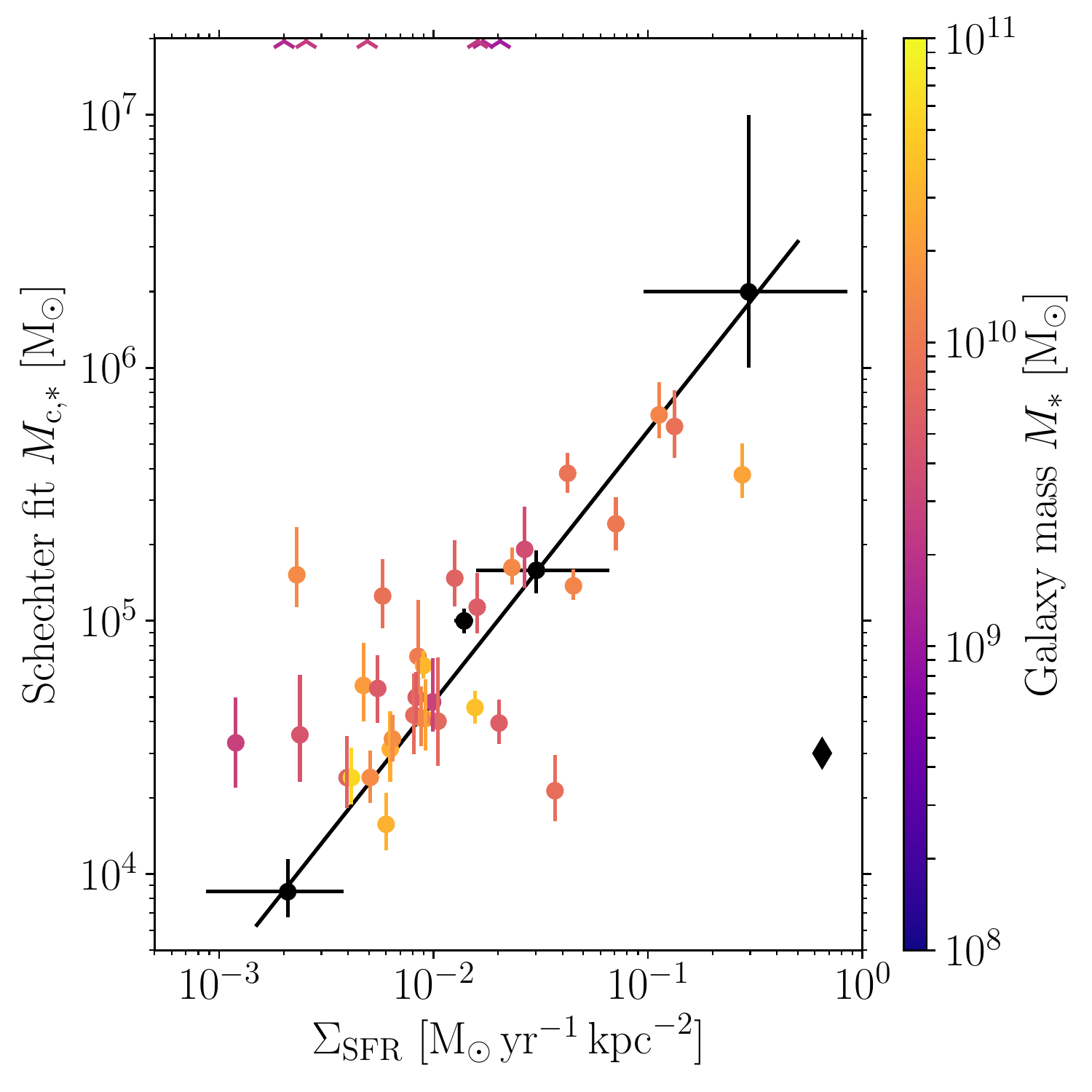}
  \caption{Mass function truncation $\Mcstar$ as a function of $\SigmaSFR$ for the simulated galaxies with $>50$ clusters younger than $300 \Myr$ at $z=0$. The points show the median of the posterior distribution from the MCMC fit, coloured by the galaxy stellar mass, while errorbars show the 16 and 84 per cent confidence intervals. Arrows at the top of the figure indicate galaxies for which $\Mcstar$ was unable to be constrained. Black points show the fits to observed cluster populations in the Antennae system \citep{Zhang_and_Fall_99, Jordan_et_al_07_XII}, M31 \citep{Johnson_et_al_17}, M51 \citep{Gieles_09, Messa_et_al_18_II} and M83 \citep{Adamo_et_al_15}. The solid line shows the relation fit to the observations by \citet{Johnson_et_al_17}. The black diamond shows the best-fitting truncation mass of YSCs in the Central Molecular Zone (CMZ) of the Milky Way \citep{Trujillo-Gomez_et_al_19} versus its $\SigmaSFR$ \citep{Barnes_et_al_17}, demonstrating that the empirical relation from \citet{Johnson_et_al_17} is not fundamental, but must have an additional dependence, most likely on the epicyclic frequency as in \citet{Reina-Campos_and_Kruijssen_17}. This decreases $\Mcstar$ towards galactic centres.}
  \label{fig:Mcstar}
\end{figure}

In this section we test the model for the upper exponential truncation of the cluster mass function ($\Mcstar$) implemented in the MOSAICS models \citep{Reina-Campos_and_Kruijssen_17}. 
As for the CFE, this is not strictly an independent prediction since the galaxy-scale version of the model has previously been tested against observations \citep{Reina-Campos_and_Kruijssen_17, Messa_et_al_18_II}, but it serves as a test and validation of the implementation in terms of local variables within the E-MOSAICS model. 
However, we also provide predictions at high redshifts which may be tested with future observations.

Using the fitting procedure described in Section \ref{sec:analysis}, for each galaxy with $>50$ clusters we fit a \citet{Schechter_76} mass function with an upper exponential truncation mass $\Mcstar$ to the initial masses of clusters younger than $300 \Myr$, using a fixed low-mass index of $\beta = -2$ (i.e. the input value used in the simulations).
In Appendix \ref{app:init_v_final}, we compare the resulting $\Mcstar$ for fitting initial cluster mass functions (with a fixed power-law index) and final (evolved) cluster mass functions (with a variable power-law index).
We find that both methods generally give consistent measurements for $\Mcstar$, with potentially a small offset to higher initial $\Mcstar$ due to stellar evolutionary mass loss (a factor of $\sim0.1$~dex).

Following the fit, we exclude galaxies for which $\Mcstar$ is larger than the most massive cluster in the population.
In such cases, $\Mcstar$ is poorly constrained since cluster formation does not fully sample up to the truncation mass. 
For galaxies with $\Mcstar < \max(M)$, fits typically have $1\sigma$ uncertainties of $<0.5$ dex.
For $\Mcstar > \max(M)$ uncertainties reach up to $\sim2$ dex, even for populations with $>100$ clusters.
Due to the initial cluster mass limit ($5 \times 10^3 \Msun$), truncation masses are typically only able to be fit above masses of a few times $10^4 \Msun$.
This limit also biases the results to galaxies ($M_\ast \gtrsim 10^9 \Msun$ at $z=0$) which have a large enough population of YSCs above the mass limit to fit a mass function.
At $z=\{0,0.5,1,2\}$ this leaves us with a sample of $\{33,51,65,60\}$ galaxies.

In Fig. \ref{fig:Mcstar} we compare the fitted $\Mcstar$ for the simulated galaxies at $z=0$ with results from observed nearby galaxies (described in caption).
The predicted $\Mcstar$ for the simulated galaxies are in good agreement with the observed galaxies, falling about the relation described by the observations \citep{Johnson_et_al_17} over the same range in $\SigmaSFR$.
More observations are needed to test whether the scatter found in $\Mcstar$ for the simulated YSCs is consistent with observed YSC populations, which is possible with (e.g.) the LEGUS survey \citep{Calzetti_et_al_15} and the PHANGS-HST survey (Lee et al., in prep.).

\begin{figure*}
  \includegraphics[width=\textwidth]{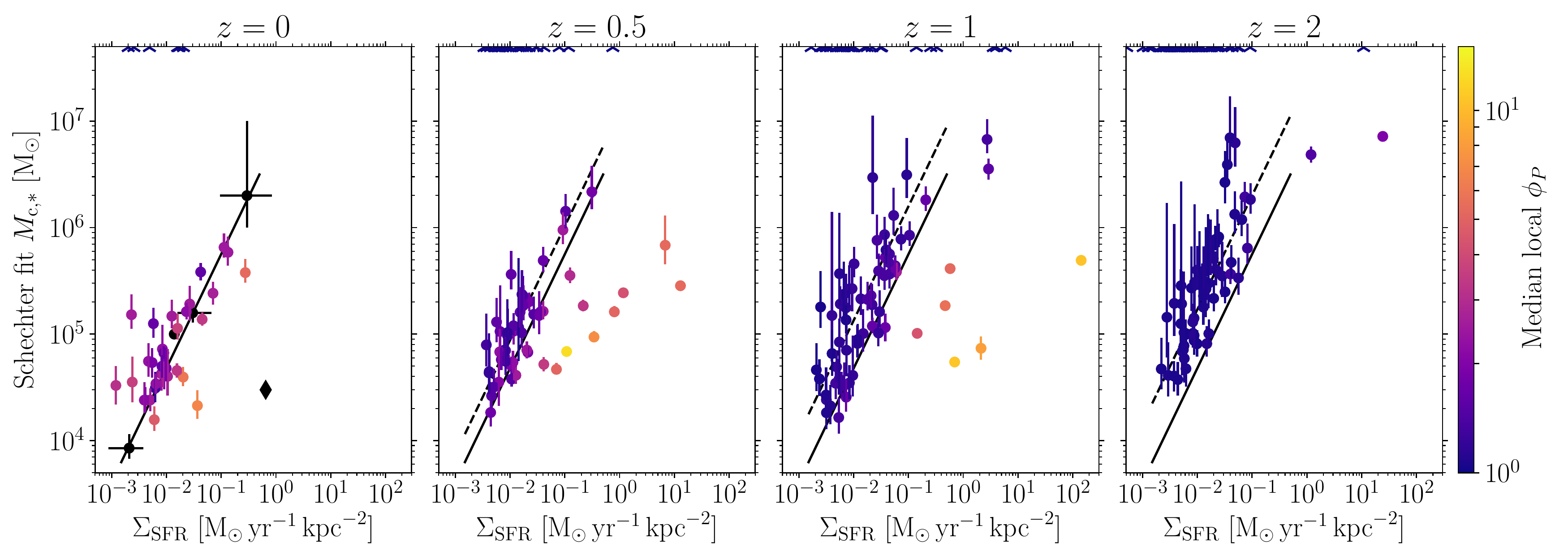}
  \caption{Mass function truncation $\Mcstar$ as a function of $\SigmaSFR$ at redshifts $z={0,0.5,1,2}$ (panels from left to right), coloured by the median $\phi_P$ of star particles younger than $300 \Myr$. Arrows at the top of the panels indicate galaxies for which $\Mcstar$ was unable to be constrained. Black points show observed galaxies at $z=0$ for reference (as in Fig. \ref{fig:Mcstar}). Dashed lines show the fit to the observed galaxies rescaled assuming the median $\phi_P$ at each redshift found for the simulations (see Appendix \ref{app:phiP}).}
  \label{fig:Mcstar_redshift}
\end{figure*}

Overall, the \citet{Reina-Campos_and_Kruijssen_17} model for $\Mcstar$, and its implementation in terms of local gas and dynamical properties in the E-MOSAICS cluster formation model, performs well in reproducing YSC populations in realistic galaxy formation simulations at $z=0$.
Therefore we can be confident in extending the formation model to more extreme environments, such as to the epochs of GC formation.
In Fig. \ref{fig:Mcstar_redshift} we perform the same comparison of $\Mcstar$ against $\SigmaSFR$ for simulated galaxies at redshifts of $z = \{0,0.5,1,2\}$. 
For reference, we also show the measurements from observed galaxies at $z=0$ in each panel. 
At a given $\SigmaSFR$ or $\Mcstar$, the typical stellar mass of galaxies forming clusters decreases with increasing redshift, implying that galaxies of a given stellar mass (at that epoch) can form higher mass clusters in the early universe compared to $z=0$.

At $z=0.5$, the $\Mcstar$-$\SigmaSFR$ distribution is similar to $z=0$. 
Due to the larger sample of galaxies at this snapshot, the distribution extends to higher $\Mcstar$ and $\SigmaSFR$, comparable with that found for the Antennae galaxies \citep[$\Mcstar \sim 2 \times 10^6 \Msun$,][]{Zhang_and_Fall_99, Jordan_et_al_07_XII}.
In fact, the simulated galaxy at $z=0.5$ closest to the Antennae measurement is one of a pair of galaxies about to undergo a major merger (with stellar masses $\sim10^{10} \Msun$), which are separated by $\approx10 \kpc$ at the time of the snapshot.
Due to the infrequency of the snapshots (we output 29 between $z=20$ and $z=0$), catching a galaxy merger during its peak is extremely unlikely.
At higher redshifts of $z=\{1,2\}$, near the peak of GC formation for $\Lstar$ galaxies in the E-MOSAICS model \citep[$z \sim 1$-$4$,][]{Reina-Campos_et_al_19}, we find in the simulated galaxies that $\Mcstar$ is higher at a given $\SigmaSFR$ than the relation at $z=0$. 
Therefore, to reach a given $\Mcstar$, galaxies require a lower $\SigmaSFR$ in the early universe compared to today (by $\sim$0.5 dex at $z=2$).

This increase in $\Mcstar$ is caused by two effects. 
At late times, a higher contribution of the gravity of stars to the mid-plane gas pressure (i.e. $\phi_P$, \citealt{Elmegreen_89, Krumholz_and_McKee_05}; equations 7 and 8 in \citealt{P18}) results in a lower gas surface density (and therefore Toomre mass) at a given pressure. 
Additionally, the density threshold for star formation increases with decreasing metallicity in the EAGLE model (which is mainly has an effect at $\SigmaSFR < 10^{-2} \Msun \pyr \kpc^{-2}$, see the discussion in Section \ref{sec:CFE}), thus resulting in a higher $\Mcstar$.
We further discuss the effect of $\phi_P$ in Appendix \ref{app:phiP}, finding that the $\phi_P$ increases from $\phi_P \approx 1$ at $z=2$ to $\phi_P \approx 2.5$ at $z=0$.
In Fig. \ref{fig:Mcstar_redshift}, we show the effect of decreasing $\phi_P$ at higher redshifts by rescaling the fit to observed local galaxies at $z=0$ assuming the median $\phi_P$ from the simulations at each redshift (dashed line).
The simulated galaxies agree well with the rescaled relations at each redshift, demonstrating the effect of $\phi_P$ on $\Mcstar$.
Note that as $\phi_P \approx 1$ at $z=2$ (right panel in Fig. \ref{fig:Mcstar_redshift}), galaxies at $z>2$ should simply follow the $z=2$ relation, since $\phi_P$ cannot be less than unity.
This result could be tested in galaxies from the local Universe by comparing cluster formation in regions of high stellar density (high $\phi_P$) and low stellar density (low $\phi_P$) at similar $\SigmaSFR$.

In Fig. \ref{fig:Mcstar_redshift} (particularly evident at $z=\{0.5, 1\}$), a number of galaxies fall well below the present-day $\Mcstar$-$\SigmaSFR$ relation, approaching the value found for the CMZ in the Milky Way \citep{Trujillo-Gomez_et_al_19}. 
This is driven by two effects, both due to star formation in regions of high stellar density (i.e. very central star formation in the galaxy). 
Firstly, such galaxies have a decreased $\Mcstar$ due to the higher contribution of the gravity of stars to the mid-plane gas pressure (high $\phi_P$, as discussed above; see Fig. \ref{fig:phiP-SigmaSFR}).
However, a high stellar density (high $\phi_P$) alone is not sufficient to account for the large decrease in $\Mcstar$.
For example, at $z=0$ a number of galaxies are significantly elevated in $\phi_P$ ($\phi_P > 3$, Fig. \ref{fig:phiP-SigmaSFR}), but fall along the present-day $\Mcstar$-$\SigmaSFR$ relation (Fig. \ref{fig:Mcstar}).
The main contributing factor is due to the high Coriolis or centrifugal forces in the region of star formation\footnote{A high $\phi_P$ does not imply high Coriolis/centrifugal forces, but high Coriolis/centrifugal forces generally occur in regions with high $\phi_P$.}, resulting in decreased Toomre masses, and therefore $\Mcstar$.
This is captured by the \citet{Reina-Campos_and_Kruijssen_17} model in terms of a dependence on the epicyclic frequency, which is not accounted for in the simple scaling with $\Sigma_{\rm SFR}$ from \citet{Johnson_et_al_17}.
Though this effect is most evident at $z=\{0.5, 1\}$ in the simulations, it may occur at any redshift and simply results from very central star formation.
Such an effect has been observed at the centre of local Universe galaxies \citep{Reina-Campos_and_Kruijssen_17, Messa_et_al_18_II, Trujillo-Gomez_et_al_19}.


\subsection{Mass function index} \label{sec:MFslope}

\begin{figure}
  \includegraphics[width=84mm]{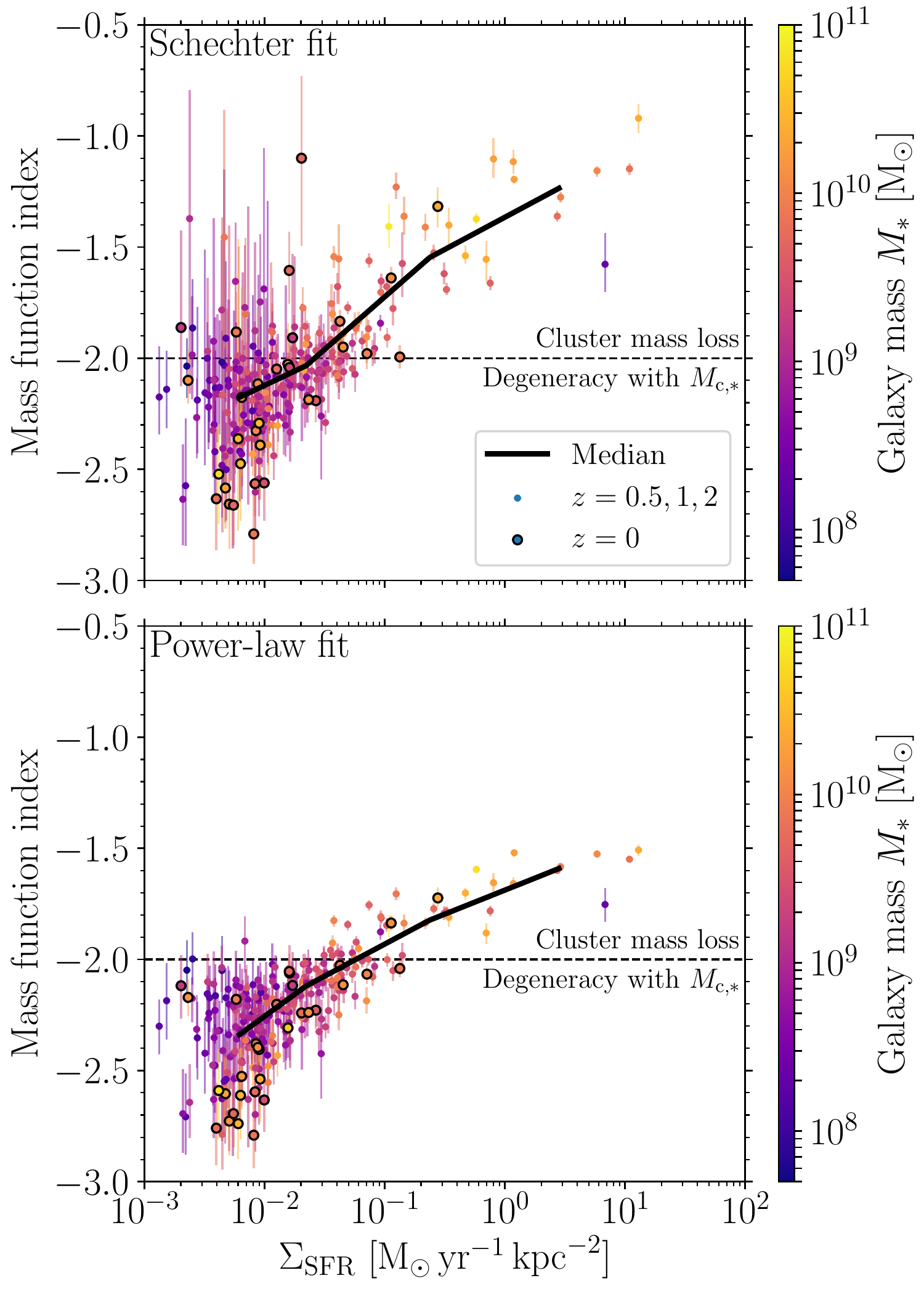}
  \caption{Power-law index of the final cluster mass function, fit with a \citealt{Schechter_76} function (upper panel) and a power-law function (lower panel), as a function of the star formation rate density of the galaxy, $\SigmaSFR$. The points show the median of the posterior distribution from the MCMC fit, while errorbars show the 16 and 84 per cent confidence intervals. Galaxies from the $z=\{0,0.5,1,2\}$ snapshots are included in the figure, with galaxies at $z=0$ highlighted with black circles. The black lines show the median mass function index and $\SigmaSFR$ in $1$~dex intervals from $10^{-3}$ to $10^{2} \Msun \pyr \kpc^{-2}$. Initially, all clusters are drawn from a Schechter function with index $\beta=-2$ (dashed line in the figure). The flatter mass functions at higher star formation rate densities are caused by dynamical mass loss of the clusters, while indices may be steeper than the initial value at low $\SigmaSFR$ due to the degeneracy between the index and mass function truncation.}
  \label{fig:MF_slope}
\end{figure}

\citet{Cook_et_al_16} found that galaxies with higher star formation rate surface densities tend to have flatter cluster luminosity function indices. They suggest that this might be a result of the cluster formation process, with higher star formation efficiencies resulting in proportionally more massive star-forming regions.
In this section, we investigate how other effects, namely increased relative mass loss towards low cluster masses or the degeneracy between the power-law index and $\Mcstar$, may instead cause this effect. 
Following the method described in Section \ref{sec:analysis}, we fit Schechter and power-law mass functions to the final (evolved) cluster populations, using a variable mass function index with a uniform prior of $-3 < \beta < -0.5$ \citep[similar to][]{Johnson_et_al_17, Messa_et_al_18_II}.

In Fig. \ref{fig:MF_slope} we compare the cluster mass function index with the star formation rate density of the galaxy, for galaxies in the $z=\{0,0.5,1,2\}$ snapshots (in order to increase the galaxy sample and extend the range in $\SigmaSFR$).
The upper panel shows the results for Schechter function fits, while the lower panel shows the results for pure power-law fits.
The figure includes all galaxies with $>50$ clusters with evolved masses $>5\times10^3 \Msun$ (regardless of how well $\Mcstar$ is constrained in the case of Schechter fits), and therefore includes the effect of the degeneracy between $\beta$ and $\Mcstar$ in the fits.
We find that the cluster mass function indices are flatter at higher star formation rate densities, similar to the effect found by \citet{Cook_et_al_16}.
This result is true for both Schechter and power-law fits, though the latter tend to find steeper mass function indices.
In the simulations this is caused by two effects. 
At high star formation rate densities ($\SigmaSFR \gtrsim 10^{-1.5} \Msun \pyr \kpc^{-2}$), the mass function indices become flatter due to increased relative mass loss towards low cluster masses, primarily due to tidal shocks from dense gas.
At low star formation rate densities ($\SigmaSFR \lesssim 10^{-1.5} \Msun \pyr \kpc^{-2}$), mass function indices may appear steeper due to low truncation masses and the degeneracy between the index and the truncation mass (galaxies with $\beta < -2$ generally have poorly constrained $\Mcstar$).
This crossing point, where galaxies typically fall above or below an index of $-2$, depends on the lower cluster mass limit; higher or lower mass limits result in higher or lower crossing points in $\SigmaSFR$, respectively.
When using a lower cluster mass limit of $10^4 \Msun$, rather than $5\times10^3 \Msun$, the crossing point shifts to higher $\SigmaSFR$ by $\approx0.3$~dex.
Observed local galaxies at low $\SigmaSFR$ ($\sim10^{-3}$-$10^{-2} \Msun \pyr \kpc^{-2}$) are also consistent with the power-law indices for the cluster mass function found in this work ($\beta \approx -2$, e.g. M31, \citealt{Johnson_et_al_17}; M51, \citealt{Messa_et_al_18_I}).

The mass function indices at a given $\SigmaSFR$ for $z=0$ and $z>0$ galaxies are generally consistent. 
However, at $\SigmaSFR \lesssim 10^{-2} \Msun \pyr \kpc^{-2}$, galaxies at $z=0$ tend to have steeper indices due to their smaller $\Mcstar$ (Section \ref{sec:Mcstar}).

Since their methods differ from ours (they fit luminosity functions and have a variable lower cluster luminosity limit between galaxies), we cannot make a direct quantitative comparison to the results from \citet{Cook_et_al_16}.
Additionally, the simulations and observations are largely biased to different star formation rate densities ($\SigmaSFR \gtrsim 10^{-2.5} \Msun \pyr \kpc^{-2}$ and $\SigmaSFR \lesssim 10^{-2} \Msun \pyr \kpc^{-2}$, respectively).
However, the results from the simulations suggest that the relation between cluster luminosity function index and $\SigmaSFR$ found by \citet{Cook_et_al_16} can be explained by the degeneracy between the mass function index and truncation $\Mcstar$ (at low $\SigmaSFR$).
Similar measurements of the mass function index should be extended to higher star formation rate densities to assess and test the impact of cluster mass loss.


\subsection{Specific $U$-band cluster luminosity} \label{sec:TLU}

\begin{figure}
  \includegraphics[width=84mm]{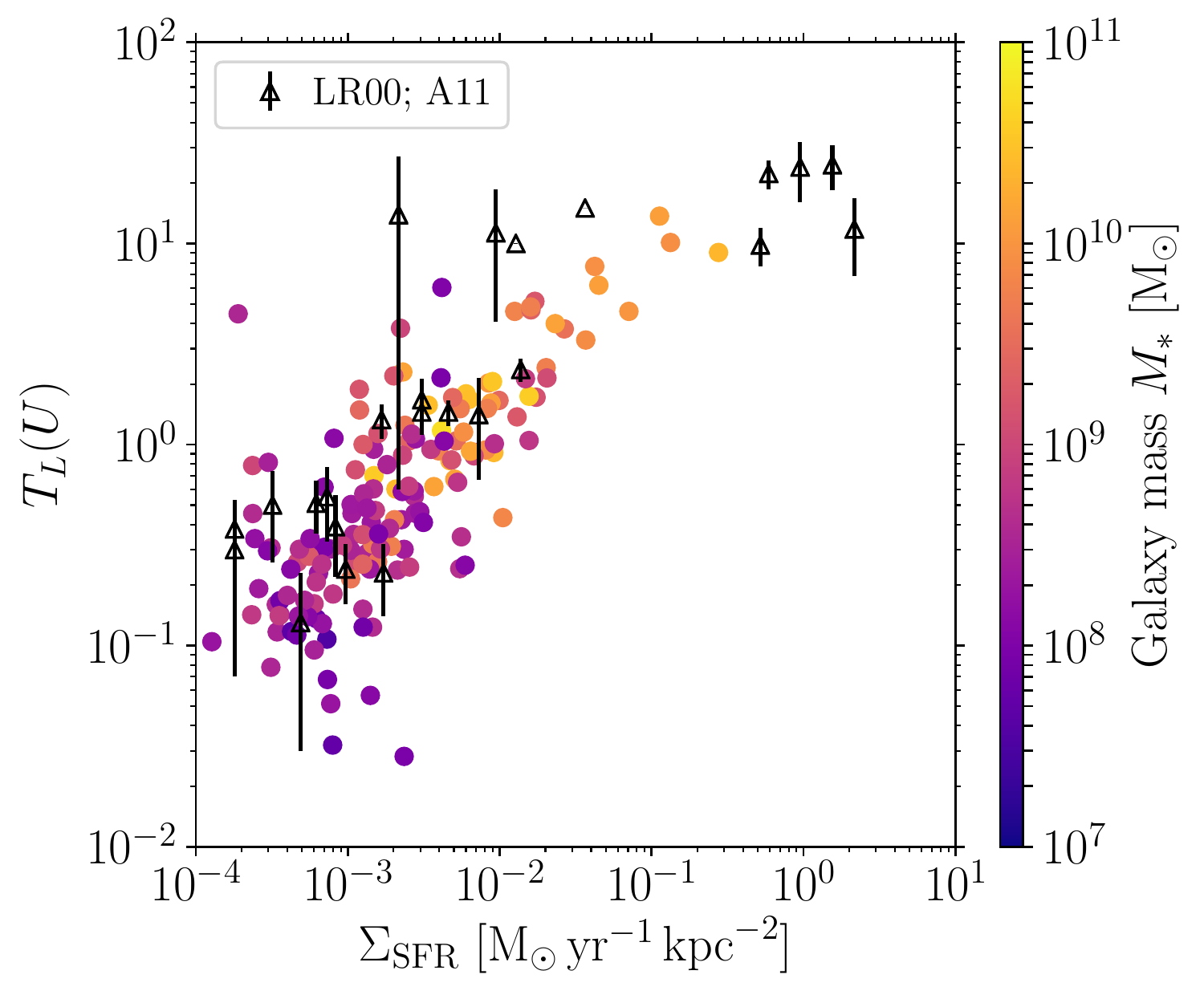}
  \caption{Specific $U$-band cluster luminosity, $T_L(U) = 100 L_\rmn{clusters} / L_\rmn{galaxy}$ (i.e. the percentage of the $U$-band light of a galaxy contributed by star clusters), as a function of $\SigmaSFR$, with points coloured by the stellar mass of the simulated galaxies. Open triangles show observed galaxies from \citet{Larsen_and_Richtler_00} and \citet{Adamo_Ostlin_and_Zackrisson_11}.}
  \label{fig:TLU}
\end{figure}

\begin{figure*}
  \centering
  \includegraphics[width=0.95\textwidth]{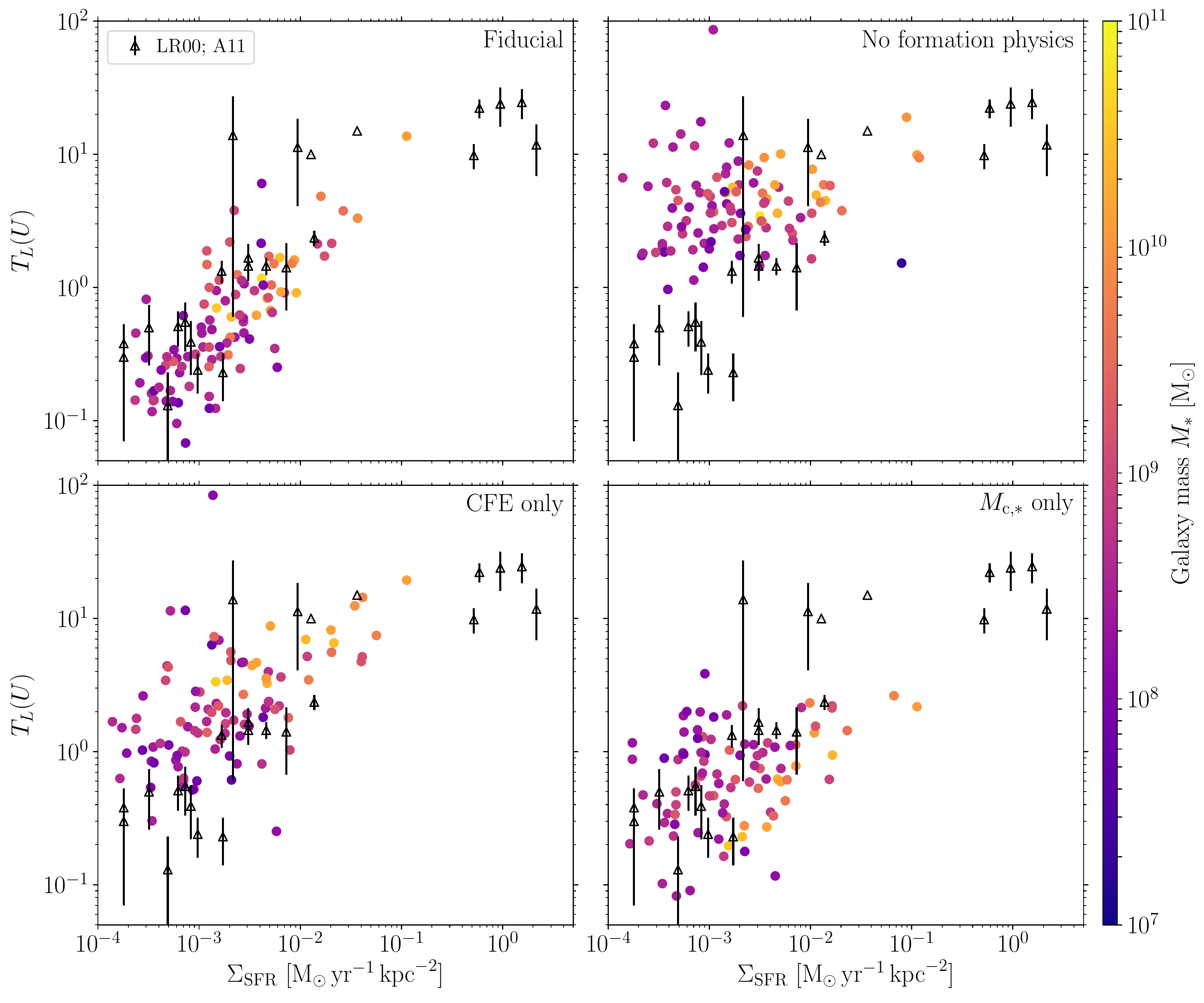}
  \caption{$T_L(U)$ for the simulated galaxies with different cluster formation physics. Open triangles show observed galaxies from \citet{Larsen_and_Richtler_00} and \citet{Adamo_Ostlin_and_Zackrisson_11}. \textit{Top left}: Fiducial E-MOSAICS model. \textit{Top right}: Constant CFE (10 per cent), $\Mcstar = \infty$. \textit{Bottom left}: environmentally-dependent CFE, $\Mcstar = \infty$. \textit{Bottom right:} environmentally-dependent $\Mcstar$, constant CFE (10 per cent).}
  \label{fig:TLU_alt_phys}
\end{figure*}

An empirical precursor to the CFE, the specific $U$-band cluster luminosity, $T_L(U) = 100 L_\rmn{clusters} / L_\rmn{galaxy}$ (i.e. the percentage of $U$-band light of a galaxy contributed by star clusters), was introduced by \citet{Larsen_and_Richtler_00}, who found it correlated strongly with $\SigmaSFR$ for observed galaxies.
While the CFE is the most relevant quantity for simulations of cluster formation, it is not a direct observable and a number of caveats apply to its estimation (e.g. assumptions for and extrapolation of the cluster mass function, uncertainties in masses and ages from stellar population modelling, SFR estimations, corrections for dust, etc.).
On the other hand, $T_L(U)$ can be directly determined from observations of galaxies (though may depend on selection criteria for YSCs) and therefore presents a useful test for models of YSC populations.

In Fig. \ref{fig:TLU} we show $T_L(U)$ for the simulated galaxies at $z=0$. 
We calculate the total luminosity of star clusters for all surviving clusters with initial masses $>5\times10^3 \Msun$. 
In order to calculate the total $U$-band luminosity of the galaxy, we assume simple stellar populations for each star particle and calculate their luminosities using the method described in Section \ref{sec:analysis}.
Both luminosities were calculated within $R_\rmn{lim}$ and assuming no extinction\footnote{Adopting a basic model for extinction where all stars and clusters are embedded within an optically thick cloud until a specific age \citep[e.g. $10 \Myr$; c.f.][]{Charlot_and_Fall_00} has no effect on the results, because extinction has the same effect on stars and clusters.}.
We compare the simulated galaxies in Fig. \ref{fig:TLU} with observed galaxies from \citet{Larsen_and_Richtler_00} and \citet{Adamo_Ostlin_and_Zackrisson_11}.
We find good agreement in the trend of $T_L(U)$ with $\SigmaSFR$ between the simulated and observed galaxies. 
At low $\SigmaSFR$ ($\sim 10^{-3} \Msun \pyr \kpc^{-2}$), $T_L(U)$ may be slightly underestimated in the simulations due to the instantaneous disruption of clusters with initial masses $M<5\times10^3 \Msun$. 
However, similar limitations also apply to the observed galaxies, depending on the distance to the galaxy and detectability of clusters.
In the simulations, $T_L(U)$ is largely determined by the CFE, such that those galaxies with a high CFE also have a high $T_L(U)$.
The scatter in $T_L(U)$ at fixed $\SigmaSFR$ (or CFE) shows no clear trends with sSFR or metallicity, and arises from temporal variations in the CFE and $\SigmaSFR$, as well as stochastic sampling at the high-mass end of the cluster mass function.

In Fig. \ref{fig:TLU_alt_phys}, we quantify the effect on $T_L(U)$ of varying the star cluster formation physics in the simulations and show the four cluster formation models described in Section \ref{sec:EMOSAICS}. 
Each model is a variation on the fiducial, environmentally-dependent model (top left panel), with environmentally-independent versions for either the truncation mass or CFE (\textit{CFE only} and \textit{$\Mcstar$ only} models, bottom left and right panels, respectively), or both (\textit{no formation physics} model, top right panel).
The figure only includes galaxies from the L012N0376 volume and the first ten zoom-in simulations (MW00-MW09), i.e. simulations with all four variations of the cluster formation physics (and thus the top left panel shows fewer galaxies than in Fig. \ref{fig:TLU}).

Fig. \ref{fig:TLU_alt_phys} clearly shows the critical role of both the CFE and $\Mcstar$ models in reproducing the observations of $T_L(U)$.
With a constant CFE ($\Gamma = 0.1$) and pure power-law mass function (upper right panel), the \textit{no formation physics} model implies a (roughly) constant $T_L(U)$, and therefore cannot simultaneously reproduce galaxies of high ($>10$) and low ($<1$) specific luminosities.
The \textit{CFE only} model (bottom left panel) assumes an environmentally-dependent CFE and a pure power-law mass function. Due to the variation of the CFE with $\SigmaSFR$, the model agrees better with the observed galaxies than for the model with constant CFE. However, a variation in CFE alone (at least in the current formulation of the \citealt{Kruijssen_12} model) is also largely unable to account for galaxies with $T_L(U) < 1$. The \textit{CFE only} model predicts higher $T_L(U)$ than observed at $\SigmaSFR \sim 10^{-3} \Msun \pyr \kpc^{-2}$, but agrees with the observed galaxies at higher $\SigmaSFR$.
The bottom right panel of Fig. \ref{fig:TLU_alt_phys} shows the \textit{$\Mcstar$ only} model, which assumes a constant $\rmn{CFE} = 0.1$ and an environmentally-dependent $\Mcstar$.
Though the model assumes the same constant CFE as for the \textit{no formation physics} model, the \textit{$\Mcstar$ only} model shows good agreement for galaxies with $\SigmaSFR \lesssim 10^{-2} \Msun \pyr \kpc^{-2}$, but underpredicts $T_L(U)$ at higher $\SigmaSFR$. In the \textit{$\Mcstar$ only} model, $T_L(U)$ is significantly lower than the assumed 10 per cent CFE due to the low $\Mcstar$ (at low $\SigmaSFR$) and the preferential formation of very low mass clusters, which are not detected\footnote{Note that $T_L(U)$ will therefore depend upon the lower initial cluster mass limit in the simulations (we adopt $5\times10^3 \Msun$). However, similar detection limits also apply for observed galaxies, depending on cluster age \citep[e.g.][]{Johnson_et_al_15, Adamo_et_al_15, Messa_et_al_18_I}.}.
Therefore, we conclude that environmental variations in both the CFE and $\Mcstar$ are necessary for reproducing the observed $T_L(U)$ relation.


\subsection{Brightest cluster-SFR relation} \label{sec:MVbright}

\subsubsection{Fiducial E-MOSAICS model} \label{sec:MVbright_fiducial}

\begin{figure*}
  \centering
  \includegraphics[width=0.95\textwidth]{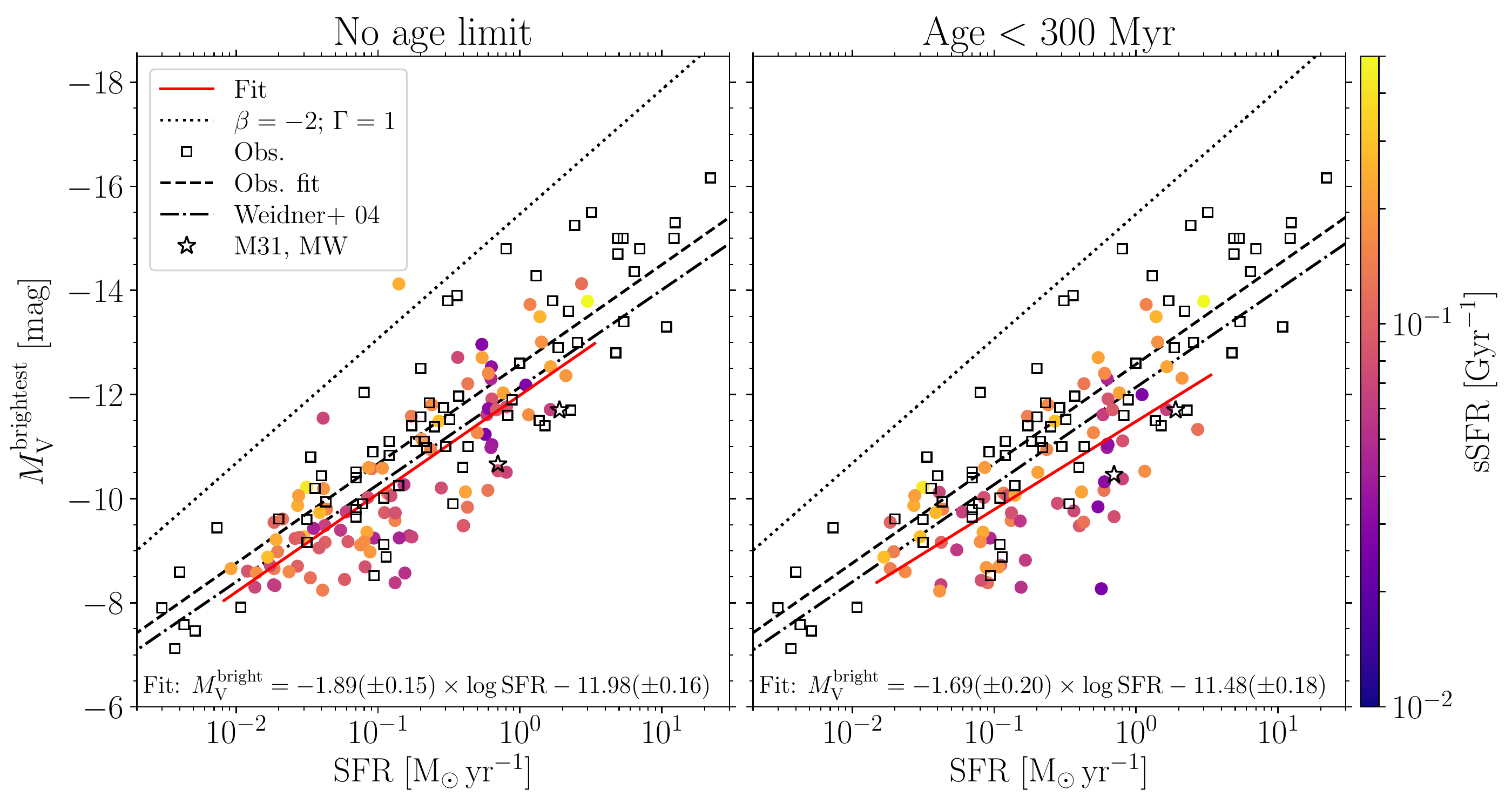}
  \caption{The $\MVbright$-SFR relation for the fiducial cluster formation model with no cluster age limit (left panel) and an age limit of $300\Myr$ (right panel). Each point represents an individual galaxy, with the symbols coloured by specific star formation rate of the galaxy. The solid red line is a linear regression fit to the simulations, with the equation shown in each panel. Black squares show the sample of observed galaxies compiled by \citet{Adamo_et_al_15} and the dashed line shows the best-fitting relation to this sample (Eq. \ref{eq:MV_fit_obs}). The dash-dotted line shows the best-fitting observed relation from \citet*[their eq. 2]{Weidner_Kroupa_and_Larsen_04}. Stars show the Milky Way ($\SFR = 1.9 \Msun \pyr$) and M31 ($\SFR = 0.7 \Msun \pyr$) for comparison (see Section \ref{sec:MVbright_fiducial} for details). The dotted line is the expectation for a $\beta=-2$ power-law mass function and constant 100 per cent CFE \citep{Bastian_08}, which is inconsistent with the observed relation. The slope of the best-fitting relation from the simulations is fully consistent with the slope of the relations from the observed galaxies.}
  \label{fig:MV}
\end{figure*}

The relation between the brightest cluster and the SFR of the galaxy is an empirical relation, of which the construction does not rely on modelled quantities such as clusters ages or masses.
The relation is sensitive to cluster formation physics and therefore presents a useful test of cluster formation models. 
Moreover, neither the star cluster nor galaxy formation physics implemented in the simulations were calibrated to reproduce the relation, thus a comparison affords an independent test of the predictions of the simulations.

In this section we consider two cases.
First, we consider the relation without applying an age limit to the simulated clusters, since in a number of cases observational measurements are obtained with single-band photometry, prohibiting age measurements \citep[e.g.][]{Larsen_02}.
Second, we apply an upper age limit of $300\Myr$, consistent with the calculation for the SFR.
For the simulations we apply a cluster luminosity limit of $M_\rmn{V} < -8.2$ (assuming a metallicity of $\log(Z/\Zsun)=-0.5$, similar to the metallicity of star-forming gas in a $M_\ast \sim 10^8 \Msun$ galaxy in the EAGLE Recal model, \citealt{S15}) to reflect the $5\times10^3 \Msun$ initial cluster mass limit, which is similar to the luminosity limit of observational programmes \citep[e.g.][]{Whitmore_et_al_14}.
V-band mass-to-light ratios for the simulated clusters are calculated using the FSPS model (see Section \ref{sec:analysis}).
Luminosities were then determined using the current cluster masses, which include cluster mass loss.

Fig. \ref{fig:MV} shows the predictions for the brightest cluster in the V-band as a function of galaxy SFR (both measured within $1.5 R_{1/2}$) for the fiducial E-MOSAICS model, with no cluster age limit (left panel) and with an upper limit of $300 \Myr$ (right panel).
The best-fitting relation for the simulated galaxies from a least-squares linear regression is given in each panel and shown as the solid red line.
The brightest cluster is generally consistent for both with and without an age limit. However, for some galaxies, the brightest young cluster is significantly fainter than the brightest cluster of any age, which results in a slightly flatter slope of the best-fitting relation for the $<300 \Myr$ age limit.
For comparison, the figure also shows the sample of observational measurements compiled by \citet{Adamo_et_al_15}.
Note that we have not attempted to match the sample selection for the observations, other than the selection in sSFR for the simulations, and therefore some bias between the simulated and observed galaxy populations may exist. 
The best-fitting relation for the observed sample of galaxies is given by
\begin{equation} \label{eq:MV_fit_obs}
\MVbright = -1.91(\pm 0.09) \times \log \frac{\rmn{SFR}}{\Msun \pyr} - 12.58(\pm 0.13), 
\end{equation} 
shown as a dashed line.
To this compilation of galaxies we also add measurements of the Milky Way and M31.
For the Milky Way we assume $\rmn{SFR} = 1.9 \Msun \pyr$ \citep{Chomiuk_and_Povich_11} and the brightest cluster to be Westerlund 1, with $\MVbright \approx -11.7$ \citep[assuming a mass of $6 \times 10^4 \Msun$ and age of 5 Myr,][and a V-band mass-to-light ratio from FSPS assuming a Solar metallicity]{Mengel_and_Tacconi-Garman_07}.
For M31 we assume $\rmn{SFR} = 0.7 \Msun \pyr$ \citep{Kang_Bianchi_and_Rey_09, Lewis_et_al_15}.
We take the brightest cluster in M31 (of any age) to be the globular cluster G1, with $\MVbright = -10.66$ \citep{Galleti_et_al_04}. 
For the brightest young cluster we use the brightest cluster from the PHAT survey, $\MVbright(<1\Gyr) \approx -10.46$ \citep{Johnson_et_al_15}, using their eq. 6 to calculate a V-band magnitude and assuming a distance modulus of $24.47$ \citep{McConnachie_et_al_05}.

The slope of the best-fitting relation for the simulated galaxies ($-1.89 \pm 0.15$ for no age limit, $-1.69 \pm 0.2$ for clusters $<300 \Myr$) is fully consistent with that of the observed galaxies ($-1.91 \pm 0.09$ for the \citealt{Adamo_et_al_15} sample; $-1.87 \pm 0.06$ for the relation from \citealt{Weidner_Kroupa_and_Larsen_04}).
The scatter around the observed relation is also very similar for the simulations and observations, with standard deviations in $\MVbright$ from the predicted and observed relations of $\approx0.95 \Mag$ and $1.01 \Mag$, respectively.
Therefore, the observed $\MVbright$-SFR relation is reproduced by the fiducial E-MOSAICS model with an environmentally varying CFE and upper mass function truncation mass, such that star formation in environments with high natal gas pressures results in more star formation in bound clusters and up to higher cluster masses.

\begin{figure}
  \includegraphics[width=84mm]{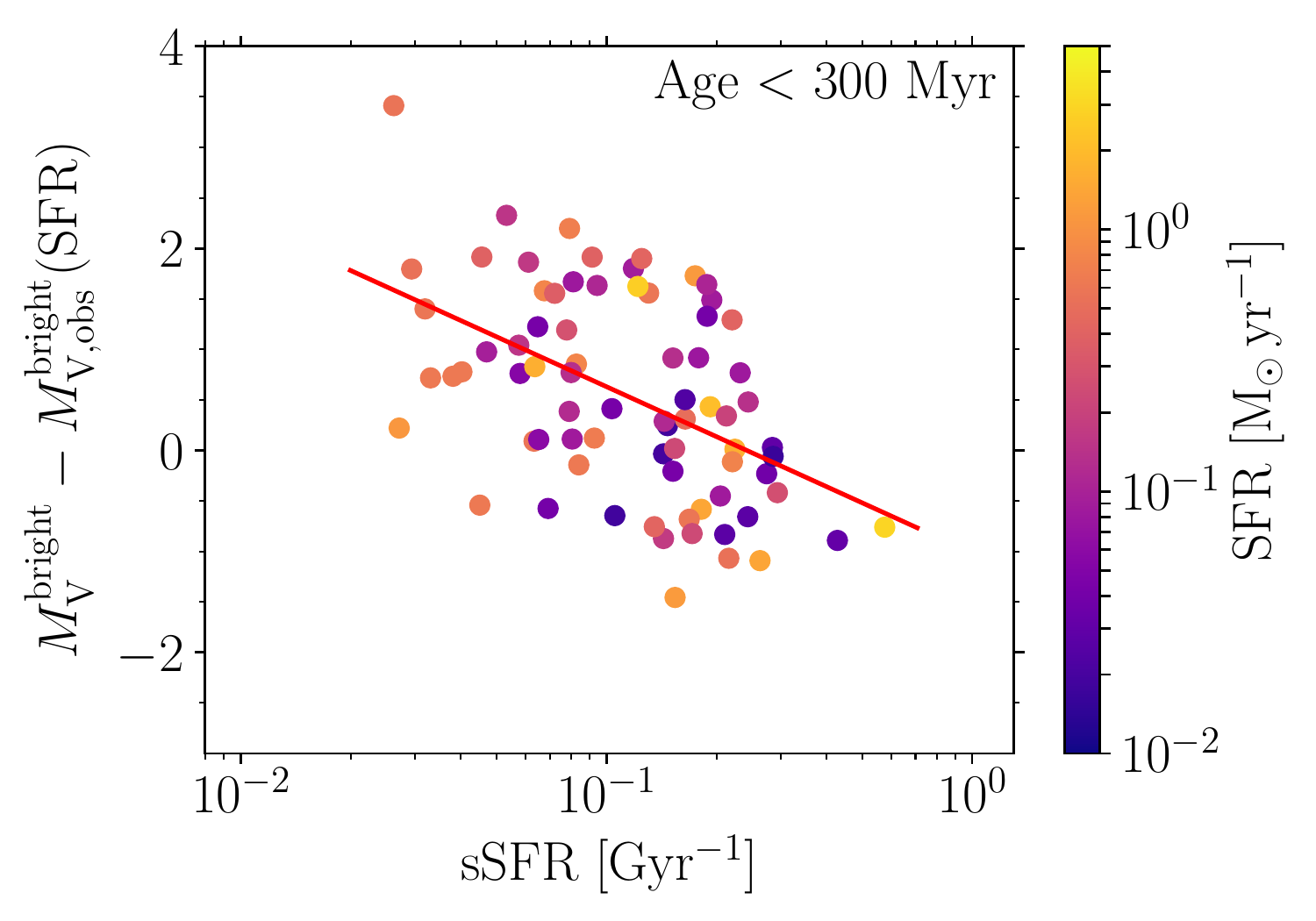}
  \caption{Difference in magnitude of the brightest cluster from the \citet{Weidner_Kroupa_and_Larsen_04} $\MVbright$-SFR relation (their eq. 2) compared with the sSFR of the galaxy, for the simulated galaxies in the right panel of Fig. \ref{fig:MV}. Solid red line shows the best-fitting relation $\Delta M_\rmn{V} = -1.64 (\pm 0.35) \times \log(\rmn{sSFR} / \rmn{Gyr}^{-1}) -1.01 (\pm 0.34)$. Galaxies with lower sSFRs typically have fainter clusters due to having lower cluster mass function truncations ($\Mcstar$).}
  \label{fig:deltaMV}
\end{figure}

To investigate the origin of scatter away from the $\MVbright$-SFR relation, in Fig. \ref{fig:MV} we colour the simulated galaxies by their sSFR.
In the right panel (ages $<300\Myr$), at a fixed SFR the simulations show a gradient in sSFR, such that the galaxies with the brightest clusters also typically have the highest sSFR (or lowest galaxy masses).
This trend is weaker in the left panel (no age limit) as cluster luminosities may be uncorrelated with the present SFR.
We explore this further in Fig. \ref{fig:deltaMV}, where we show the magnitude difference from the observed $\MVbright$-SFR relation \citep{Weidner_Kroupa_and_Larsen_04} compared with the sSFR for the simulated galaxies {for cluster ages $<300\Myr$}.
Simulated galaxies with brightest clusters that are significantly fainter than the observed relation typically have the lowest sSFRs.
The fitted relation crosses the zero-point (in $\Delta \MVbright$) at $\rmn{sSFR} \approx 0.2 \Gyr^{-1}$, similar to the typical sSFR (which is a weak function of galaxy mass) for galaxies on the star-forming main sequence \citep{Schiminovich_et_al_07}.
By selecting galaxy populations at different constant sSFRs, a prediction of our fiducial model is that we expect to find (age-limited) $\MVbright$-SFR relations that are offset to fainter luminosities at lower sSFRs.
Additionally, galaxies at higher redshifts, with higher sSFRs, should be offset to brighter $\MVbright$.
The physical cause of this effect in the simulations is lower cluster mass function truncations in higher mass galaxies (which at fixed SFR have lower sSFR). 
These galaxies typically have lower gas mass fractions, which result in a higher $\phi_P$, while larger galaxy masses result in an increased importance of Coriolis/centrifugal forces in setting the Toomre mass.
Both factors result in a lower cluster mass function truncation.
This result could be tested with future observations in galaxies of different masses at fixed sSFR.

\subsubsection{Alternative cluster formation models} \label{sec:MVbright_alt_phys}

\begin{figure*}
  \centering
  \includegraphics[width=0.95\textwidth]{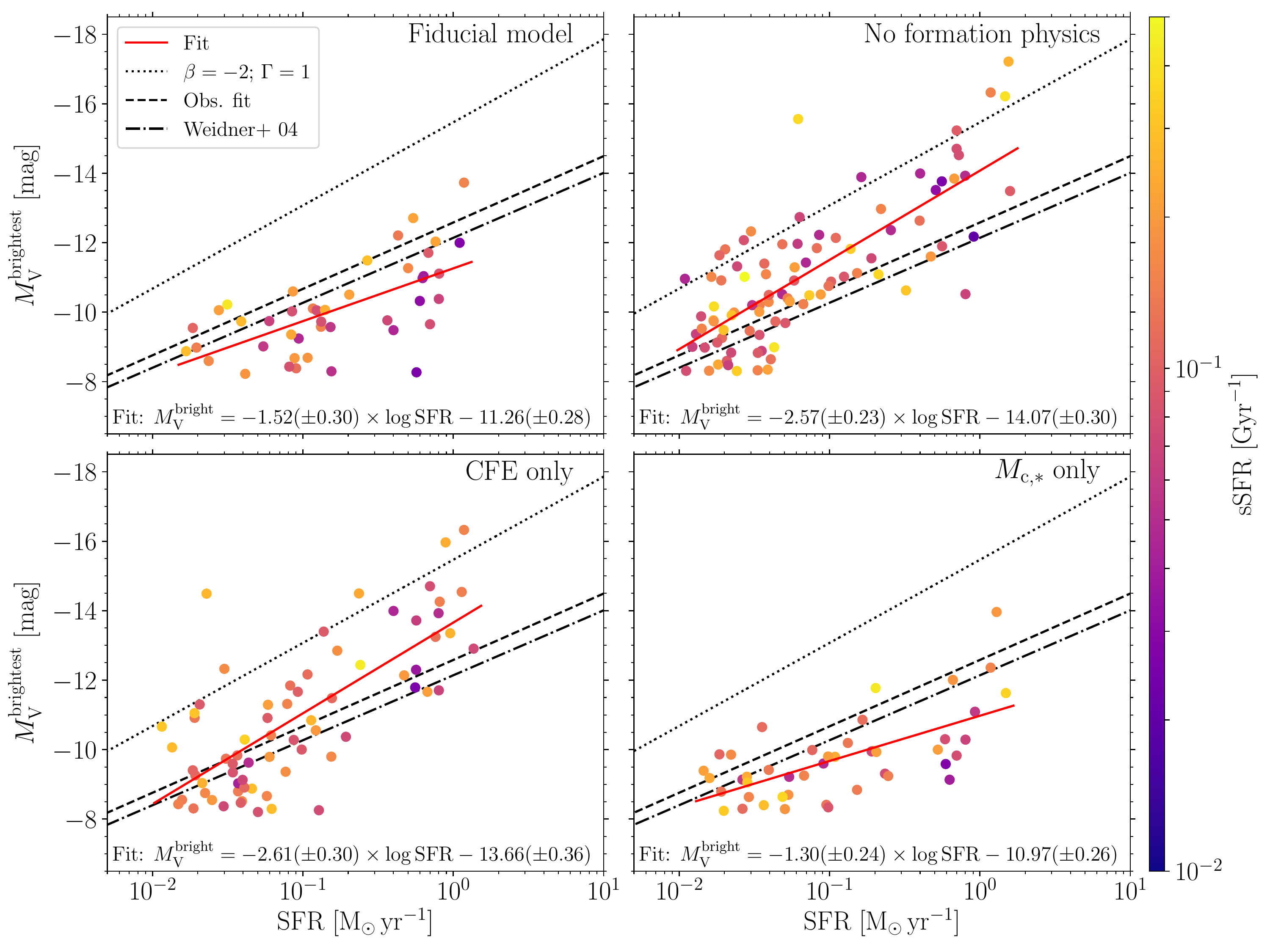}
  \caption{$\MVbright$-SFR relation (for clusters $<300\Myr$) for the simulated galaxies with different cluster formation physics. \textit{Top left}: Fiducial E-MOSAICS model. \textit{Top right}: Constant CFE (10 per cent), $\Mcstar = \infty$. \textit{Bottom left}: environmentally-dependent CFE, $\Mcstar = \infty$. \textit{Bottom right:} environmentally-dependent $\Mcstar$, constant CFE (10 per cent). Dashed, dash-dotted and dotted lines as in Fig. \ref{fig:MV}.}
  \label{fig:MV_alt_phys}
\end{figure*}

The result from Fig. \ref{fig:MV}, that environmentally-dependent cluster formation reproduces the $\MVbright$-SFR relation, can be further tested by considering the alternative cluster formation model variations in the E-MOSAICS suite. 
In Fig. \ref{fig:MV_alt_phys}, we compare the $\MVbright$-SFR relation for the four cluster formation models described in Section \ref{sec:EMOSAICS}. 
As in Fig. \ref{fig:TLU_alt_phys} (Section \ref{sec:TLU}), each model is a variation on the fiducial, environmentally-dependent model (top left panel), with environmentally-independent versions for either the truncation mass or CFE (\textit{CFE only} and \textit{$\Mcstar$ only} models, bottom left and right panels, respectively), or both (\textit{no formation physics} model, top right panel) and only includes galaxies from simulations with all four versions of the cluster formation physics (the L012N0376 volume and first ten zoom-in simulations, MW00-MW09).
Fig. \ref{fig:MV_alt_phys} highlights the importance of cluster formation physics in governing the $\MVbright$-SFR relation. 

In the \textit{no formation physics} model (top right panel), which assumes $\Gamma = 0.1$ and a $\beta=-2$ power-law mass function (i.e. with our standard cluster formation physics disabled), the simulations are inconsistent with the observed relation and recover the slope of the expected relation for a pure power-law mass function and constant CFE \citep[dotted line in the figure;][]{Bastian_08}.
The relation is therefore determined by a size-of-sample effect, with larger cluster populations more likely to have brighter clusters. The stochasticity at the high-mass end of the mass function induces a large scatter between galaxies at a given SFR. 
Also for this reason, the absolute offset in the relation is determined by the choice of CFE.
However, for any choice of constant CFE, the slope of the relation will remain inconsistent with the observations.

In the \textit{CFE only} model (bottom left panel), which assumes an environmentally-dependent CFE and a $\beta=-2$ power-law mass function, the simulations are again inconsistent with the observed relation. 
Due to the correlation of the CFE with galaxy mass (and thus SFR; see also Fig. \ref{fig:CFE}), the \textit{CFE only} model yields a relation that is even steeper than the \textit{no formation physics} model, since low-mass galaxies with low SFRs typically have CFEs below 10 per cent. 
Thus at low SFRs in the figure ($\SFR < 0.1 \Msun \pyr$) many galaxies are consistent with the observed relation, while those at higher SFRs are inconsistent with observed counterparts due to the lack of a truncation in the cluster mass function.
Again, there is large galaxy-to-galaxy scatter at a given SFR due to the stochasticity at the high-mass end of the mass function.

Finally, the \textit{$\Mcstar$ only} model, with a constant $\Gamma = 0.1$ and an environmentally-dependent exponential truncation mass to the cluster mass function ($\Mcstar$), is shown in the bottom right panel of the figure. 
At low SFRs, the simulations are consistent with observed galaxies.
However, since the CFE does not vary between galaxies, this model predicts a flatter slope than is observed, and thus at higher SFRs the brightest clusters are under-luminous compared to observed galaxies.

These results therefore show that an environmental dependence of both the CFE and cluster truncation mass is necessary for reproducing the observed properties of YSC populations.


\subsection{Cluster age distributions} \label{sec:age_dist}

\begin{figure}
\centering
\includegraphics[width=84mm]{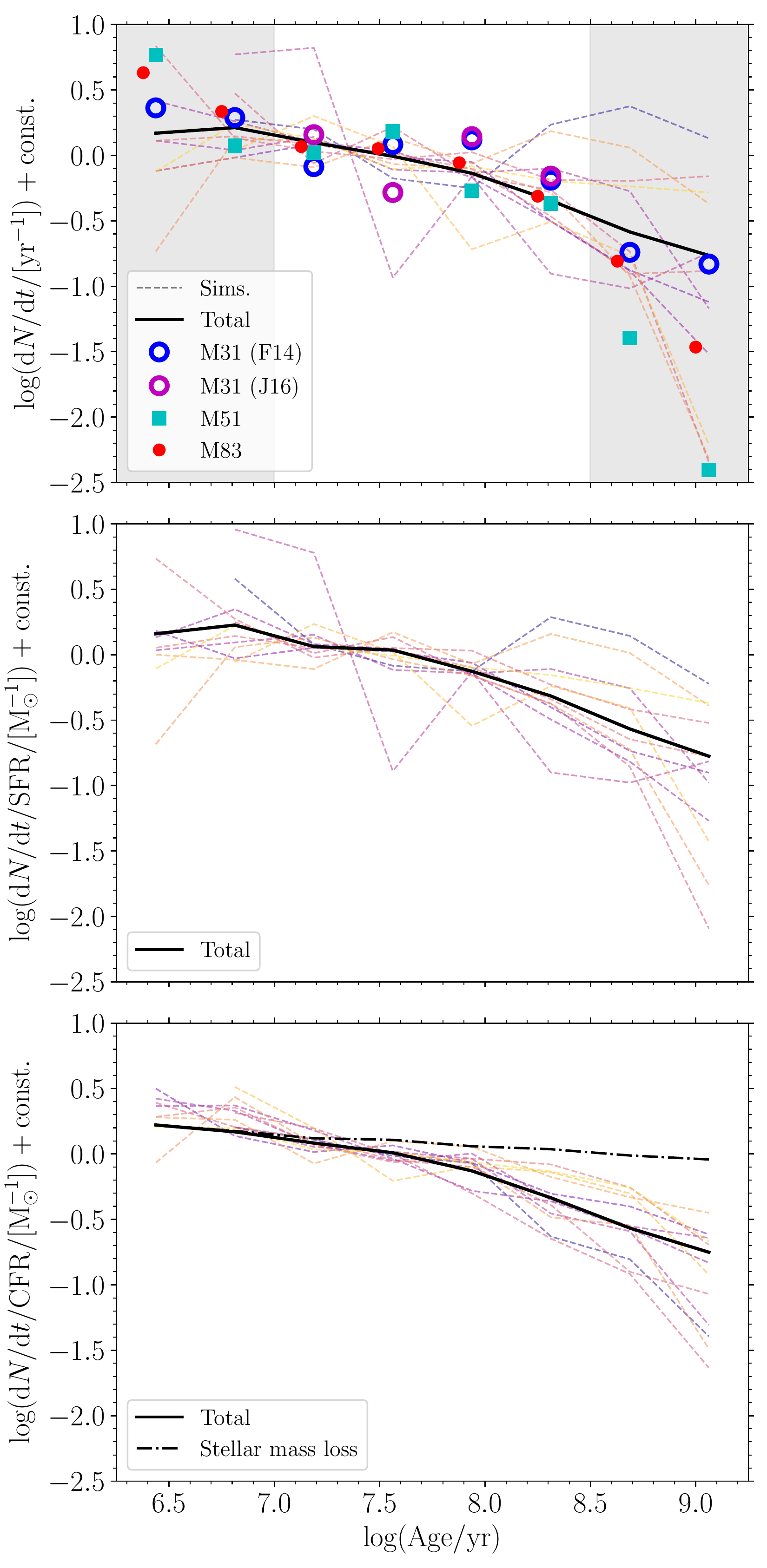}
  \caption{\textit{Upper:} Cluster age distributions for 12 of the target $\Lstar$ galaxies in the Milky Way-mass zooms (dashed lines, with a different colour for each galaxy). For comparison we show cluster age distributions from M83 \citep[using the distribution from \citealt{Adamo_and_Bastian_18}]{Silva-Villa_et_al_14}, M31 (\citealt{Fouesneau_et_al_14} and \citealt{Johnson_et_al_16}) and M51 \citep{Messa_et_al_18_I, Messa_et_al_18_II}, all using a cluster mass limit of $>5\times10^3 \Msun$. The shaded regions show ages where observed cluster populations may be contaminated by unbound associations ($<10 \Myr$) and are typically incomplete due to luminosity limits ($>300 \Myr$, for masses $>5\times10^3 \Msun$). \textit{Middle:} Cluster age distributions normalised by the star formation rate (SFR). \textit{Lower:} Cluster age distributions normalised by the cluster formation rate (CFR) for masses $>5\times10^3 \Msun$. The dash-dotted line shows the total cluster age distribution if only cluster mass loss due to stellar evolution is included.}
  \label{fig:age_dist}
\end{figure}

In this section we investigate the effect of time-varying star and cluster formation rates (CFRs) on cluster age distributions.
The age distribution of clusters is considered to be a strong test of cluster mass-loss models, potentially enabling the discrimination between mass/environmentally dependent or independent cluster disruption \citep[e.g.][see \citealt{Adamo_and_Bastian_18} for a review]{Gieles_et_al_05, Lamers_et_al_05b, Whitmore_Chandar_and_Fall_07, Bastian_et_al_09, Lamers_09, Chandar_Fall_and_Whitmore_10, Kruijssen_et_al_11, Kruijssen_et_al_12, Bastian_et_al_12, Miholics_Kruijssen_and_Sills_17}.
However, the effect of time-varying SFRs and CFRs on the interpretation of cluster age distributions has not been investigated in realistic galaxy simulations in cosmological environments.

In the upper panel of Fig. \ref{fig:age_dist}, we show the cluster age distributions of galaxies from the zoom-in simulations of Milky-Way mass galaxies.
We apply a cluster mass limit of $>5\times10^3 \Msun$ at all ages and use clusters within $R_\rmn{lim}$ from the centre of the potential.
We limit the sample to galaxies with at least $100$ surviving clusters younger than $10^{9.25} \yr$ \citep[similar to the number of clusters in M31 more massive than $5\times10^3 \Msun$,][]{Fouesneau_et_al_14, Johnson_et_al_16}, leaving 12 galaxies.
For the age distributions we use eight bins with widths of 0.375 dex between $10^{6.25}$ and $10^{9.25} \yr$, where the distributions are normalised at $10^{7.5} \yr$ by fitting a power-law relation to the four bins between $10^7$ and $10^{8.5} \yr$.
The typical timesteps for stellar particles at $z \approx 0$ are $\sim 1 \Myr$, and therefore even in the youngest age bins the numerical sampling of disruption by tidal heating is adequate.
We find a large scatter between the simulations, with some galaxies showing very flat age distributions and others where the number of clusters decreases rapidly at ages $>10^8 \yr$.
We discuss the cause of this scatter between galaxies below.
The solid line in the figure shows the total cluster age distribution for all of the galaxies combined.
For the four bins in the range $\log(\rmn{age}/\yr)=\{7,8.5\}$ (where observations are complete) we find a slope of $-0.39 \pm 0.04$ for the total population, with a range between $-1.18$ and $0.17$ for individual galaxies. 

We compare the cluster age distributions from the simulations with the observed distributions in M31 \citep{Fouesneau_et_al_14, Johnson_et_al_16}, M83 \citep{Silva-Villa_et_al_14, Adamo_and_Bastian_18} and M51 \citep{Messa_et_al_18_I, Messa_et_al_18_II}, also using a mass limit of $>5\times10^3 \Msun$. 
This sample comprises observed galaxies most similar to our sample of simulated Milky Way-mass galaxies.
At ages $<10^7 \yr$, the observations may be contaminated by unbound stellar associations \citep{Gieles_and_Portegies_Zwart_11, Bastian_et_al_12, Ward_and_Kruijssen_18}, while at ages $>10^{8.5} \yr$ cluster populations are incomplete due to luminosity limits.
Within the region for which observations are complete ($10^7$-$10^{8.5} \yr$), the age distributions of the simulations and observations are in good agreement, showing a similar level of scatter between different galaxies and consistent slopes of the age distributions ($-0.07$ and $-0.14$ for the \citealt{Fouesneau_et_al_14} and \citealt{Johnson_et_al_16} M31 catalogues, respectively, $-0.43$ for M51 and $-0.33$ for M83). 
We note that mass loss by tidal shocks is underestimated in the simulations, due to the lack of an explicit model for the cold, dense phase of the star-forming interstellar medium in the EAGLE model \citep[see][]{P18}.
The extent to which this affects the age distribution predictions from the simulations depends upon what cluster age this mechanism becomes important at $z \approx 0$ (which likely depends upon the local environment within which young clusters reside).
However, a similar result, where cluster mass loss is mainly important after a few hundred megayears, was found in simulations of isolated galaxies by \citet[using the same MOSAICS dynamical evolution model, but using a galaxy formation model with a simple model for the cold, dense phase of the interstellar medium, see \citealt{Pelupessy_et_al_04}]{Kruijssen_et_al_11}.

The cluster age distribution is a function of both cluster formation and evolution.
Therefore, variations in the SFR or CFR will also lead to variations in the age distributions.
We assess the impact of these effects in the bottom two panels of Fig. \ref{fig:age_dist}.
In the middle panel of Fig. \ref{fig:age_dist}, we show the cluster age distribution normalised by the total SFR in each temporal bin for each simulated galaxy.
Overall, the distributions for the galaxies show similar behaviour to the standard distributions (upper panel), with little reduction in scatter between the galaxies (the scatter about the total relation decreased from $0.43$~dex to $0.34$~dex).
This is expected, since the SFR in each galaxy is relatively constant over the period investigated. Over $10^{9.2} \yr$, the SFR in each temporal bin typically varies by less than a factor of two from the median for all bins.
The majority of the impact of variations in the SFR occurs at $>10^{8.5} \yr$, as can be seen by the slight reduction of scatter in the $10^9 \yr$ bin between the upper and middle panels in the figure.
The slope of the SFR-normalised age distribution is also very similar to the standard age distribution ($-0.35 \pm 0.07$, with a range between $-1.14$ and $0.15$).
Therefore, for the galaxies investigated, any variations in SFR induce little impact in the cluster age distributions. 

In the bottom panel of Fig. \ref{fig:age_dist} we show the cluster age distributions normalised by the CFR (for masses $>5\times10^3 \Msun$) in each temporal bin for the simulated galaxies.
Again, the slope of the CFR-normalised age distribution is similar to the standard age distribution, but with a reduced range ($-0.37 \pm 0.06$, with a range between $-0.75$ and $-0.09$).
Unlike the SFR, normalising by the CFR has a large effect on the age distributions, significantly reducing the galaxy-to-galaxy scatter in the distributions from 0.43 dex to 0.21 dex about the total distribution.
When normalised by the CFR, the age distributions follow a very similar evolution for the galaxies, with some divergence at $>10^8 \yr$ reflecting the varying cluster mass-loss rates between the galaxies (due to, e.g., differing potentials, gas densities or galactocentric radii of cluster formation).
Though the SFR varies by up to a factor of two between bins, the CFR can vary by up to a factor of six, and therefore the CFR is not simply following changes in the SFR\footnote{A similar level of variation occurs when comparing the initial age distribution, $\rmn{d} N_\rmn{init} / \rmn{d}t$ (i.e. without including cluster mass loss), rather than $\rmn{CFR}(M>5\times10^3 \Msun)$, thus it is not simply due to stochasticity in the masses of clusters that form.}.
In addition to the SFR, the CFE (i.e. the fraction of star formation in bound clusters) may also vary with time by up to a factor of two, and thus together account for about half of the variance in the $\rmn{CFR}(M>5\times10^3 \Msun)$.
The rest of the variation (factor of three) can be attributed to the stochasticity of cluster formation at the high-mass end of the cluster mass function (seven of the galaxies have $\Mcstar < 10^5 \Msun$, and therefore in these galaxies only the high-mass end of the cluster population satisfies our mass cut of $M>5\times 10^3 \Msun$).
This effect is lessened in galaxies with larger cluster populations, although even for those with $>1000$ clusters above the mass limit, the variation in the mass-limited CFR, in addition to that of the SFR and CFE, is a factor of two.
For the observed galaxies in Fig. \ref{fig:age_dist}, M31 may be most affected by this effect since it has the fewest clusters with $M>5\times10^3 \Msun$, though this may be alleviated somewhat by including lower mass clusters \citep[i.e. the completeness limit in the PHAT survey for clusters younger than $300 \Myr$ is $\sim1000 \Msun$,][]{Johnson_et_al_15}.

In Fig. \ref{fig:age_dist}, we also show the contribution of stellar-evolutionary mass loss (dash-dotted line) to the total cluster age distribution (solid line).
The stellar mass loss relation has a significantly flatter slope ($\approx -0.1$) than the total cluster age distribution.
Stellar mass loss dominates for ages $\lesssim 10^{7} \yr$, with dynamical mass loss (two-body relaxation and tidal shocks) becoming dominant only at older ages.

Therefore, temporal variations in the CFE and CFR should also be considered when using cluster age distributions to test models for cluster mass loss, since they may impart non-negligible variations in the cluster age distributions.

\section{Discussion and conclusions} \label{sec:conclusions}

In this paper, we present a comparison of the YSC populations in the E-MOSAICS simulations with observed populations in nearby galaxies. 
The aim of this work is to both test the cluster formation models in realistic simulations of galaxy formation and, by varying the cluster formation physics (the CFE and exponential truncation of the cluster mass function), obtain an insight into the formation processes at work in observed cluster populations.

We find that, due to a combination of spatially varying (non-uniform) star formation and the metallicity-dependent density threshold for star formation in the EAGLE model, the CFE in the simulated galaxies is elevated above the (input) \citet{Kruijssen_12} relation at $\SigmaSFR < 10^{-2.5} \Msun \pyr \kpc^{-2}$ (Section \ref{sec:CFE}; Fig. \ref{fig:CFE}).
In fact, our sample of simulated galaxies could be fit by a power-law relation in CFE-$\SigmaSFR$ \citep[also see][]{Goddard_Bastian_and_Kennicutt_10}, despite the input relation.
A similar effect might be present in observed cluster populations, though this could also be caused by a non-linear $\SigmaG$-$\SigmaSFR$ relation \citep{Johnson_et_al_16}.

For cluster populations in the simulated galaxies that can be fit with a \citet{Schechter_76} mass function, the formulation of the \citet{Reina-Campos_and_Kruijssen_17} model implemented in E-MOSAICS performs well in reproducing observed truncation masses (Section \ref{sec:Mcstar}; Fig. \ref{fig:Mcstar}).
However, more observations are needed to test if the scatter in $\Mcstar$ seen between galaxies (i.e. at a given $\SigmaSFR$) in the simulations is consistent with observed galaxies.
Due to the higher gas fractions of galaxies at higher redshifts, the E-MOSAICS model predicts that, at the epochs of GC formation ($z \gtrsim 1$), the $\Mcstar$-$\SigmaSFR$ relation should be elevated above the present day relation (Fig. \ref{fig:Mcstar_redshift}).

In Section \ref{sec:MFslope}, we investigate the power-law index of the cluster mass function at low cluster masses, for evolved cluster populations. We find that in galaxies with high SFR surface densities ($\SigmaSFR \gtrsim 10^{-1} \Msun \pyr \kpc^{-2}$), mass function indices become flatter ($\beta>-2$) due to cluster disruption and increased relative mass loss towards low cluster masses. In galaxies with lower SFR surface densities, mass function indices are similar to the initial index ($\beta=-2$) or may potentially become steeper ($\beta < -2$) due to the degeneracy between the mass function index and upper truncation mass ($\Mcstar$).
The results are consistent with the findings of \citet{Cook_et_al_16}, who investigated observed cluster luminosity functions, suggesting that variable cluster mass function index is not necessary to explain their findings.

In Section \ref{sec:MVbright}, we test the relation between the brightest cluster within a galaxy and its SFR.
This observed relation presents a useful test for the simulations since it sensitive to cluster formation physics, yet was not used to test or calibrate the cluster formation models.
We find that the fiducial E-MOSAICS cluster formation model reproduces both the slope of the $\MVbright$-SFR relation and the scatter around the relation (Fig. \ref{fig:MV}).
If an upper age limit ($<300 \Myr$) is applied to the clusters, we find that a number of the simulated galaxies fall significantly below the observed relation, which tend to be those with low sSFRs.
This result implies that the $\MVbright$-SFR relation may partially arise due to the preferential selection of galaxies on the star-forming main sequence of galaxies.  
Selecting samples of galaxies at fixed sSFRs may therefore result in offset $\MVbright$-SFR sequences, which is a prediction offering a new avenue to test cluster formation models.

Some previous work has suggested that young cluster formation proceeds in an environmentally-independent manner \citep{Whitmore_Chandar_and_Fall_07, Fall_and_Chandar_12, Chandar_Fall_and_Whitmore_15, Chandar_et_al_17}.
However, in Sections \ref{sec:TLU} and \ref{sec:MVbright_alt_phys} we find that only the fully environmentally-dependent cluster formation (fiducial) model can explain both the $T_L(U)$-$\SigmaSFR$ and $\MVbright$-SFR relations. 
If cluster formation is environmentally-independent (constant CFE and a power-law mass function with $\beta = -2$), both the slope of the $\MVbright$-SFR relation and its absolute offset are inconsistent with the observed relation.
Similarly, a constant CFE cannot simultaneously explain galaxies at both high and low specific $U$-band cluster luminosities [$T_L(U)$], while an environmental-dependence in $\Mcstar$ is required to explain galaxies with $T_L(U) < 1$.
We therefore conclude that an environmental dependence in cluster formation, both for the CFE and the upper truncation of the cluster mass function ($\Mcstar$), is required to reproduce observed young cluster populations.
This shows that models in which a constant fraction of stars form in clusters are inconsistent with observations.
The importance of environmentally-dependent cluster formation in reproducing GC populations has similarly been discussed in previous work with the E-MOSAICS simulations \citep{P18, Usher_et_al_18, Reina-Campos_et_al_19}.

Finally, we compare the cluster age distributions in 12 (out of 25) of our sample of Milky Way-mass galaxies with those in observed nearby disc galaxies (Section \ref{sec:age_dist}).
We find that, for ages where observed cluster populations are generally complete (between $10^7$ and $10^{8.5} \yr$), the scatter in the age distributions between the simulated galaxies is similar to that in the observed galaxies.
However, for the environmentally-dependent cluster formation model, the variation of the CFR (not SFR) over Myr timescales may impart significant variation in the cluster age distributions (Fig. \ref{fig:age_dist}).
Therefore, the use of cluster ages as a discriminator of cluster mass-loss mechanisms should be approached with caution unless the variation of the CFR with time can be accounted for.
The variation of the CFR is mainly driven by an underlying time-variation of the CFE.
However, even for galaxies with $>1000$ clusters, additional (potentially dominant) variations may be imparted simply due to the stochasticity of cluster formation in time, with larger variations for less numerous cluster populations.

The results presented in this paper reinforce the understanding of cluster formation as an environmentally-dependent process.
We make predictions for future comparisons to observations ($\Mcstar$ at high and low gas fractions, Section \ref{sec:Mcstar}; the $\MVbright$-SFR relation at varying sSFRs, Section \ref{sec:MVbright}), which will be useful in both further testing the models implemented in E-MOSAICS, as well as cluster formation theories in general.
This work also highlights the importance of realistic simulations of galaxies with a diverse range of properties and environments when testing models of cluster formation.
Comparing the simulations and observations at the extremes of the galaxy population will enable a strong test of cluster formation theories, by testing them outside of the range in which they were developed, and allow further insight into the star and galaxy formation mechanisms which shape young star cluster populations across cosmic history.

\section*{Acknowledgements}

We thank Angela Adamo and S{\o}ren Larsen for helpful comments and suggestions, David Cook for providing data for the dwarf galaxy sample and the anonymous referee for a constructive report which improved the paper.
JP, NB and CU gratefully acknowledge funding from a European Research Council consolidator grant (ERC-CoG-646928-Multi-Pop). 
NB and RAC are Royal Society University Research Fellows.
JMDK acknowledges funding from the German Research Foundation (DFG - Emmy Noether Research Group KR4801/1-1).
JMDK and MRC acknowledge funding from the European Research Council (ERC-StG-714907, MUSTANG).
MRC is supported by a PhD Fellowship from the International Max Planck Research School at the University of Heidelberg (IMPRS-HD). 
This work used the DiRAC Data Centric system at Durham University, operated by the Institute for Computational Cosmology on behalf of the STFC DiRAC HPC Facility (\url{www.dirac.ac.uk}). This equipment was funded by BIS National E-infrastructure capital grant ST/K00042X/1, STFC capital grants ST/H008519/1 and ST/K00087X/1, STFC DiRAC Operations grant ST/K003267/1 and Durham University. DiRAC is part of the National E-Infrastructure.
The work also made use of high performance computing facilities at Liverpool John Moores University, partly funded by the Royal Society and LJMU's Faculty of Engineering and Technology.



\bibliographystyle{mnras}
\bibliography{emosaics} 

\begin{thebibliography}{}
\makeatletter
\relax
\def\mn@urlcharsother{\let\do\@makeother \do\$\do\&\do\#\do\^\do\_\do\%\do\~}
\def\mn@doi{\begingroup\mn@urlcharsother \@ifnextchar [ {\mn@doi@}
  {\mn@doi@[]}}
\def\mn@doi@[#1]#2{\def\@tempa{#1}\ifx\@tempa\@empty \href
  {http://dx.doi.org/#2} {doi:#2}\else \href {http://dx.doi.org/#2} {#1}\fi
  \endgroup}
\def\mn@eprint#1#2{\mn@eprint@#1:#2::\@nil}
\def\mn@eprint@arXiv#1{\href {http://arxiv.org/abs/#1} {{\tt arXiv:#1}}}
\def\mn@eprint@dblp#1{\href {http://dblp.uni-trier.de/rec/bibtex/#1.xml}
  {dblp:#1}}
\def\mn@eprint@#1:#2:#3:#4\@nil{\def\@tempa {#1}\def\@tempb {#2}\def\@tempc
  {#3}\ifx \@tempc \@empty \let \@tempc \@tempb \let \@tempb \@tempa \fi \ifx
  \@tempb \@empty \def\@tempb {arXiv}\fi \@ifundefined
  {mn@eprint@\@tempb}{\@tempb:\@tempc}{\expandafter \expandafter \csname
  mn@eprint@\@tempb\endcsname \expandafter{\@tempc}}}

\bibitem[\protect\citeauthoryear{{Adamo} \& {Bastian}}{{Adamo} \&
  {Bastian}}{2018}]{Adamo_and_Bastian_18}
{Adamo} A.,  {Bastian} N.,  2018, in {Stahler} S.,  ed.,  Astrophysics and
  Space Science Library Vol. 424, The Birth of Star Clusters. p.~91,
  \mn@doi{10.1007/978-3-319-22801-3_4}

\bibitem[\protect\citeauthoryear{{Adamo}, {{\"O}stlin}, {Zackrisson}, {Hayes},
  {Cumming}  \& {Micheva}}{{Adamo} et~al.}{2010}]{Adamo_et_al_10}
{Adamo} A.,  {{\"O}stlin} G.,  {Zackrisson} E.,  {Hayes} M.,  {Cumming} R.~J.,
   {Micheva} G.,  2010, \mn@doi [\mnras] {10.1111/j.1365-2966.2010.16983.x},
  \href {http://adsabs.harvard.edu/abs/2010MNRAS.407..870A} {407, 870}

\bibitem[\protect\citeauthoryear{{Adamo}, {{\"O}stlin}  \&
  {Zackrisson}}{{Adamo} et~al.}{2011}]{Adamo_Ostlin_and_Zackrisson_11}
{Adamo} A.,  {{\"O}stlin} G.,   {Zackrisson} E.,  2011, \mn@doi [\mnras]
  {10.1111/j.1365-2966.2011.19377.x}, \href
  {http://adsabs.harvard.edu/abs/2011MNRAS.417.1904A} {417, 1904}

\bibitem[\protect\citeauthoryear{{Adamo}, {Kruijssen}, {Bastian}, {Silva-Villa}
   \& {Ryon}}{{Adamo} et~al.}{2015}]{Adamo_et_al_15}
{Adamo} A.,  {Kruijssen} J.~M.~D.,  {Bastian} N.,  {Silva-Villa} E.,   {Ryon}
  J.,  2015, \mn@doi [\mnras] {10.1093/mnras/stv1203}, \href
  {http://adsabs.harvard.edu/abs/2015MNRAS.452..246A} {452, 246}

\bibitem[\protect\citeauthoryear{{Annibali}, {Tosi}, {Aloisi}  \& {van der
  Marel}}{{Annibali} et~al.}{2011}]{Annibali_et_al_11}
{Annibali} F.,  {Tosi} M.,  {Aloisi} A.,   {van der Marel} R.~P.,  2011,
  \mn@doi [\aj] {10.1088/0004-6256/142/4/129}, \href
  {http://adsabs.harvard.edu/abs/2011AJ....142..129A} {142, 129}

\bibitem[\protect\citeauthoryear{{Bah{\'e}} et~al.,}{{Bah{\'e}}
  et~al.}{2016}]{Bahe_et_al_16}
{Bah{\'e}} Y.~M.,  et~al., 2016, \mn@doi [\mnras] {10.1093/mnras/stv2674},
  \href {http://adsabs.harvard.edu/abs/2016MNRAS.456.1115B} {456, 1115}

\bibitem[\protect\citeauthoryear{{Barnes}, {Longmore}, {Battersby}, {Bally},
  {Kruijssen}, {Henshaw}  \& {Walker}}{{Barnes} et~al.}{2017}]{Barnes_et_al_17}
{Barnes} A.~T.,  {Longmore} S.~N.,  {Battersby} C.,  {Bally} J.,  {Kruijssen}
  J.~M.~D.,  {Henshaw} J.~D.,   {Walker} D.~L.,  2017, \mn@doi [\mnras]
  {10.1093/mnras/stx941}, \href
  {http://adsabs.harvard.edu/abs/2017MNRAS.469.2263B} {469, 2263}

\bibitem[\protect\citeauthoryear{{Bastian}}{{Bastian}}{2008}]{Bastian_08}
{Bastian} N.,  2008, \mn@doi [\mnras] {10.1111/j.1365-2966.2008.13775.x}, \href
  {http://adsabs.harvard.edu/abs/2008MNRAS.390..759B} {390, 759}

\bibitem[\protect\citeauthoryear{{Bastian}, {Trancho}, {Konstantopoulos}  \&
  {Miller}}{{Bastian} et~al.}{2009}]{Bastian_et_al_09}
{Bastian} N.,  {Trancho} G.,  {Konstantopoulos} I.~S.,   {Miller} B.~W.,  2009,
  \mn@doi [\apj] {10.1088/0004-637X/701/1/607}, \href
  {http://adsabs.harvard.edu/abs/2009ApJ...701..607B} {701, 607}

\bibitem[\protect\citeauthoryear{{Bastian} et~al.,}{{Bastian}
  et~al.}{2012}]{Bastian_et_al_12}
{Bastian} N.,  et~al., 2012, \mn@doi [\mnras]
  {10.1111/j.1365-2966.2011.19909.x}, \href
  {http://adsabs.harvard.edu/abs/2012MNRAS.419.2606B} {419, 2606}

\bibitem[\protect\citeauthoryear{{Baumgardt}, {Parmentier}, {Anders}  \&
  {Grebel}}{{Baumgardt} et~al.}{2013}]{Baumgardt_et_al_13}
{Baumgardt} H.,  {Parmentier} G.,  {Anders} P.,   {Grebel} E.~K.,  2013,
  \mn@doi [\mnras] {10.1093/mnras/sts667}, \href
  {http://adsabs.harvard.edu/abs/2013MNRAS.430..676B} {430, 676}

\bibitem[\protect\citeauthoryear{{Bigiel}, {Leroy}, {Walter}, {Brinks}, {de
  Blok}, {Madore}  \& {Thornley}}{{Bigiel} et~al.}{2008}]{Bigiel_et_al_08}
{Bigiel} F.,  {Leroy} A.,  {Walter} F.,  {Brinks} E.,  {de Blok} W.~J.~G.,
  {Madore} B.,   {Thornley} M.~D.,  2008, \mn@doi [\aj]
  {10.1088/0004-6256/136/6/2846}, \href
  {http://adsabs.harvard.edu/abs/2008AJ....136.2846B} {136, 2846}

\bibitem[\protect\citeauthoryear{{Bik}, {Lamers}, {Bastian}, {Panagia}  \&
  {Romaniello}}{{Bik} et~al.}{2003}]{Bik_et_al_03}
{Bik} A.,  {Lamers} H.~J.~G.~L.~M.,  {Bastian} N.,  {Panagia} N.,
  {Romaniello} M.,  2003, \mn@doi [\aap] {10.1051/0004-6361:20021384}, \href
  {http://adsabs.harvard.edu/abs/2003A%26A...397..473B} {397, 473}

\bibitem[\protect\citeauthoryear{{Billett}, {Hunter}  \& {Elmegreen}}{{Billett}
  et~al.}{2002}]{Billett_Hunter_and_Elmegreen_02}
{Billett} O.~H.,  {Hunter} D.~A.,   {Elmegreen} B.~G.,  2002, \mn@doi [\aj]
  {10.1086/339181}, \href {http://adsabs.harvard.edu/abs/2002AJ....123.1454B}
  {123, 1454}

\bibitem[\protect\citeauthoryear{{Booth} \& {Schaye}}{{Booth} \&
  {Schaye}}{2009}]{Booth_and_Schaye_09}
{Booth} C.~M.,  {Schaye} J.,  2009, \mn@doi [\mnras]
  {10.1111/j.1365-2966.2009.15043.x}, \href
  {http://adsabs.harvard.edu/abs/2009MNRAS.398...53B} {398, 53}

\bibitem[\protect\citeauthoryear{{Bournaud}, {Duc}  \& {Emsellem}}{{Bournaud}
  et~al.}{2008}]{Bournaud_Duc_and_Emsellem_08}
{Bournaud} F.,  {Duc} P.-A.,   {Emsellem} E.,  2008, \mn@doi [\mnras]
  {10.1111/j.1745-3933.2008.00511.x}, \href
  {http://adsabs.harvard.edu/abs/2008MNRAS.389L...8B} {389, L8}

\bibitem[\protect\citeauthoryear{{Bourne} et~al.,}{{Bourne}
  et~al.}{2017}]{Bourne_et_al_17}
{Bourne} N.,  et~al., 2017, \mn@doi [\mnras] {10.1093/mnras/stx031}, \href
  {https://ui.adsabs.harvard.edu/abs/2017MNRAS.467.1360B} {467, 1360}

\bibitem[\protect\citeauthoryear{{Calzetti} et~al.,}{{Calzetti}
  et~al.}{2015}]{Calzetti_et_al_15}
{Calzetti} D.,  et~al., 2015, \mn@doi [\aj] {10.1088/0004-6256/149/2/51}, \href
  {http://adsabs.harvard.edu/abs/2015AJ....149...51C} {149, 51}

\bibitem[\protect\citeauthoryear{{Chabrier}}{{Chabrier}}{2003}]{Chabrier_03}
{Chabrier} G.,  2003, \pasp, \href
  {http://adsabs.harvard.edu/abs/2003PASP..115..763C} {115, 763}

\bibitem[\protect\citeauthoryear{{Chandar}, {Fall}  \& {Whitmore}}{{Chandar}
  et~al.}{2010}]{Chandar_Fall_and_Whitmore_10}
{Chandar} R.,  {Fall} S.~M.,   {Whitmore} B.~C.,  2010, \mn@doi [\apj]
  {10.1088/0004-637X/711/2/1263}, \href
  {http://adsabs.harvard.edu/abs/2010ApJ...711.1263C} {711, 1263}

\bibitem[\protect\citeauthoryear{{Chandar}, {Fall}  \& {Whitmore}}{{Chandar}
  et~al.}{2015}]{Chandar_Fall_and_Whitmore_15}
{Chandar} R.,  {Fall} S.~M.,   {Whitmore} B.~C.,  2015, \mn@doi [\apj]
  {10.1088/0004-637X/810/1/1}, \href
  {http://adsabs.harvard.edu/abs/2015ApJ...810....1C} {810, 1}

\bibitem[\protect\citeauthoryear{{Chandar}, {Fall}, {Whitmore}  \&
  {Mulia}}{{Chandar} et~al.}{2017}]{Chandar_et_al_17}
{Chandar} R.,  {Fall} S.~M.,  {Whitmore} B.~C.,   {Mulia} A.~J.,  2017, \mn@doi
  [\apj] {10.3847/1538-4357/aa92ce}, \href
  {http://adsabs.harvard.edu/abs/2017ApJ...849..128C} {849, 128}

\bibitem[\protect\citeauthoryear{{Charlot} \& {Fall}}{{Charlot} \&
  {Fall}}{2000}]{Charlot_and_Fall_00}
{Charlot} S.,  {Fall} S.~M.,  2000, \mn@doi [\apj] {10.1086/309250}, \href
  {http://adsabs.harvard.edu/abs/2000ApJ...539..718C} {539, 718}

\bibitem[\protect\citeauthoryear{{Chomiuk} \& {Povich}}{{Chomiuk} \&
  {Povich}}{2011}]{Chomiuk_and_Povich_11}
{Chomiuk} L.,  {Povich} M.~S.,  2011, \mn@doi [\aj]
  {10.1088/0004-6256/142/6/197}, \href
  {http://adsabs.harvard.edu/abs/2011AJ....142..197C} {142, 197}

\bibitem[\protect\citeauthoryear{{Conroy} \& {Gunn}}{{Conroy} \&
  {Gunn}}{2010}]{Conroy_and_Gunn_10}
{Conroy} C.,  {Gunn} J.~E.,  2010, \mn@doi [\apj]
  {10.1088/0004-637X/712/2/833}, \href
  {http://adsabs.harvard.edu/abs/2010ApJ...712..833C} {712, 833}

\bibitem[\protect\citeauthoryear{{Conroy}, {Gunn}  \& {White}}{{Conroy}
  et~al.}{2009}]{Conroy_Gunn_and_White_09}
{Conroy} C.,  {Gunn} J.~E.,   {White} M.,  2009, \mn@doi [\apj]
  {10.1088/0004-637X/699/1/486}, \href
  {http://adsabs.harvard.edu/abs/2009ApJ...699..486C} {699, 486}

\bibitem[\protect\citeauthoryear{{Cook} et~al.,}{{Cook}
  et~al.}{2012}]{Cook_et_al_12}
{Cook} D.~O.,  et~al., 2012, \mn@doi [\apj] {10.1088/0004-637X/751/2/100},
  \href {https://ui.adsabs.harvard.edu/abs/2012ApJ...751..100C} {751, 100}

\bibitem[\protect\citeauthoryear{{Cook}, {Dale}, {Lee}, {Thilker}, {Calzetti}
  \& {Kennicutt}}{{Cook} et~al.}{2016}]{Cook_et_al_16}
{Cook} D.~O.,  {Dale} D.~A.,  {Lee} J.~C.,  {Thilker} D.,  {Calzetti} D.,
  {Kennicutt} R.~C.,  2016, \mn@doi [\mnras] {10.1093/mnras/stw1694}, \href
  {http://adsabs.harvard.edu/abs/2016MNRAS.462.3766C} {462, 3766}

\bibitem[\protect\citeauthoryear{{Cook} et~al.}{{Cook}
  et~al.}{2019}]{Cook_et_al_19}
{Cook} D.~O.,  et~al., 2019, \mn@doi [\mnras] {10.1093/mnras/stz331}, \href
  {https://ui.adsabs.harvard.edu/abs/2019MNRAS.484.4897C} {484, 4897}

\bibitem[\protect\citeauthoryear{{Crain} et~al.,}{{Crain} et~al.}{2015}]{C15}
{Crain} R.~A.,  et~al., 2015, \mn@doi [\mnras] {10.1093/mnras/stv725}, \href
  {http://adsabs.harvard.edu/abs/2015MNRAS.450.1937C} {450, 1937}

\bibitem[\protect\citeauthoryear{{Crain} et~al.,}{{Crain}
  et~al.}{2017}]{Crain_et_al_17}
{Crain} R.~A.,  et~al., 2017, \mn@doi [\mnras] {10.1093/mnras/stw2586}, \href
  {http://adsabs.harvard.edu/abs/2017MNRAS.464.4204C} {464, 4204}

\bibitem[\protect\citeauthoryear{{Dolag}, {Borgani}, {Murante}  \&
  {Springel}}{{Dolag} et~al.}{2009}]{Dolag_et_al_09}
{Dolag} K.,  {Borgani} S.,  {Murante} G.,   {Springel} V.,  2009, \mn@doi
  [\mnras] {10.1111/j.1365-2966.2009.15034.x}, \href
  {http://adsabs.harvard.edu/abs/2009MNRAS.399..497D} {399, 497}

\bibitem[\protect\citeauthoryear{{Dowell}, {Buckalew}  \& {Tan}}{{Dowell}
  et~al.}{2008}]{Dowell_Buckalew_and_Tan_08}
{Dowell} J.~D.,  {Buckalew} B.~A.,   {Tan} J.~C.,  2008, \mn@doi [\aj]
  {10.1088/0004-6256/135/3/823}, \href
  {http://adsabs.harvard.edu/abs/2008AJ....135..823D} {135, 823}

\bibitem[\protect\citeauthoryear{{Efremov} \& {Elmegreen}}{{Efremov} \&
  {Elmegreen}}{1998}]{Efremov_and_Elmegreen_98}
{Efremov} Y.~N.,  {Elmegreen} B.~G.,  1998, \mn@doi [\mnras]
  {10.1046/j.1365-8711.1998.01819.x}, \href
  {http://adsabs.harvard.edu/abs/1998MNRAS.299..588E} {299, 588}

\bibitem[\protect\citeauthoryear{{Elmegreen}}{{Elmegreen}}{1989}]{Elmegreen_89}
{Elmegreen} B.~G.,  1989, \mn@doi [\apj] {10.1086/167192}, \href
  {http://adsabs.harvard.edu/abs/1989ApJ...338..178E} {338, 178}

\bibitem[\protect\citeauthoryear{{Elmegreen}}{{Elmegreen}}{2002}]{Elmegreen_02a}
{Elmegreen} B.~G.,  2002, \mn@doi [\apj] {10.1086/324384}, \href
  {https://ui.adsabs.harvard.edu/abs/2002ApJ...564..773E} {564, 773}

\bibitem[\protect\citeauthoryear{{Elmegreen} \& {Efremov}}{{Elmegreen} \&
  {Efremov}}{1997}]{Elmegreen_and_Efremov_97}
{Elmegreen} B.~G.,  {Efremov} Y.~N.,  1997, \mn@doi [\apj] {10.1086/303966},
  \href {http://adsabs.harvard.edu/abs/1997ApJ...480..235E} {480, 235}

\bibitem[\protect\citeauthoryear{{Elmegreen} \& {Elmegreen}}{{Elmegreen} \&
  {Elmegreen}}{2001}]{Elmegreen_and_Elmegreen_01}
{Elmegreen} B.~G.,  {Elmegreen} D.~M.,  2001, \mn@doi [\aj] {10.1086/319416},
  \href {http://adsabs.harvard.edu/abs/2001AJ....121.1507E} {121, 1507}

\bibitem[\protect\citeauthoryear{{Elmegreen} \& {Falgarone}}{{Elmegreen} \&
  {Falgarone}}{1996}]{Elmegreen_and_Falgarone_96}
{Elmegreen} B.~G.,  {Falgarone} E.,  1996, \mn@doi [\apj] {10.1086/178009},
  \href {https://ui.adsabs.harvard.edu/abs/1996ApJ...471..816E} {471, 816}

\bibitem[\protect\citeauthoryear{{Fall} \& {Chandar}}{{Fall} \&
  {Chandar}}{2012}]{Fall_and_Chandar_12}
{Fall} S.~M.,  {Chandar} R.,  2012, \mn@doi [\apj]
  {10.1088/0004-637X/752/2/96}, \href
  {http://adsabs.harvard.edu/abs/2012ApJ...752...96F} {752, 96}

\bibitem[\protect\citeauthoryear{{Fedotov}, {Gallagher}, {Konstantopoulos},
  {Chandar}, {Bastian}, {Charlton}, {Whitmore}  \& {Trancho}}{{Fedotov}
  et~al.}{2011}]{Fedotov_et_al_11}
{Fedotov} K.,  {Gallagher} S.~C.,  {Konstantopoulos} I.~S.,  {Chandar} R.,
  {Bastian} N.,  {Charlton} J.~C.,  {Whitmore} B.,   {Trancho} G.,  2011,
  \mn@doi [\aj] {10.1088/0004-6256/142/2/42}, \href
  {http://adsabs.harvard.edu/abs/2011AJ....142...42F} {142, 42}

\bibitem[\protect\citeauthoryear{{Fonnesbeck}, {Patil}, {Huard}  \&
  {Salvatier}}{{Fonnesbeck} et~al.}{2015}]{PYMC}
{Fonnesbeck} C.,  {Patil} A.,  {Huard} D.,   {Salvatier} J.,  2015, {PyMC:
  Bayesian Stochastic Modelling in Python}, Astrophysics Source Code Library
  (\mn@eprint {ascl} {1506.005})

\bibitem[\protect\citeauthoryear{{Forbes} et~al.,}{{Forbes}
  et~al.}{2018}]{Forbes_et_al_18_review}
{Forbes} D.~A.,  et~al., 2018, \mn@doi [Proceedings of the Royal Society of
  London Series A] {10.1098/rspa.2017.0616}, \href
  {http://adsabs.harvard.edu/abs/2018RSPSA.47470616F} {474, 20170616}

\bibitem[\protect\citeauthoryear{{Fouesneau} et~al.,}{{Fouesneau}
  et~al.}{2014}]{Fouesneau_et_al_14}
{Fouesneau} M.,  et~al., 2014, \mn@doi [\apj] {10.1088/0004-637X/786/2/117},
  \href {http://adsabs.harvard.edu/abs/2014ApJ...786..117F} {786, 117}

\bibitem[\protect\citeauthoryear{{Furlong} et~al.,}{{Furlong}
  et~al.}{2015}]{Furlong_et_al_15}
{Furlong} M.,  et~al., 2015, \mn@doi [\mnras] {10.1093/mnras/stv852}, \href
  {http://adsabs.harvard.edu/abs/2015MNRAS.450.4486F} {450, 4486}

\bibitem[\protect\citeauthoryear{{Furlong} et~al.,}{{Furlong}
  et~al.}{2017}]{Furlong_et_al_17}
{Furlong} M.,  et~al., 2017, \mn@doi [\mnras] {10.1093/mnras/stw2740}, \href
  {http://adsabs.harvard.edu/abs/2017MNRAS.465..722F} {465, 722}

\bibitem[\protect\citeauthoryear{{Galleti}, {Federici}, {Bellazzini}, {Fusi
  Pecci}  \& {Macrina}}{{Galleti} et~al.}{2004}]{Galleti_et_al_04}
{Galleti} S.,  {Federici} L.,  {Bellazzini} M.,  {Fusi Pecci} F.,   {Macrina}
  S.,  2004, \mn@doi [\aap] {10.1051/0004-6361:20035632}, \href
  {http://adsabs.harvard.edu/abs/2004A%26A...416..917G} {416, 917}

\bibitem[\protect\citeauthoryear{{Gieles}}{{Gieles}}{2009}]{Gieles_09}
{Gieles} M.,  2009, \mn@doi [\mnras] {10.1111/j.1365-2966.2009.14473.x}, \href
  {http://adsabs.harvard.edu/abs/2009MNRAS.394.2113G} {394, 2113}

\bibitem[\protect\citeauthoryear{{Gieles} \& {Portegies Zwart}}{{Gieles} \&
  {Portegies Zwart}}{2011}]{Gieles_and_Portegies_Zwart_11}
{Gieles} M.,  {Portegies Zwart} S.~F.,  2011, \mn@doi [\mnras]
  {10.1111/j.1745-3933.2010.00967.x}, \href
  {https://ui.adsabs.harvard.edu/abs/2011MNRAS.410L...6G} {410, L6}

\bibitem[\protect\citeauthoryear{{Gieles}, {Bastian}, {Lamers}  \&
  {Mout}}{{Gieles} et~al.}{2005}]{Gieles_et_al_05}
{Gieles} M.,  {Bastian} N.,  {Lamers} H.~J.~G.~L.~M.,   {Mout} J.~N.,  2005,
  \mn@doi [\aap] {10.1051/0004-6361:20052914}, \href
  {http://adsabs.harvard.edu/abs/2005A%26A...441..949G} {441, 949}

\bibitem[\protect\citeauthoryear{{Gieles}, {Larsen}, {Scheepmaker}, {Bastian},
  {Haas}  \& {Lamers}}{{Gieles} et~al.}{2006a}]{Gieles_et_al_06a}
{Gieles} M.,  {Larsen} S.~S.,  {Scheepmaker} R.~A.,  {Bastian} N.,  {Haas}
  M.~R.,   {Lamers} H.~J.~G.~L.~M.,  2006a, \mn@doi [\aap]
  {10.1051/0004-6361:200500224}, \href
  {http://adsabs.harvard.edu/abs/2006A%26A...446L...9G} {446, L9}

\bibitem[\protect\citeauthoryear{{Gieles}, {Larsen}, {Bastian}  \&
  {Stein}}{{Gieles} et~al.}{2006b}]{Gieles_et_al_06b}
{Gieles} M.,  {Larsen} S.~S.,  {Bastian} N.,   {Stein} I.~T.,  2006b, \mn@doi
  [\aap] {10.1051/0004-6361:20053589}, \href
  {http://adsabs.harvard.edu/abs/2006A%26A...450..129G} {450, 129}

\bibitem[\protect\citeauthoryear{{Ginsburg} \& {Kruijssen}}{{Ginsburg} \&
  {Kruijssen}}{2018}]{Ginsburg_and_Kruijssen_18}
{Ginsburg} A.,  {Kruijssen} J.~M.~D.,  2018, \mn@doi [\apjl]
  {10.3847/2041-8213/aada89}, \href
  {http://adsabs.harvard.edu/abs/2018ApJ...864L..17G} {864, L17}

\bibitem[\protect\citeauthoryear{{Girardi}, {Bressan}, {Bertelli}  \&
  {Chiosi}}{{Girardi} et~al.}{2000}]{Girardi_et_al_00}
{Girardi} L.,  {Bressan} A.,  {Bertelli} G.,   {Chiosi} C.,  2000, \mn@doi
  [\aaps] {10.1051/aas:2000126}, \href
  {http://adsabs.harvard.edu/abs/2000A%26AS..141..371G} {141, 371}

\bibitem[\protect\citeauthoryear{{Gnedin}, {Hernquist}  \& {Ostriker}}{{Gnedin}
  et~al.}{1999}]{Gnedin_Hernquist_and_Ostriker_99}
{Gnedin} O.~Y.,  {Hernquist} L.,   {Ostriker} J.~P.,  1999, \mn@doi [\apj]
  {10.1086/306910}, \href {http://adsabs.harvard.edu/abs/1999ApJ...514..109G}
  {514, 109}

\bibitem[\protect\citeauthoryear{{Goddard}, {Bastian}  \&
  {Kennicutt}}{{Goddard} et~al.}{2010}]{Goddard_Bastian_and_Kennicutt_10}
{Goddard} Q.~E.,  {Bastian} N.,   {Kennicutt} R.~C.,  2010, \mn@doi [\mnras]
  {10.1111/j.1365-2966.2010.16511.x}, \href
  {http://adsabs.harvard.edu/abs/2010MNRAS.405..857G} {405, 857}

\bibitem[\protect\citeauthoryear{{Haydon}, {Kruijssen}, {Hygate}, {Schruba},
  {Krumholz}, {Chevance}  \& {Longmore}}{{Haydon}
  et~al.}{2018}]{Haydon_et_al_18}
{Haydon} D.~T.,  {Kruijssen} J.~M.~D.,  {Hygate} A. e. P.~S.,  {Schruba} A.,
  {Krumholz} M.~R.,  {Chevance} M.,   {Longmore} S.~N.,  2018, arXiv e-prints,
  \href {https://ui.adsabs.harvard.edu/abs/2018arXiv181010897H} {p.
  arXiv:1810.10897}

\bibitem[\protect\citeauthoryear{{Hollyhead}, {Adamo}, {Bastian}, {Gieles}  \&
  {Ryon}}{{Hollyhead} et~al.}{2016}]{Hollyhead_et_al_16}
{Hollyhead} K.,  {Adamo} A.,  {Bastian} N.,  {Gieles} M.,   {Ryon} J.~E.,
  2016, \mn@doi [\mnras] {10.1093/mnras/stw1142}, \href
  {https://ui.adsabs.harvard.edu/abs/2016MNRAS.460.2087H} {460, 2087}

\bibitem[\protect\citeauthoryear{{Hughes}, {Pfeffer}, {Martig}, {Bastian},
  {Crain}, {Kruijssen}  \& {Reina-Campos}}{{Hughes}
  et~al.}{2019}]{Hughes_et_al_19}
{Hughes} M.~E.,  {Pfeffer} J.,  {Martig} M.,  {Bastian} N.,  {Crain} R.~A.,
  {Kruijssen} J.~M.~D.,   {Reina-Campos} M.,  2019, \mn@doi [\mnras]
  {10.1093/mnras/sty2889}, \href
  {http://adsabs.harvard.edu/abs/2019MNRAS.482.2795H} {482, 2795}

\bibitem[\protect\citeauthoryear{{Johnson} et~al.,}{{Johnson}
  et~al.}{2015}]{Johnson_et_al_15}
{Johnson} L.~C.,  et~al., 2015, \mn@doi [\apj] {10.1088/0004-637X/802/2/127},
  \href {http://adsabs.harvard.edu/abs/2015ApJ...802..127J} {802, 127}

\bibitem[\protect\citeauthoryear{{Johnson} et~al.,}{{Johnson}
  et~al.}{2016}]{Johnson_et_al_16}
{Johnson} L.~C.,  et~al., 2016, \mn@doi [\apj] {10.3847/0004-637X/827/1/33},
  \href {http://adsabs.harvard.edu/abs/2016ApJ...827...33J} {827, 33}

\bibitem[\protect\citeauthoryear{{Johnson} et~al.,}{{Johnson}
  et~al.}{2017}]{Johnson_et_al_17}
{Johnson} L.~C.,  et~al., 2017, \mn@doi [\apj] {10.3847/1538-4357/aa6a1f},
  \href {http://adsabs.harvard.edu/abs/2017ApJ...839...78J} {839, 78}

\bibitem[\protect\citeauthoryear{{Jord{\'a}n} et~al.,}{{Jord{\'a}n}
  et~al.}{2007}]{Jordan_et_al_07_XII}
{Jord{\'a}n} A.,  et~al., 2007, \mn@doi [\apjs] {10.1086/516840}, \href
  {http://adsabs.harvard.edu/abs/2007ApJS..171..101J} {171, 101}

\bibitem[\protect\citeauthoryear{{Kang}, {Bianchi}  \& {Rey}}{{Kang}
  et~al.}{2009}]{Kang_Bianchi_and_Rey_09}
{Kang} Y.,  {Bianchi} L.,   {Rey} S.-C.,  2009, \mn@doi [\apj]
  {10.1088/0004-637X/703/1/614}, \href
  {http://adsabs.harvard.edu/abs/2009ApJ...703..614K} {703, 614}

\bibitem[\protect\citeauthoryear{{Kennicutt}}{{Kennicutt}}{1998}]{Kennicutt_98}
{Kennicutt} Jr. R.~C.,  1998, \mn@doi [\apj] {10.1086/305588}, \href
  {http://adsabs.harvard.edu/abs/1998ApJ...498..541K} {498, 541}

\bibitem[\protect\citeauthoryear{{Kennicutt} \& {Evans}}{{Kennicutt} \&
  {Evans}}{2012}]{Kennicutt_and_Evans_12}
{Kennicutt} R.~C.,  {Evans} N.~J.,  2012, \mn@doi [\araa]
  {10.1146/annurev-astro-081811-125610}, \href
  {http://adsabs.harvard.edu/abs/2012ARA%26A..50..531K} {50, 531}

\bibitem[\protect\citeauthoryear{{Kim} et~al.,}{{Kim}
  et~al.}{2018}]{Kim_et_al_18}
{Kim} J.-h.,  et~al., 2018, \mn@doi [\mnras] {10.1093/mnras/stx2994}, \href
  {http://adsabs.harvard.edu/abs/2018MNRAS.474.4232K} {474, 4232}

\bibitem[\protect\citeauthoryear{{Kruijssen}}{{Kruijssen}}{2012}]{Kruijssen_12}
{Kruijssen} J.~M.~D.,  2012, \mn@doi [\mnras]
  {10.1111/j.1365-2966.2012.21923.x}, \href
  {http://adsabs.harvard.edu/abs/2012MNRAS.426.3008K} {426, 3008}

\bibitem[\protect\citeauthoryear{{Kruijssen}}{{Kruijssen}}{2014}]{Kruijssen_14}
{Kruijssen} J.~M.~D.,  2014, \mn@doi [Classical and Quantum Gravity]
  {10.1088/0264-9381/31/24/244006}, \href
  {http://adsabs.harvard.edu/abs/2014CQGra..31x4006K} {31, 244006}

\bibitem[\protect\citeauthoryear{{Kruijssen}}{{Kruijssen}}{2015}]{Kruijssen_15}
{Kruijssen} J.~M.~D.,  2015, \mn@doi [\mnras] {10.1093/mnras/stv2026}, \href
  {http://adsabs.harvard.edu/abs/2015MNRAS.454.1658K} {454, 1658}

\bibitem[\protect\citeauthoryear{{Kruijssen} \& {Bastian}}{{Kruijssen} \&
  {Bastian}}{2016}]{Kruijssen_and_Bastian_16}
{Kruijssen} J.~M.~D.,  {Bastian} N.,  2016, \mn@doi [\mnras]
  {10.1093/mnrasl/slv182}, \href
  {http://adsabs.harvard.edu/abs/2016MNRAS.457L..24K} {457, L24}

\bibitem[\protect\citeauthoryear{{Kruijssen} \& {Longmore}}{{Kruijssen} \&
  {Longmore}}{2014}]{Kruijssen_and_Longmore_14}
{Kruijssen} J.~M.~D.,  {Longmore} S.~N.,  2014, \mn@doi [\mnras]
  {10.1093/mnras/stu098}, \href
  {http://adsabs.harvard.edu/abs/2014MNRAS.439.3239K} {439, 3239}

\bibitem[\protect\citeauthoryear{{Kruijssen}, {Pelupessy}, {Lamers}, {Portegies
  Zwart}  \& {Icke}}{{Kruijssen} et~al.}{2011}]{Kruijssen_et_al_11}
{Kruijssen} J.~M.~D.,  {Pelupessy} F.~I.,  {Lamers} H.~J.~G.~L.~M.,  {Portegies
  Zwart} S.~F.,   {Icke} V.,  2011, \mn@doi [\mnras]
  {10.1111/j.1365-2966.2011.18467.x}, \href
  {http://adsabs.harvard.edu/abs/2011MNRAS.414.1339K} {414, 1339}

\bibitem[\protect\citeauthoryear{{Kruijssen}, {Pelupessy}, {Lamers}, {Portegies
  Zwart}, {Bastian}  \& {Icke}}{{Kruijssen} et~al.}{2012}]{Kruijssen_et_al_12}
{Kruijssen} J.~M.~D.,  {Pelupessy} F.~I.,  {Lamers} H.~J.~G.~L.~M.,  {Portegies
  Zwart} S.~F.,  {Bastian} N.,   {Icke} V.,  2012, \mn@doi [\mnras]
  {10.1111/j.1365-2966.2012.20322.x}, \href
  {http://adsabs.harvard.edu/abs/2012MNRAS.421.1927K} {421, 1927}

\bibitem[\protect\citeauthoryear{{Kruijssen}, {Pfeffer}, {Crain}  \&
  {Bastian}}{{Kruijssen} et~al.}{2019a}]{K19}
{Kruijssen} J.~M.~D.,  {Pfeffer} J.~L.,  {Crain} R.~A.,   {Bastian} N.,  2019a,
  \mn@doi [\mnras] {10.1093/mnras/stz968}, \href
  {https://ui.adsabs.harvard.edu/abs/2019MNRAS.486.3134K} {486, 3134}

\bibitem[\protect\citeauthoryear{{Kruijssen}, {Pfeffer}, {Reina-Campos},
  {Crain}  \& {Bastian}}{{Kruijssen} et~al.}{2019b}]{K19b}
{Kruijssen} J.~M.~D.,  {Pfeffer} J.~L.,  {Reina-Campos} M.,  {Crain} R.~A.,
  {Bastian} N.,  2019b, \mn@doi [\mnras] {10.1093/mnras/sty1609}, \href
  {https://ui.adsabs.harvard.edu/abs/2019MNRAS.486.3180K} {486, 3180}

\bibitem[\protect\citeauthoryear{{Krumholz} \& {McKee}}{{Krumholz} \&
  {McKee}}{2005}]{Krumholz_and_McKee_05}
{Krumholz} M.~R.,  {McKee} C.~F.,  2005, \mn@doi [\apj] {10.1086/431734}, \href
  {http://adsabs.harvard.edu/abs/2005ApJ...630..250K} {630, 250}

\bibitem[\protect\citeauthoryear{{Krumholz}, {McKee}  \& {Bland
  -Hawthorn}}{{Krumholz} et~al.}{2018}]{Krumholz_McKee_and_Bland-Hawthorn_18}
{Krumholz} M.~R.,  {McKee} C.~F.,   {Bland -Hawthorn} J.,  2018, arXiv
  e-prints, \href {https://ui.adsabs.harvard.edu/abs/2018arXiv181201615K} {p.
  arXiv:1812.01615}

\bibitem[\protect\citeauthoryear{{Lagos} et~al.}{{Lagos}
  et~al.}{2015}]{Lagos_et_al_15_short}
{Lagos} C.~d.~P.,  et~al., 2015, \mn@doi [\mnras] {10.1093/mnras/stv1488},
  \href {http://adsabs.harvard.edu/abs/2015MNRAS.452.3815L} {452, 3815}

\bibitem[\protect\citeauthoryear{{Lagos} et~al.,}{{Lagos}
  et~al.}{2016}]{Lagos_et_al_16}
{Lagos} C.~d.~P.,  et~al., 2016, \mn@doi [\mnras] {10.1093/mnras/stw717}, \href
  {http://adsabs.harvard.edu/abs/2016MNRAS.459.2632D} {459, 2632}

\bibitem[\protect\citeauthoryear{{Lamers}}{{Lamers}}{2009}]{Lamers_09}
{Lamers} H.~J.~G.~L.~M.,  2009, \mn@doi [\apss] {10.1007/s10509-009-0125-4},
  \href {http://adsabs.harvard.edu/abs/2009Ap%26SS.324..183L} {324, 183}

\bibitem[\protect\citeauthoryear{{Lamers}, {Gieles}, {Bastian}, {Baumgardt},
  {Kharchenko}  \& {Portegies Zwart}}{{Lamers} et~al.}{2005}]{Lamers_et_al_05b}
{Lamers} H.~J.~G.~L.~M.,  {Gieles} M.,  {Bastian} N.,  {Baumgardt} H.,
  {Kharchenko} N.~V.,   {Portegies Zwart} S.,  2005, \mn@doi [\aap]
  {10.1051/0004-6361:20042241}, \href
  {http://adsabs.harvard.edu/abs/2005A%26A...441..117L} {441, 117}

\bibitem[\protect\citeauthoryear{{Larsen}}{{Larsen}}{2002}]{Larsen_02}
{Larsen} S.~S.,  2002, \mn@doi [\aj] {10.1086/342381}, \href
  {http://adsabs.harvard.edu/abs/2002AJ....124.1393L} {124, 1393}

\bibitem[\protect\citeauthoryear{{Larsen}}{{Larsen}}{2009}]{Larsen_09}
{Larsen} S.~S.,  2009, \mn@doi [\aap] {10.1051/0004-6361:200811212}, \href
  {http://adsabs.harvard.edu/abs/2009A%26A...494..539L} {494, 539}

\bibitem[\protect\citeauthoryear{{Larsen} \& {Richtler}}{{Larsen} \&
  {Richtler}}{1999}]{Larsen_and_Richtler_99}
{Larsen} S.~S.,  {Richtler} T.,  1999, \aap, \href
  {http://adsabs.harvard.edu/abs/1999A%26A...345...59L} {345, 59}

\bibitem[\protect\citeauthoryear{{Larsen} \& {Richtler}}{{Larsen} \&
  {Richtler}}{2000}]{Larsen_and_Richtler_00}
{Larsen} S.~S.,  {Richtler} T.,  2000, \aap, \href
  {http://adsabs.harvard.edu/abs/2000A%26A...354..836L} {354, 836}

\bibitem[\protect\citeauthoryear{{Lewis} et~al.,}{{Lewis}
  et~al.}{2015}]{Lewis_et_al_15}
{Lewis} A.~R.,  et~al., 2015, \mn@doi [\apj] {10.1088/0004-637X/805/2/183},
  \href {http://adsabs.harvard.edu/abs/2015ApJ...805..183L} {805, 183}

\bibitem[\protect\citeauthoryear{{Li}, {Mac Low}  \& {Klessen}}{{Li}
  et~al.}{2004}]{Li_MacLow_Klessen_04}
{Li} Y.,  {Mac Low} M.-M.,   {Klessen} R.~S.,  2004, \mn@doi [\apjl]
  {10.1086/425320}, \href {http://adsabs.harvard.edu/abs/2004ApJ...614L..29L}
  {614, L29}

\bibitem[\protect\citeauthoryear{{Li}, {Gnedin}, {Gnedin}, {Meng}, {Semenov}
  \& {Kravtsov}}{{Li} et~al.}{2017}]{Li_et_al_17}
{Li} H.,  {Gnedin} O.~Y.,  {Gnedin} N.~Y.,  {Meng} X.,  {Semenov} V.~A.,
  {Kravtsov} A.~V.,  2017, \mn@doi [\apj] {10.3847/1538-4357/834/1/69}, \href
  {http://adsabs.harvard.edu/abs/2017ApJ...834...69L} {834, 69}

\bibitem[\protect\citeauthoryear{{Longmore} et~al.,}{{Longmore}
  et~al.}{2014}]{Longmore_et_al_14}
{Longmore} S.~N.,  et~al., 2014, \mn@doi [Protostars and Planets VI]
  {10.2458/azu_uapress_9780816531240-ch013}, \href
  {http://adsabs.harvard.edu/abs/2014prpl.conf..291L} {pp 291--314}

\bibitem[\protect\citeauthoryear{{Maji}, {Zhu}, {Li}, {Charlton}, {Hernquist}
  \& {Knebe}}{{Maji} et~al.}{2017}]{Maji_et_al_17}
{Maji} M.,  {Zhu} Q.,  {Li} Y.,  {Charlton} J.,  {Hernquist} L.,   {Knebe} A.,
  2017, \mn@doi [\apj] {10.3847/1538-4357/aa7aa1}, \href
  {http://adsabs.harvard.edu/abs/2017ApJ...844..108M} {844, 108}

\bibitem[\protect\citeauthoryear{{Marasco}, {Crain}, {Schaye}, {Bah{\'e}}, {van
  der Hulst}, {Theuns}  \& {Bower}}{{Marasco} et~al.}{2016}]{Marasco_et_al_16}
{Marasco} A.,  {Crain} R.~A.,  {Schaye} J.,  {Bah{\'e}} Y.~M.,  {van der Hulst}
  T.,  {Theuns} T.,   {Bower} R.~G.,  2016, \mn@doi [\mnras]
  {10.1093/mnras/stw1498}, \href
  {http://adsabs.harvard.edu/abs/2016MNRAS.461.2630M} {461, 2630}

\bibitem[\protect\citeauthoryear{{Marigo} \& {Girardi}}{{Marigo} \&
  {Girardi}}{2007}]{Marigo_and_Girardi_07}
{Marigo} P.,  {Girardi} L.,  2007, \mn@doi [\aap] {10.1051/0004-6361:20066772},
  \href {http://adsabs.harvard.edu/abs/2007A%26A...469..239M} {469, 239}

\bibitem[\protect\citeauthoryear{{Marigo}, {Girardi}, {Bressan}, {Groenewegen},
  {Silva}  \& {Granato}}{{Marigo} et~al.}{2008}]{Marigo_et_al_08}
{Marigo} P.,  {Girardi} L.,  {Bressan} A.,  {Groenewegen} M.~A.~T.,  {Silva}
  L.,   {Granato} G.~L.,  2008, \mn@doi [\aap] {10.1051/0004-6361:20078467},
  \href {http://adsabs.harvard.edu/abs/2008A%26A...482..883M} {482, 883}

\bibitem[\protect\citeauthoryear{{McConnachie}, {Irwin}, {Ferguson}, {Ibata},
  {Lewis}  \& {Tanvir}}{{McConnachie} et~al.}{2005}]{McConnachie_et_al_05}
{McConnachie} A.~W.,  {Irwin} M.~J.,  {Ferguson} A.~M.~N.,  {Ibata} R.~A.,
  {Lewis} G.~F.,   {Tanvir} N.,  2005, \mn@doi [\mnras]
  {10.1111/j.1365-2966.2004.08514.x}, \href
  {http://adsabs.harvard.edu/abs/2005MNRAS.356..979M} {356, 979}

\bibitem[\protect\citeauthoryear{{McCrady} \& {Graham}}{{McCrady} \&
  {Graham}}{2007}]{McCrady_and_Graham_07}
{McCrady} N.,  {Graham} J.~R.,  2007, \mn@doi [\apj] {10.1086/518357}, \href
  {http://adsabs.harvard.edu/abs/2007ApJ...663..844M} {663, 844}

\bibitem[\protect\citeauthoryear{{Mengel} \& {Tacconi-Garman}}{{Mengel} \&
  {Tacconi-Garman}}{2007}]{Mengel_and_Tacconi-Garman_07}
{Mengel} S.,  {Tacconi-Garman} L.~E.,  2007, \mn@doi [\aap]
  {10.1051/0004-6361:20066717}, \href
  {http://adsabs.harvard.edu/abs/2007A%26A...466..151M} {466, 151}

\bibitem[\protect\citeauthoryear{{Messa} et~al.,}{{Messa}
  et~al.}{2018a}]{Messa_et_al_18_I}
{Messa} M.,  et~al., 2018a, \mn@doi [\mnras] {10.1093/mnras/stx2403}, \href
  {http://adsabs.harvard.edu/abs/2018MNRAS.473..996M} {473, 996}

\bibitem[\protect\citeauthoryear{{Messa} et~al.,}{{Messa}
  et~al.}{2018b}]{Messa_et_al_18_II}
{Messa} M.,  et~al., 2018b, \mn@doi [\mnras] {10.1093/mnras/sty577}, \href
  {http://adsabs.harvard.edu/abs/2018MNRAS.477.1683M} {477, 1683}

\bibitem[\protect\citeauthoryear{{Miholics}, {Kruijssen}  \&
  {Sills}}{{Miholics} et~al.}{2017}]{Miholics_Kruijssen_and_Sills_17}
{Miholics} M.,  {Kruijssen} J.~M.~D.,   {Sills} A.,  2017, \mn@doi [\mnras]
  {10.1093/mnras/stx1312}, \href
  {http://adsabs.harvard.edu/abs/2017MNRAS.470.1421M} {470, 1421}

\bibitem[\protect\citeauthoryear{{Oppenheimer} et~al.,}{{Oppenheimer}
  et~al.}{2016}]{Oppenheimer_et_al_16}
{Oppenheimer} B.~D.,  et~al., 2016, \mn@doi [\mnras] {10.1093/mnras/stw1066},
  \href {http://adsabs.harvard.edu/abs/2016MNRAS.460.2157O} {460, 2157}

\bibitem[\protect\citeauthoryear{{Pelupessy}, {van der Werf}  \&
  {Icke}}{{Pelupessy} et~al.}{2004}]{Pelupessy_et_al_04}
{Pelupessy} F.~I.,  {van der Werf} P.~P.,   {Icke} V.,  2004, \mn@doi [\aap]
  {10.1051/0004-6361:20047071}, \href
  {http://adsabs.harvard.edu/abs/2004A%26A...422...55P} {422, 55}

\bibitem[\protect\citeauthoryear{{Pfeffer}, {Kruijssen}, {Crain}  \&
  {Bastian}}{{Pfeffer} et~al.}{2018}]{P18}
{Pfeffer} J.,  {Kruijssen} J.~M.~D.,  {Crain} R.~A.,   {Bastian} N.,  2018,
  \mn@doi [\mnras] {10.1093/mnras/stx3124}, \href
  {http://adsabs.harvard.edu/abs/2018MNRAS.475.4309P} {475, 4309}

\bibitem[\protect\citeauthoryear{{Pfeffer}, {Bastian}, {Crain}, {Kruijssen},
  {Hughes}  \& {Reina-Campos}}{{Pfeffer} et~al.}{2019}]{Pfeffer_et_al_19}
{Pfeffer} J.,  {Bastian} N.,  {Crain} R.~A.,  {Kruijssen} J.~M.~D.,  {Hughes}
  M.~E.,   {Reina-Campos} M.,  2019, \mn@doi [\mnras] {10.1093/mnras/stz1592},
  \href {https://ui.adsabs.harvard.edu/abs/2019MNRAS.487.4550P} {487, 4550}

\bibitem[\protect\citeauthoryear{{Planck Collaboration} et~al.,}{{Planck
  Collaboration} et~al.}{2014}]{Planck_2014_paperI}
{Planck Collaboration} et~al., 2014, \mn@doi [\aap]
  {10.1051/0004-6361/201321529}, \href
  {http://adsabs.harvard.edu/abs/2014A%26A...571A...1P} {571, A1}

\bibitem[\protect\citeauthoryear{{Portegies Zwart}, {McMillan}  \&
  {Gieles}}{{Portegies Zwart}
  et~al.}{2010}]{Portegies-Zwart_McMillan_and_Gieles_10}
{Portegies Zwart} S.~F.,  {McMillan} S.~L.~W.,   {Gieles} M.,  2010, \mn@doi
  [\araa] {10.1146/annurev-astro-081309-130834}, \href
  {http://adsabs.harvard.edu/abs/2010ARA%26A..48..431P} {48, 431}

\bibitem[\protect\citeauthoryear{{Prieto} \& {Gnedin}}{{Prieto} \&
  {Gnedin}}{2008}]{Prieto_and_Gnedin_08}
{Prieto} J.~L.,  {Gnedin} O.~Y.,  2008, \mn@doi [\apj] {10.1086/591777}, \href
  {http://adsabs.harvard.edu/abs/2008ApJ...689..919P} {689, 919}

\bibitem[\protect\citeauthoryear{{Rahmati}, {Schaye}, {Bower}, {Crain},
  {Furlong}, {Schaller}  \& {Theuns}}{{Rahmati}
  et~al.}{2015}]{Rahmati_et_al_15}
{Rahmati} A.,  {Schaye} J.,  {Bower} R.~G.,  {Crain} R.~A.,  {Furlong} M.,
  {Schaller} M.,   {Theuns} T.,  2015, \mn@doi [\mnras]
  {10.1093/mnras/stv1414}, \href
  {http://adsabs.harvard.edu/abs/2015MNRAS.452.2034R} {452, 2034}

\bibitem[\protect\citeauthoryear{{Rahmati}, {Schaye}, {Crain}, {Oppenheimer},
  {Schaller}  \& {Theuns}}{{Rahmati} et~al.}{2016}]{Rahmati_et_al_16}
{Rahmati} A.,  {Schaye} J.,  {Crain} R.~A.,  {Oppenheimer} B.~D.,  {Schaller}
  M.,   {Theuns} T.,  2016, \mn@doi [\mnras] {10.1093/mnras/stw453}, \href
  {http://adsabs.harvard.edu/abs/2016MNRAS.459..310R} {459, 310}

\bibitem[\protect\citeauthoryear{{Reina-Campos} \& {Kruijssen}}{{Reina-Campos}
  \& {Kruijssen}}{2017}]{Reina-Campos_and_Kruijssen_17}
{Reina-Campos} M.,  {Kruijssen} J.~M.~D.,  2017, \mn@doi [\mnras]
  {10.1093/mnras/stx790}, \href
  {http://adsabs.harvard.edu/abs/2017MNRAS.469.1282R} {469, 1282}

\bibitem[\protect\citeauthoryear{{Reina-Campos}, {Kruijssen}, {Pfeffer},
  {Bastian}  \& {Crain}}{{Reina-Campos} et~al.}{2018}]{Reina-Campos_et_al_18}
{Reina-Campos} M.,  {Kruijssen} J.~M.~D.,  {Pfeffer} J.,  {Bastian} N.,
  {Crain} R.~A.,  2018, \mn@doi [\mnras] {10.1093/mnras/sty2451}, \href
  {http://adsabs.harvard.edu/abs/2018MNRAS.481.2851R} {481, 2851}

\bibitem[\protect\citeauthoryear{{Reina-Campos}, {Kruijssen}, {Pfeffer},
  {Bastian}  \& {Crain}}{{Reina-Campos} et~al.}{2019}]{Reina-Campos_et_al_19}
{Reina-Campos} M.,  {Kruijssen} J.~M.~D.,  {Pfeffer} J.~L.,  {Bastian} N.,
  {Crain} R.~A.,  2019, \mn@doi [\mnras] {10.1093/mnras/stz1236}, \href
  {https://ui.adsabs.harvard.edu/abs/2019MNRAS.486.5838R} {486, 5838}

\bibitem[\protect\citeauthoryear{{Renaud}, {Bournaud}  \& {Duc}}{{Renaud}
  et~al.}{2015}]{Renaud_Bournaud_and_Duc_15}
{Renaud} F.,  {Bournaud} F.,   {Duc} P.-A.,  2015, \mn@doi [\mnras]
  {10.1093/mnras/stu2208}, \href
  {http://adsabs.harvard.edu/abs/2015MNRAS.446.2038R} {446, 2038}

\bibitem[\protect\citeauthoryear{{Rosas-Guevara} et~al.,}{{Rosas-Guevara}
  et~al.}{2015}]{Rosas_Guevara_et_al_15}
{Rosas-Guevara} Y.~M.,  et~al., 2015, \mn@doi [\mnras] {10.1093/mnras/stv2056},
  \href {http://adsabs.harvard.edu/abs/2015MNRAS.454.1038R} {454, 1038}

\bibitem[\protect\citeauthoryear{{Ryon} et~al.,}{{Ryon}
  et~al.}{2014}]{Ryon_et_al_14}
{Ryon} J.~E.,  et~al., 2014, \mn@doi [\aj] {10.1088/0004-6256/148/2/33}, \href
  {http://adsabs.harvard.edu/abs/2014AJ....148...33R} {148, 33}

\bibitem[\protect\citeauthoryear{{Salpeter}}{{Salpeter}}{1955}]{Salpeter_55}
{Salpeter} E.~E.,  1955, \apj, \href
  {http://ukads.nottingham.ac.uk/abs/1955ApJ...121..161S} {121, 161}

\bibitem[\protect\citeauthoryear{{S{\'a}nchez-Bl{\'a}zquez}
  et~al.,}{{S{\'a}nchez-Bl{\'a}zquez} et~al.}{2006}]{Sanchez-Blazquez_et_al_06}
{S{\'a}nchez-Bl{\'a}zquez} P.,  et~al., 2006, \mn@doi [\mnras]
  {10.1111/j.1365-2966.2006.10699.x}, \href
  {http://adsabs.harvard.edu/abs/2006MNRAS.371..703S} {371, 703}

\bibitem[\protect\citeauthoryear{{Schaye}}{{Schaye}}{2004}]{Schaye_04}
{Schaye} J.,  2004, \mn@doi [\apj] {10.1086/421232}, \href
  {http://adsabs.harvard.edu/abs/2004ApJ...609..667S} {609, 667}

\bibitem[\protect\citeauthoryear{{Schaye} \& {Dalla Vecchia}}{{Schaye} \&
  {Dalla Vecchia}}{2008}]{Schaye_and_Dalla_Vecchia_08}
{Schaye} J.,  {Dalla Vecchia} C.,  2008, \mn@doi [\mnras]
  {10.1111/j.1365-2966.2007.12639.x}, \href
  {http://adsabs.harvard.edu/abs/2008MNRAS.383.1210S} {383, 1210}

\bibitem[\protect\citeauthoryear{{Schaye} et~al.,}{{Schaye} et~al.}{2015}]{S15}
{Schaye} J.,  et~al., 2015, \mn@doi [\mnras] {10.1093/mnras/stu2058}, \href
  {http://adsabs.harvard.edu/abs/2015MNRAS.446..521S} {446, 521}

\bibitem[\protect\citeauthoryear{{Schechter}}{{Schechter}}{1976}]{Schechter_76}
{Schechter} P.,  1976, \mn@doi [\apj] {10.1086/154079}, \href
  {http://adsabs.harvard.edu/abs/1976ApJ...203..297S} {203, 297}

\bibitem[\protect\citeauthoryear{{Schiminovich} et~al.,}{{Schiminovich}
  et~al.}{2007}]{Schiminovich_et_al_07}
{Schiminovich} D.,  et~al., 2007, \mn@doi [\apjs] {10.1086/524659}, \href
  {http://adsabs.harvard.edu/abs/2007ApJS..173..315S} {173, 315}

\bibitem[\protect\citeauthoryear{{Silva-Villa} \& {Larsen}}{{Silva-Villa} \&
  {Larsen}}{2011}]{Silva-Villa_and_Larsen_11}
{Silva-Villa} E.,  {Larsen} S.~S.,  2011, \mn@doi [\aap]
  {10.1051/0004-6361/201016206}, \href
  {https://ui.adsabs.harvard.edu/abs/2011A%26A...529A..25S} {529, A25}

\bibitem[\protect\citeauthoryear{{Silva-Villa}, {Adamo}  \&
  {Bastian}}{{Silva-Villa} et~al.}{2013}]{Silva-Villa_et_al_13}
{Silva-Villa} E.,  {Adamo} A.,   {Bastian} N.,  2013, \mn@doi [\mnras]
  {10.1093/mnrasl/slt115}, \href
  {http://adsabs.harvard.edu/abs/2013MNRAS.436L..69S} {436, L69}

\bibitem[\protect\citeauthoryear{{Silva-Villa}, {Adamo}, {Bastian}, {Fouesneau}
   \& {Zackrisson}}{{Silva-Villa} et~al.}{2014}]{Silva-Villa_et_al_14}
{Silva-Villa} E.,  {Adamo} A.,  {Bastian} N.,  {Fouesneau} M.,   {Zackrisson}
  E.,  2014, \mn@doi [\mnras] {10.1093/mnrasl/slu028}, \href
  {http://adsabs.harvard.edu/abs/2014MNRAS.440L.116S} {440, L116}

\bibitem[\protect\citeauthoryear{{Springel}}{{Springel}}{2005}]{Springel_05}
{Springel} V.,  2005, \mn@doi [\mnras] {10.1111/j.1365-2966.2005.09655.x},
  \href
  {http://adsabs.harvard.edu/cgi-bin/nph-bib_query?bibcode=2005MNRAS.364.1105S&db_key=AST}
  {364, 1105}

\bibitem[\protect\citeauthoryear{{Springel}, {White}, {Tormen}  \&
  {Kauffmann}}{{Springel} et~al.}{2001}]{Springel_et_al_01}
{Springel} V.,  {White} S.~D.~M.,  {Tormen} G.,   {Kauffmann} G.,  2001,
  \mn@doi [\mnras] {10.1046/j.1365-8711.2001.04912.x}, \href
  {http://adsabs.harvard.edu/abs/2001MNRAS.328..726S} {328, 726}

\bibitem[\protect\citeauthoryear{{Toomre}}{{Toomre}}{1964}]{Toomre_64}
{Toomre} A.,  1964, \mn@doi [\apj] {10.1086/147861}, \href
  {http://adsabs.harvard.edu/abs/1964ApJ...139.1217T} {139, 1217}

\bibitem[\protect\citeauthoryear{{Trayford} et~al.}{{Trayford}
  et~al.}{2015}]{Trayford_et_al_15_short}
{Trayford} J.~W.,  et~al., 2015, \mn@doi [\mnras] {10.1093/mnras/stv1461},
  \href {http://adsabs.harvard.edu/abs/2015MNRAS.452.2879T} {452, 2879}

\bibitem[\protect\citeauthoryear{{Trujillo-Gomez}, {Reina-Campos}  \&
  {Kruijssen}}{{Trujillo-Gomez} et~al.}{2019}]{Trujillo-Gomez_et_al_19}
{Trujillo-Gomez} S.,  {Reina-Campos} M.,   {Kruijssen} J.~M.~D.,  2019,
  \mnras~in~press, arXiv:1907.04861

\bibitem[\protect\citeauthoryear{{Turner}, {Schaye}, {Crain}, {Theuns}  \&
  {Wendt}}{{Turner} et~al.}{2016}]{Turner_et_al_16}
{Turner} M.~L.,  {Schaye} J.,  {Crain} R.~A.,  {Theuns} T.,   {Wendt} M.,
  2016, \mn@doi [\mnras] {10.1093/mnras/stw1816}, \href
  {http://adsabs.harvard.edu/abs/2016MNRAS.462.2440T} {462, 2440}

\bibitem[\protect\citeauthoryear{{Turner}, {Schaye}, {Crain}, {Rudie},
  {Steidel}, {Strom}  \& {Theuns}}{{Turner} et~al.}{2017}]{Turner_et_al_17}
{Turner} M.~L.,  {Schaye} J.,  {Crain} R.~A.,  {Rudie} G.,  {Steidel} C.~C.,
  {Strom} A.,   {Theuns} T.,  2017, \mn@doi [\mnras] {10.1093/mnras/stx1616},
  \href {http://adsabs.harvard.edu/abs/2017MNRAS.471..690T} {471, 690}

\bibitem[\protect\citeauthoryear{{Usher}, {Pfeffer}, {Bastian}, {Kruijssen},
  {Crain}  \& {Reina-Campos}}{{Usher} et~al.}{2018}]{Usher_et_al_18}
{Usher} C.,  {Pfeffer} J.,  {Bastian} N.,  {Kruijssen} J.~M.~D.,  {Crain}
  R.~A.,   {Reina-Campos} M.,  2018, \mn@doi [\mnras] {10.1093/mnras/sty1895},
  \href {http://adsabs.harvard.edu/abs/2018MNRAS.480.3279U} {480, 3279}

\bibitem[\protect\citeauthoryear{{Ward} \& {Kruijssen}}{{Ward} \&
  {Kruijssen}}{2018}]{Ward_and_Kruijssen_18}
{Ward} J.~L.,  {Kruijssen} J.~M.~D.,  2018, \mn@doi [\mnras]
  {10.1093/mnras/sty117}, \href
  {https://ui.adsabs.harvard.edu/abs/2018MNRAS.475.5659W} {475, 5659}

\bibitem[\protect\citeauthoryear{{Weidner}, {Kroupa}  \& {Larsen}}{{Weidner}
  et~al.}{2004}]{Weidner_Kroupa_and_Larsen_04}
{Weidner} C.,  {Kroupa} P.,   {Larsen} S.~S.,  2004, \mn@doi [\mnras]
  {10.1111/j.1365-2966.2004.07758.x}, \href
  {http://adsabs.harvard.edu/abs/2004MNRAS.350.1503W} {350, 1503}

\bibitem[\protect\citeauthoryear{{Whitmore}, {Chandar}  \& {Fall}}{{Whitmore}
  et~al.}{2007}]{Whitmore_Chandar_and_Fall_07}
{Whitmore} B.~C.,  {Chandar} R.,   {Fall} S.~M.,  2007, \mn@doi [\aj]
  {10.1086/510288}, \href {http://adsabs.harvard.edu/abs/2007AJ....133.1067W}
  {133, 1067}

\bibitem[\protect\citeauthoryear{{Whitmore}, {Chandar}, {Bowers}, {Larsen},
  {Lindsay}, {Ansari}  \& {Evans}}{{Whitmore} et~al.}{2014}]{Whitmore_et_al_14}
{Whitmore} B.~C.,  {Chandar} R.,  {Bowers} A.~S.,  {Larsen} S.,  {Lindsay} K.,
  {Ansari} A.,   {Evans} J.,  2014, \mn@doi [\aj] {10.1088/0004-6256/147/4/78},
  \href {http://adsabs.harvard.edu/abs/2014AJ....147...78W} {147, 78}

\bibitem[\protect\citeauthoryear{{Wiersma}, {Schaye}  \& {Smith}}{{Wiersma}
  et~al.}{2009a}]{Wiersma_Schaye_and_Smith_09}
{Wiersma} R.~P.~C.,  {Schaye} J.,   {Smith} B.~D.,  2009a, \mn@doi [\mnras]
  {10.1111/j.1365-2966.2008.14191.x}, \href
  {http://adsabs.harvard.edu/abs/2009MNRAS.393...99W} {393, 99}

\bibitem[\protect\citeauthoryear{{Wiersma}, {Schaye}, {Theuns}, {Dalla Vecchia}
   \& {Tornatore}}{{Wiersma} et~al.}{2009b}]{Wiersma_et_al_09}
{Wiersma} R.~P.~C.,  {Schaye} J.,  {Theuns} T.,  {Dalla Vecchia} C.,
  {Tornatore} L.,  2009b, \mn@doi [\mnras] {10.1111/j.1365-2966.2009.15331.x},
  \href {http://adsabs.harvard.edu/abs/2009MNRAS.399..574W} {399, 574}

\bibitem[\protect\citeauthoryear{{Zhang} \& {Fall}}{{Zhang} \&
  {Fall}}{1999}]{Zhang_and_Fall_99}
{Zhang} Q.,  {Fall} S.~M.,  1999, \mn@doi [\apjl] {10.1086/312412}, \href
  {http://adsabs.harvard.edu/abs/1999ApJ...527L..81Z} {527, L81}

\makeatother
\end{thebibliography}



\appendix

\section{Relationships between the CFE, pressure and $\SigmaSFR$} \label{app:CFE-P-SigmaSFR}

\begin{figure}
  \centering
  \includegraphics[width=80mm]{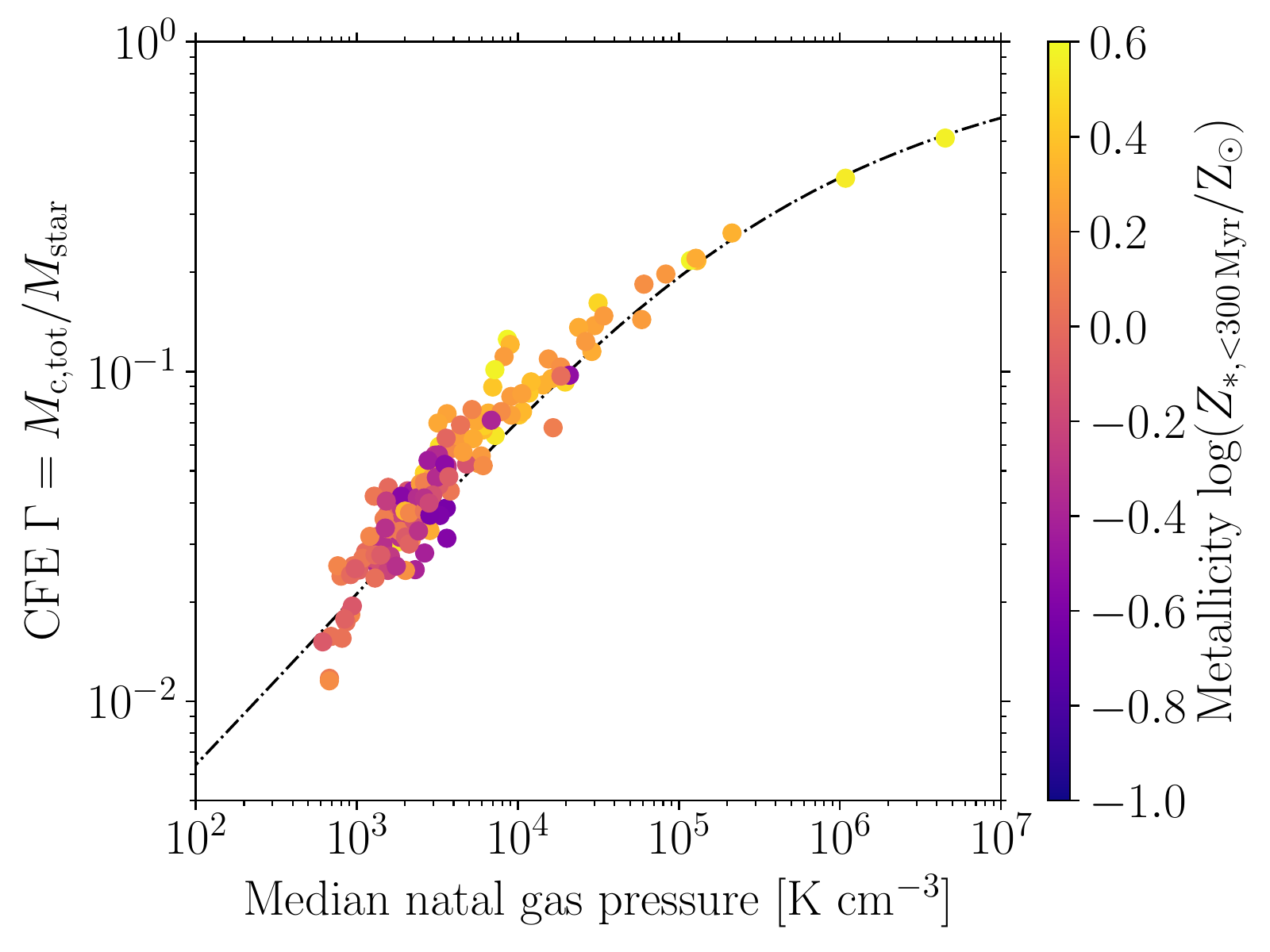}
  \includegraphics[width=80mm]{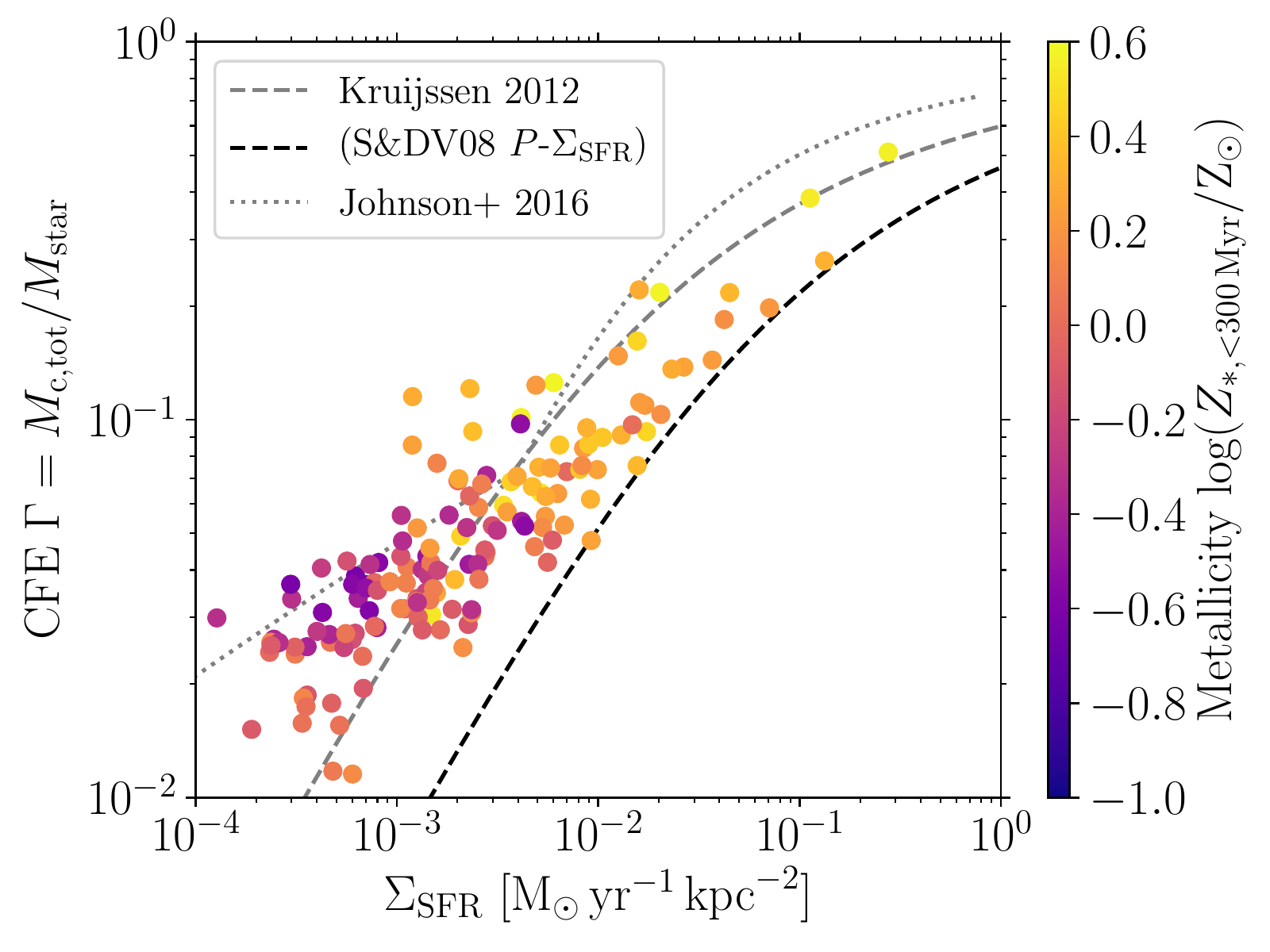}
  \includegraphics[width=80mm]{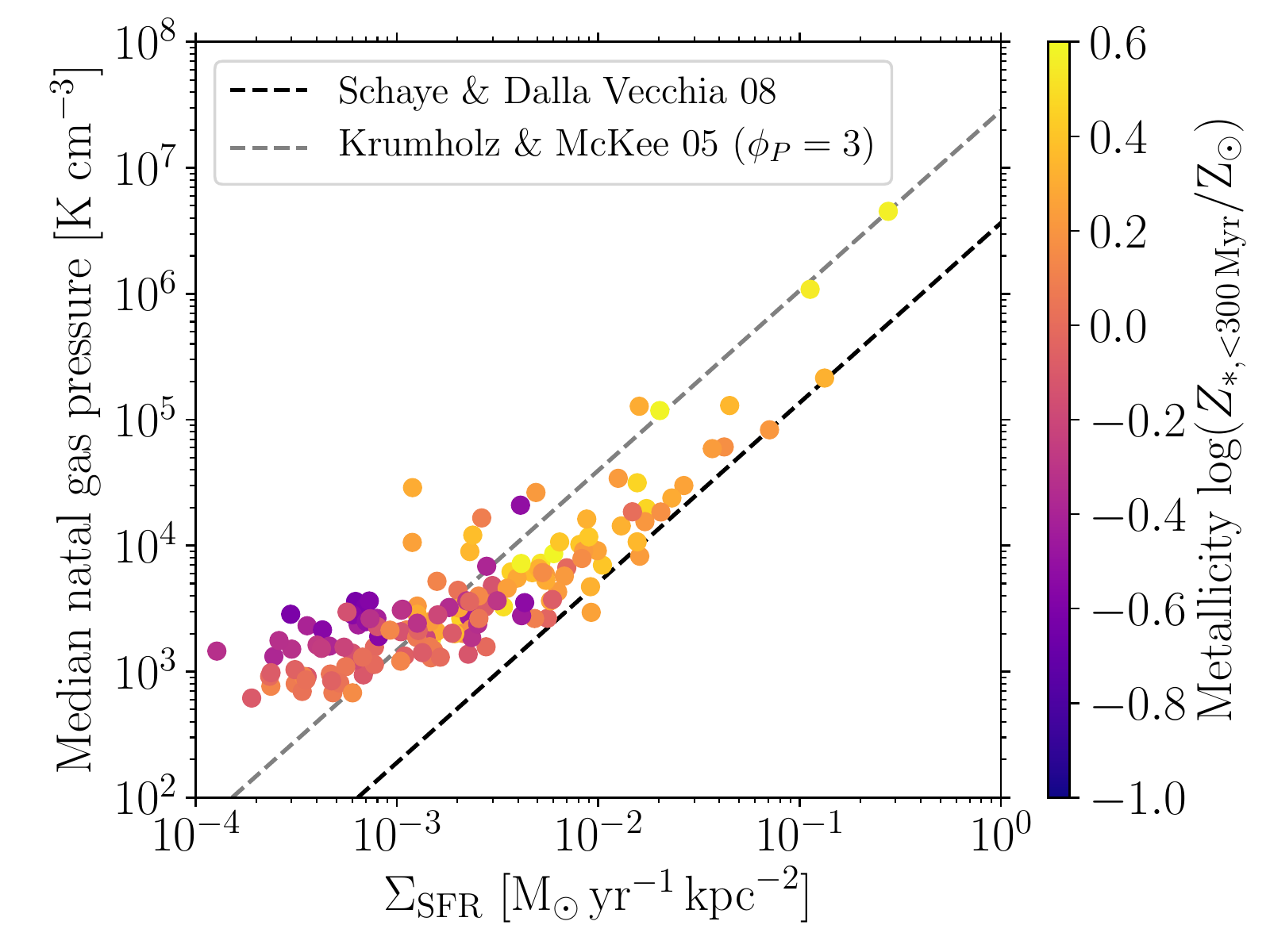}
\caption{Relationships between the CFE, median natal gas pressure of star formation and star formation rate surface density ($\SigmaSFR$) of the simulated galaxies, for star particles younger than $300 \Myr$. Symbols in each panel are coloured by the median metallicity of star particles younger than $300 \Myr$.
\textit{Top:} CFE as a function of the median natal gas pressure. The dash-dotted line shows the relation used at the particle level in the simulations \citep[see figure 3 of][]{P18}.
\textit{Middle:} CFE as a function of the star formation rate surface density ($\SigmaSFR$). The line styles are as in Fig. \ref{fig:CFE}, showing the predictions of the fiducial \citet{Kruijssen_12} model (grey dashed line), the relation shifted to match the pressure-$\SigmaSFR$ relation adopted in the EAGLE simulations \citep{Schaye_and_Dalla_Vecchia_08, S15} and the same model with a modified $\SigmaG$-$\SigmaSFR$ relation \citep{Johnson_et_al_16} (grey dotted line).
\textit{Bottom:} The relationship between the median natal pressure and $\SigmaSFR$ for the galaxies. The grey dashed line shows the relation adopted for the star formation law in EAGLE \citep[i.e. the expected relation for the simulations]{Schaye_and_Dalla_Vecchia_08}, while the black dashed line shows the relation adopted in the fiducial \citet{Kruijssen_12} CFE model \citep[assuming $\phi_P = 3$]{Krumholz_and_McKee_05}.}
  \label{fig:CFE-P-SigmaSFR}
\end{figure}

\begin{figure}
  \centering
  \includegraphics[width=80mm]{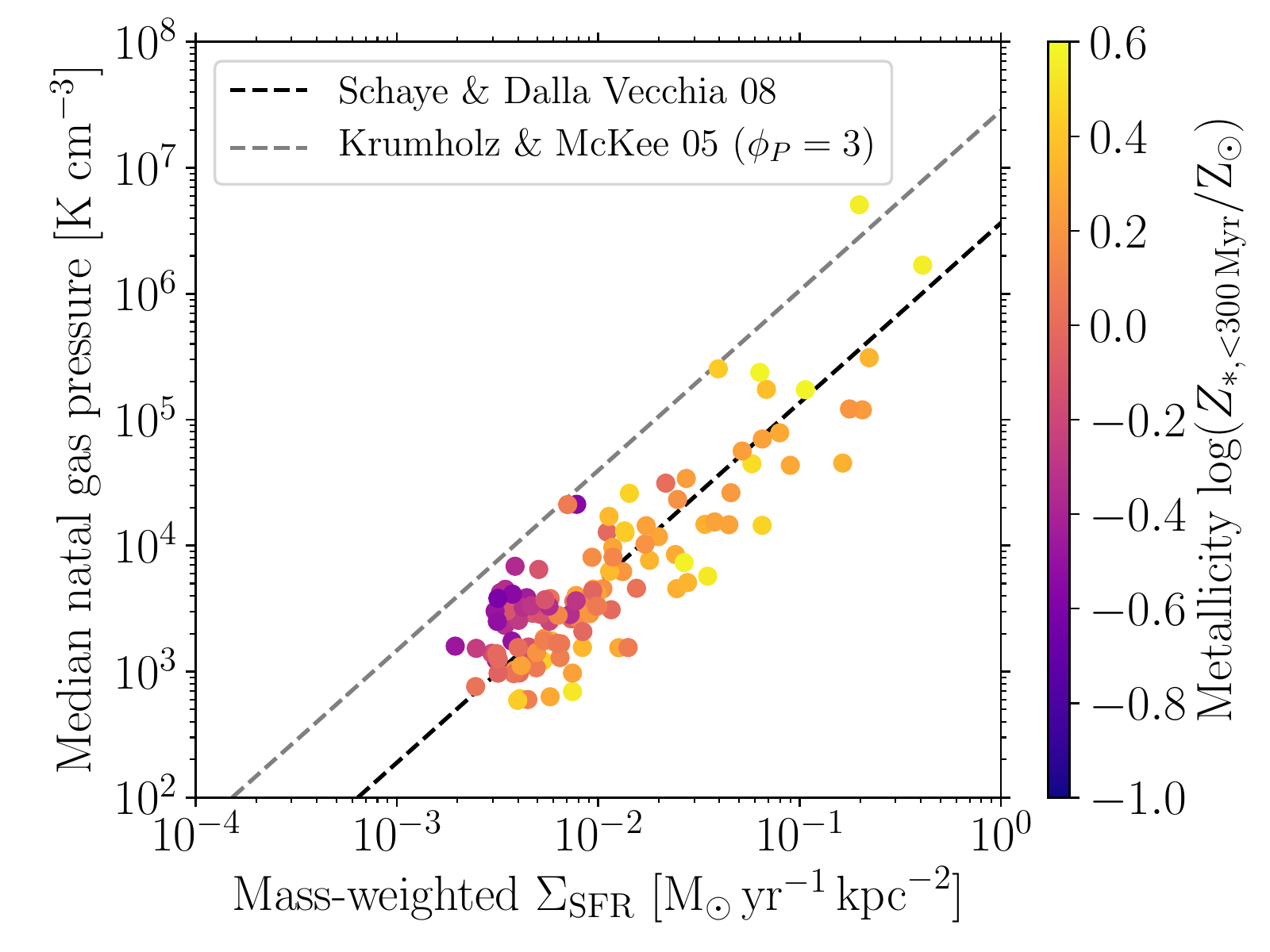}
  \caption{The relationship between the median natal pressure and $\SigmaSFR$ computed using a mass-weighted surface density. The mass-weighted $\SigmaSFR$ is calculated in a grid with regions of $0.7 \times 0.7 \kpc^2$. The dashed lines showing the \citet{Schaye_and_Dalla_Vecchia_08} and \citet{Krumholz_and_McKee_05} relations are as in the lower panel of Fig. \ref{fig:CFE-P-SigmaSFR}.}
  \label{fig:P-SigmaSFR_Mweight}
\end{figure}

In Fig. \ref{fig:CFE-P-SigmaSFR} we compare the relationships between the CFE, natal gas pressure and $\SigmaSFR$ for the simulations.
The top panel of the figure shows the CFE as a function of the natal gas pressure, which is the fundamental relationship underpinning the models. 
The galaxies closely trace the relation between CFE and pressure used at the particle level in the simulations \citep[dash-dotted line in the figure, see][]{P18}, with a small amount of scatter caused by pressure fluctuations within the galaxies. In this parameter space, metallicity plays no role and galaxies of all metallicities follow the relation.

In the middle panel of Fig. \ref{fig:CFE-P-SigmaSFR} we again compare the CFE as a function of $\SigmaSFR$ as in Fig. \ref{fig:CFE}, but with symbols coloured by metallicity. The scatter in this parameter space is significantly larger than for CFE versus pressure (top panel), demonstrating that scatter in the relation between $\SigmaSFR$ and pressure largely determines the scatter in the CFE-$\SigmaSFR$ relation.
We show this explicitly in the bottom panel of Fig. \ref{fig:CFE-P-SigmaSFR}, where we compare the natal gas pressure with $\SigmaSFR$ for the galaxies. 
The black dashed line in this panel shows the expected relationship, which describes the star formation relation used in the EAGLE model \citep{Schaye_and_Dalla_Vecchia_08,S15}.
At $\SigmaSFR \gtrsim 10^{-2} \Msun \pyr \kpc^{-2}$, most galaxies follow the \citet{Schaye_and_Dalla_Vecchia_08} relation. 
However at $\SigmaSFR \lesssim 10^{-2.5} \Msun \pyr \kpc^{-2}$, the galaxies are systematically offset from the relation (and this is also the case for a smaller fraction of galaxies at higher $\SigmaSFR$).
This offset does not depend on the definition for $R_\rmn{lim}$ (Section~\ref{sec:selection}) and occurs for both limits of $1.5 R_{1/2}$ and the radius containing 68 per cent of the recent star formation.
This deviation is largely caused by highly concentrated or substructured star formation, such that star formation occurs in a much smaller area compared to the area for which $\SigmaSFR$ is calculated.
The inverse is not the case (too high pressures for a given $\SigmaSFR$), since smaller apertures can always be chosen to bring the galaxies into better agreement with the expected $P$-$\SigmaSFR$ relation (at the expense of star particle and cluster numbers).
Such an effect can be mitigated by calculating a mass-weighted surface density \citep[see][]{Johnson_et_al_16}.
In Fig. \ref{fig:P-SigmaSFR_Mweight}, we show the natal gas pressure compared with the mass-weighted $\SigmaSFR$ for the simulated galaxies (computed within $R_\rmn{lim}$, as for Fig. \ref{fig:CFE-P-SigmaSFR}).
We calculate the mass-weighted $\SigmaSFR$ in a grid of $0.7 \times 0.7 \kpc^2$ regions \citep[large enough such that incomplete sampling of star-forming regions is not important,][]{Kruijssen_and_Longmore_14}, with each region weighted by the mass of young ($<300 \Myr$) stars.
Using a mass-weighted $\SigmaSFR$ brings the low pressure ($P/k \lesssim 5 \times 10^3 \K \cmcubed$) galaxies in line with the expected $P$-$\SigmaSFR$ relation (with the exception of low-metallicity galaxies, see below), demonstrating the effectiveness of the method in accounting for spatially non-uniform star formation in the galaxies.

A secondary effect causing offset in the pressure-$\SigmaSFR$ relation is due to the metallicity of star formation.
The EAGLE model includes a metallicity-dependent density threshold for star formation, such that star formation must occur at higher densities at lower metallicity \citep{S15}.
This threshold is included to model the effect of the thermogravitational collapse of warm, photoionized interstellar gas into a cold, dense phase, which is expected to occur at lower densities and pressures in metal-rich gas \citep{Schaye_04}.
This higher density threshold at lower metallicities results (through the lower density limit for star formation imposed by the polytropic equation-of-state implemented at high gas densities) in higher pressures of star formation at a given $\SigmaSFR$, which is evident at $\SigmaSFR \lesssim 5 \times 10^{-3} \Msun \pyr \kpc^{-2}$ in the lower panel of Fig. \ref{fig:CFE-P-SigmaSFR} and Fig. \ref{fig:P-SigmaSFR_Mweight}.
The threshold is also responsible for the apparent pressure floor in the figures (at $P/k \sim 10^3 \K \cmcubed$).
Note that this effect of increasing pressure with metallicity is not expected to occur at high $\SigmaSFR$, since the densities of star-forming gas in this regime are well above the metallicity-dependent threshold.
The effect is also more apparent at higher redshifts in the simulations, where the metallicities of star-forming gas in the galaxies are lower.

In the bottom panel of Fig. \ref{fig:CFE-P-SigmaSFR}, we also show the pressure-$\SigmaSFR$ adopted in the fiducial \citet{Kruijssen_12} model \citep{Krumholz_and_McKee_05}.
This relation is offset to lower $\SigmaSFR$ by $\approx0.6$~dex when compared to the relation used in the EAGLE model.
In the middle panel of Fig. \ref{fig:CFE-P-SigmaSFR}, we also show the \citet{Kruijssen_12} relation shifted to higher $\SigmaSFR$ to account for this offset (black dashed line).
The fiducial (grey dashed line) and shifted (black dashed line) relations for the \citet{Kruijssen_12} model thus give an indication of how the uncertainty in the pressure-$\SigmaSFR$ relation results in an uncertainty in the CFE-$\SigmaSFR$ relation.

\section{Truncations of initial and final cluster mass functions} \label{app:init_v_final}

\begin{figure}
  \includegraphics[width=84mm]{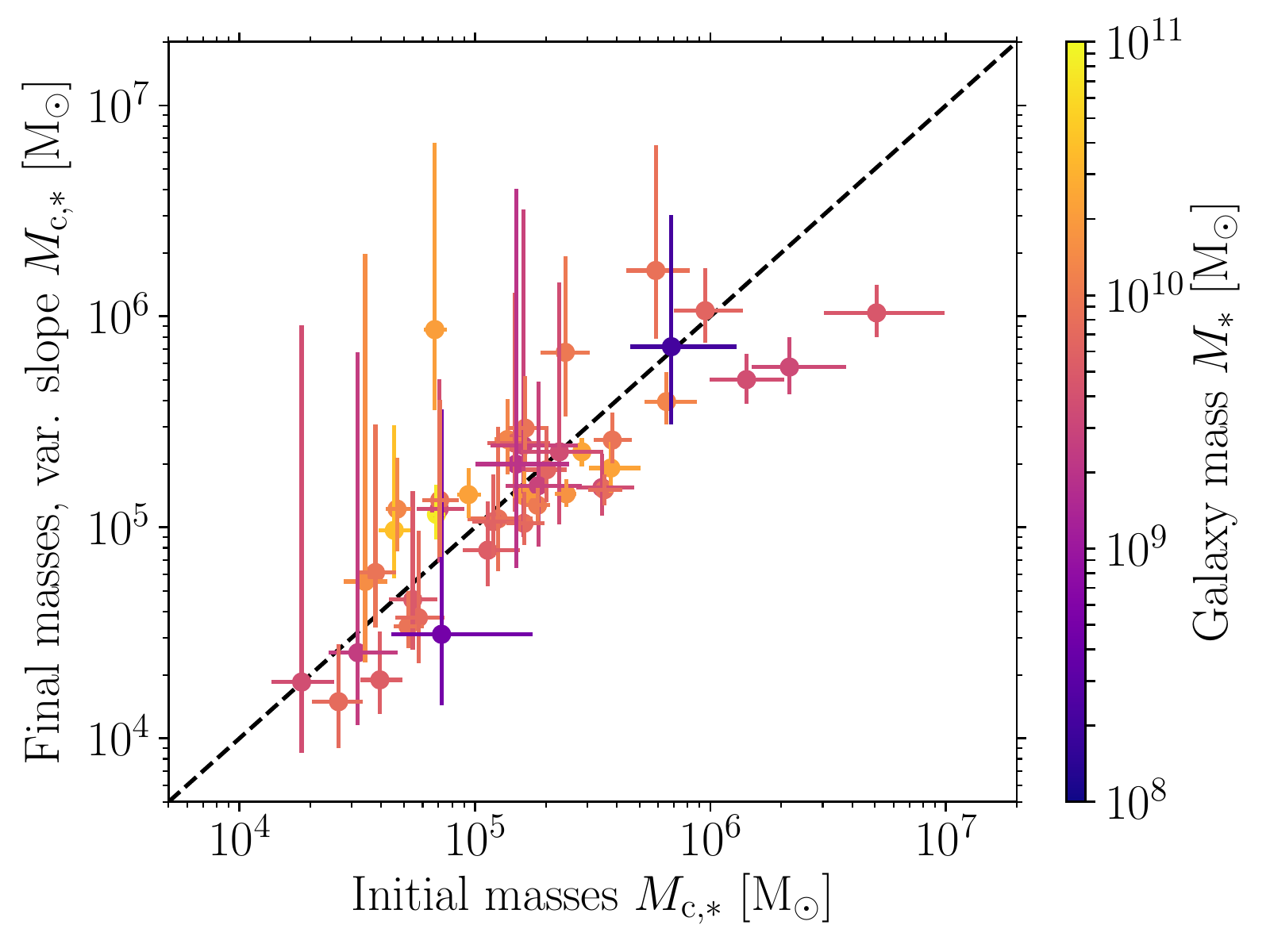}
  \caption{Cluster mass function truncation, $\Mcstar$, when fitting initial versus final (evolved) cluster masses. The points show the median of the posterior distribution from the MCMC fit, while errorbars show the 16 and 84 per cent confidence intervals. Initial cluster masses are fit with a \citet{Schechter_76} function with a constant lower-mass power-law index of $\beta = -2$. Final cluster masses are fit with a Schechter function with a variable lower-mass power-law index. Galaxies from both the $z=0$ and $z=0.5$ snapshots are included to increase galaxy numbers. The dashed line shows the one-to-one relation.}
  \label{fig:Mcstar_init_v_final}
\end{figure}

In Fig. \ref{fig:Mcstar_init_v_final}, we test how our fits of the cluster mass function truncation are affected by cluster mass loss in old, evolved cluster populations. 
We compare the upper exponential truncation mass ($\Mcstar$) resulting from \citet{Schechter_76} function fits to the initial and evolved cluster populations in the simulated galaxies.
For fits to the initial cluster masses, we fix the lower-mass power-law index to $\beta = -2$ (the initial value set in the simulations).
For fits to the final cluster masses, we use a prior on the index $\beta$ between $-3$ and $-0.5$. 
The figure only shows galaxies where the most massive cluster is more massive than $\Mcstar$ (the median of the posterior distribution), for both initial and final cluster mass fits (see the discussion in Section \ref{sec:Mcstar}).
Overall, we find good agreement in $\Mcstar$ between fitting initial and evolved cluster masses, with potentially an offset to higher initial $\Mcstar$. 
However, a small offset to higher initial $\Mcstar$ is expected simply due to stellar-evolutionary mass loss (a factor $\sim 1.3$-$1.4$ for clusters with ages 100-300 Myr, or $0.1$~dex, based on the mass loss in the simulations).
A number of galaxies have large errorbars for the fits to the final cluster masses due to the degeneracy between the truncation mass $\Mcstar$ and index $\beta$.

\section{Relation of $\phi_P$ to galactic properties} \label{app:phiP}

\begin{figure*}
  \includegraphics[width=\textwidth]{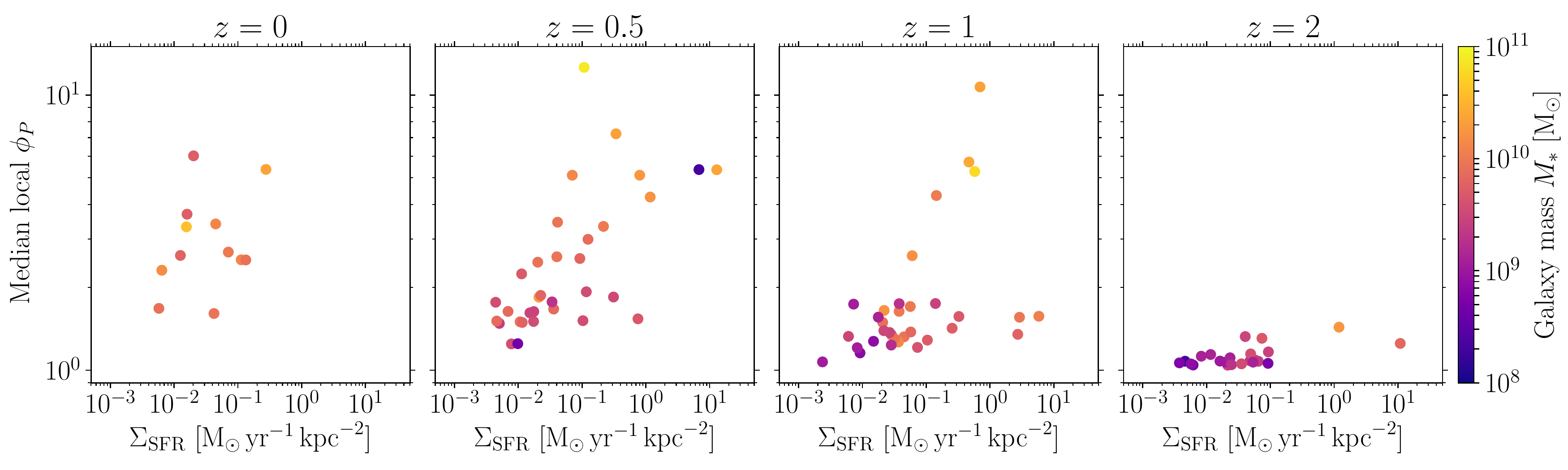}
  \caption{$\phi_P$ as a function of $\SigmaSFR$ at different redshifts. Variations in $\phi_P$ become more important at lower redshifts and at high $\SigmaSFR$. At $z = \{0,0.5,1,2\}$ we find medians of $\phi_P = \{2.5, 1.8, 1.3, 1.1\}$, respectively.}
  \label{fig:phiP-SigmaSFR}
\end{figure*}

\begin{figure*}
  \includegraphics[width=\textwidth]{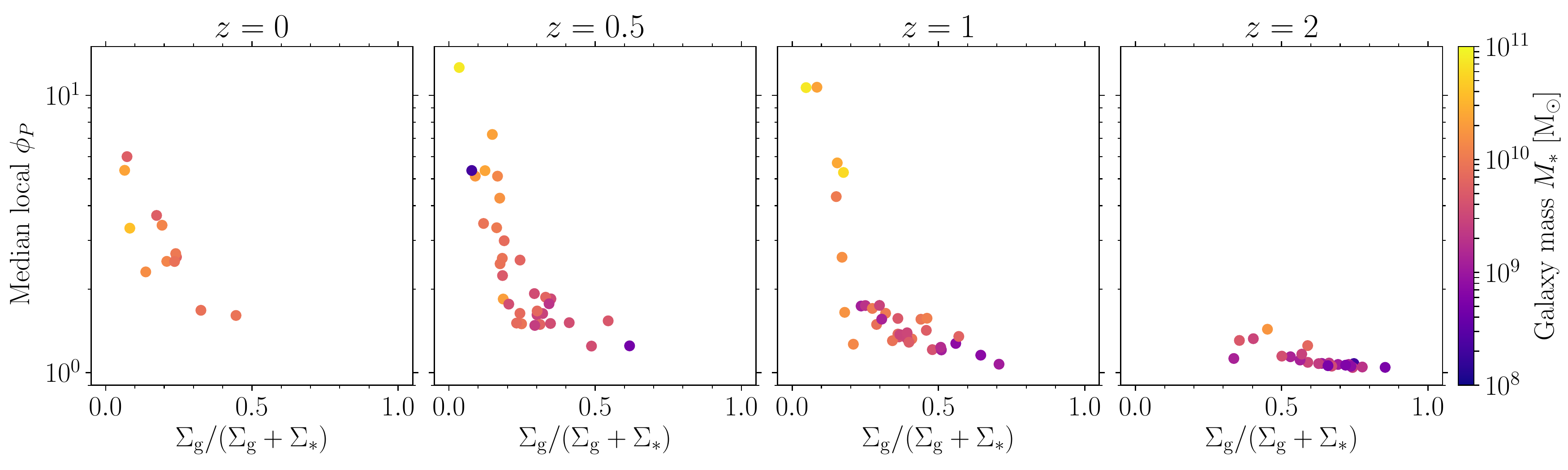}
  \caption{$\phi_P$ as a function of the gas mass fraction $\Sigma_\rmn{g}/(\Sigma_\rmn{g}+\Sigma_\ast)$ at redshifts $z = \{0,0.5,1,2\}$. $\phi_P$ increases at lower redshifts as stellar densities increase relative to gas densities.}
  \label{fig:phiP-gasFrac}
\end{figure*}

In Section \ref{sec:Mcstar} we find that, at a given $\SigmaSFR$, $\Mcstar$ is elevated above the $z=0$ relation at higher redshifts.
We suggest that this effect may be caused by the higher gas fractions in galaxies at higher redshifts, through the parameter $\phi_P$ \citep{Krumholz_and_McKee_05}, and therefore in this section we directly test that suggestion.

In Fig. \ref{fig:phiP-SigmaSFR}, we compare the median local $\phi_P$ for recently formed stars ($<300 \Myr$) with $\SigmaSFR$ for all galaxies in Figs. \ref{fig:Mcstar} and \ref{fig:Mcstar_redshift}.
We find that $\phi_P$ does not directly correlate with $\SigmaSFR$, but instead with the gas fraction, which we show in Fig. \ref{fig:phiP-gasFrac}.
Therefore, at high redshift or low $\SigmaSFR$, we typically find $\phi_P \approx 1$, since the galaxies have low (stellar) mass with high gas fractions.
The median $\phi_P$ increases from $\phi_P \approx 1$ at $z=2$ to $\phi_P \approx 2.5$ at $z=0$, implying a factor of four increase in the Toomre mass \citep[since $\Mtoomre \propto \phi_P^{-1.5}$, see equations 6 and 7 in][]{P18}, and thus in $\Mcstar$ (at the same pressure and epicyclic frequency $\kappa$).
We show this effect in Fig. \ref{fig:Mcstar_redshift} as dashed lines, by scaling the fit to observed local galaxies \citep{Johnson_et_al_16} by the median $\phi_P$ at each redshift.
Therefore, the decrease of the typical $\phi_P$ with increasing redshift explains the elevation of $\Mcstar$ at higher redshifts found for the simulations.


\bsp	
\label{lastpage}
\end{document}